\documentclass[twocolumn]{aastex62}

\usepackage{graphicx}
\usepackage{amsmath}
\usepackage{color}
\usepackage{ulem}
\usepackage{float}
\graphicspath{{fig/}}

\newcommand\hcnone{HCN $J=1 \rightarrow 0$}
\newcommand\hcnthree{HCN $J=3 \rightarrow 2$}
\newcommand\hcnfour{HCN $J=4 \rightarrow 3$}

\newcommand\hcopfour{HCO$^+\ J=4 \rightarrow 3$}
\newcommand\csfive{CS $J=5 \rightarrow 4$}
\newcommand\csseven{CS $J=7 \rightarrow 6$}
\newcommand\coone{CO $J=1 \rightarrow 0$}
\newcommand\cothree{CO $J=3 \rightarrow 2$}
\newcommand\jone{$J=1 \rightarrow 0$}

\newcommand\jthree{$J=3 \rightarrow 2$}
\newcommand\jfour{$J=4 \rightarrow 3$}
\newcommand\jfive{$J=5 \rightarrow 4$}
\newcommand\jsix{$J=6 \rightarrow 5$}
\newcommand\kms{km s$^{-1}$}

\shorttitle{The $L_{\rm dense}-L_{\rm IR}$ correlation on sub-kpc scale in galaxies}
\shortauthors{Tan et al.}
\begin{document}

\title{The MALATANG Survey: the $L_{\rm gas}-L_{\rm IR}$ correlation on sub-kiloparsec scale in six nearby star-forming galaxies as traced by \hcnfour\  and \hcopfour}

\correspondingauthor{Qing-Hua Tan}
\email{qhtan@pmo.ac.cn}

\author{Qing-Hua Tan}
\affil{Purple Mountain Observatory \& Key Laboratory for Radio Astronomy, Chinese Academy of Sciences, 8 Yuanhua Road, Nanjing 210034, China}
%\email{qhtan@pmo.ac.cn}

\author{Yu Gao}
\affiliation{Purple Mountain Observatory \& Key Laboratory for Radio Astronomy, Chinese Academy of Sciences, 8 Yuanhua Road, Nanjing 210034, China}

\author{Zhi-Yu Zhang}
\affiliation{Institute for Astronomy, University of Edinburgh, Royal Observatory, Blackford Hill, Edinburgh EH9 3HJ, UK}
\affiliation{ESO, Karl-Schwarzschild-Str.2, D-85748 Garching, Germany}

\author{Thomas R. Greve}
\affiliation{Department of Physics and Astronomy, University College London, Gower Street, London WC1E6BT, UK}

\author{Xue-Jian Jiang}
\affiliation{Purple Mountain Observatory \& Key Laboratory for Radio Astronomy, Chinese Academy of Sciences, 8 Yuanhua Road, Nanjing 210034, China}

\author{Christine D. Wilson}
\affiliation{Department of Physics and Astronomy, McMaster University, Hamilton, Ontario L8S 4M1, Canada}

\author{Chen-Tao Yang}
\affiliation{Purple Mountain Observatory \& Key Laboratory for Radio Astronomy, Chinese Academy of Sciences, 8 Yuanhua Road, Nanjing 210034, China}
\affiliation{ Institut d'Astrophysique Spatiale, CNRS, Univ. Paris-Sud, Universit$\acute{e}$ Paris-Saclay, B$\hat{a}$t. 121, 91405, Orsay Cedex, France}
\affiliation{European Southern Observatory, Alonso de C{\'o}rdova 3107, Casilla 19001,  Vitacura, Santiago, Chile}

\author{Ashley Bemis}
\affiliation{Department of Physics and Astronomy, McMaster University, Hamilton, Ontario L8S 4M1, Canada}

\author{Aeree Chung}
\affiliation{Department of Astronomy, Yonsei University, 50 Yonsei-ro, Seodaemun-gu, Seoul 03722, Korea}

\author{Satoki Matsushita}
\affiliation{Academia Sinica Institute of Astronomy and Astrophysics, P.O. Box 23-141, Taipei 10617, Taiwan}

\author{Yong Shi}
\affiliation{School of Astronomy and Space Science, Nanjing University, Nanjing 210093, China}

\author{Yi-Ping Ao}
\affiliation{National Astronomical Observatory of Japan, 2-21-1 Osawa, Mitaka, Tokyo 181-8588, Japan}
\affiliation{Purple Mountain Observatory \& Key Laboratory for Radio Astronomy, Chinese Academy of Sciences, 8 Yuanhua Road, Nanjing 210034, China}

\author{Elias Brinks}
\affiliation{Centre for Astrophysics Research, University of Hertfordshire, College Lane, Hatfield AL10 9AB, UK}

\author{Malcolm J. Currie}
\affiliation{RAL Space,
Rutherford Appleton Laboratory,
Harwell Campus​,
Didcot,
Oxfordshire,
OX11 0QX,
United Kingdom}

\author{Timothy A. Davis}
\affiliation{School of Physics and Astronomy, Cardiff University, Queen's Buildings, The Parade, Cardiff CF24 3AA, UK}

\author{Richard de Grijs}
\affiliation{Kavli Institute for Astronomy \& Astrophysics and Department of Astronomy, Peking University, Yi He Yuan Lu 5, Hai Dian District, Beijing 100871, China}
\affiliation{Department of Physics and Astronomy, Macquarie University, Balaclava Road, North Ryde, NSW 2109, Australia}
\affiliation{International Space Science Institute--Beijing, 1 Nanertiao, Zhongguancun, Hai Dian District, Beijing 100190, China}

\author{Luis C. Ho}
\affiliation{Kavli Institute for Astronomy \& Astrophysics and Department of Astronomy, Peking University, Yi He Yuan Lu 5, Hai Dian District, Beijing 100871, China}

\author{Masatoshi Imanishi}
\affiliation{National Astronomical Observatory of Japan, 2-21-1 Osawa, Mitaka, Tokyo 181-8588, Japan}

\author{Kotaro Kohno}
\affiliation{Institute of Astronomy, Graduate School of Science, The University of Tokyo, Osawa, Mitaka, Tokyo 181-0015, Japan}
\affiliation{Research Center for the Early Universe, Graduate School of Science, The University of Tokyo, 7-3-1 Hongo, Bunkyo-ku, Tokyo 113-0033, Japan}

\author{Bumhyun Lee}
\affiliation{Department of Astronomy, Yonsei University, 50 Yonsei-ro, Seodaemun-gu, Seoul 03722, Korea}

%\author{Padelis P. Papadopoulos}
%\affiliation{School of Physics and Astronomy, Cardiff University, Queen's Buildings, The Parade, Cardiff CF24 3AA, UK}
%\affiliation{Research Center for Astronomy, Academy of Athens, Soranou Efesiou 4, GR-115 27 Athens, Greece}

\author{Harriet Parsons}
\affiliation{East Asian Observatory, 660 N. A'oh$\bar{o}$k$\bar{u}$ Place, University Park, Hilo, Hawaii 96720-2700, USA}

\author{Mark G. Rawlings}
\affiliation{East Asian Observatory, 660 N. A'oh$\bar{o}$k$\bar{u}$ Place, University Park, Hilo, Hawaii 96720-2700, USA}

\author{Dimitra Rigopoulou}
\affiliation{Department of Physics, University of Oxford, Oxford, OX1 3RH, UK}

\author{Erik Rosolowsky}
\affiliation{University of Alberta, 116 St \& 85 Ave, Edmonton, AB T6G 2R3, Canada}

\author{Joanna Bulger}
\affiliation{Subaru Telescope, NAOJ, 650 N. A'oh$\bar{o}$k$\bar{u}$ Place, Hilo, HI 96720, USA}
\affiliation{Institute for Astronomy Maui, University of Hawaii, 34 Ohia Ku St., Pukalani, HI, 96768, USA}

\author{Hao Chen}
\affiliation{School of Astronomy and Space Science, Nanjing University, Nanjing 210093, China}

\author{Scott C. Chapman}
\affiliation{Department of Physics and Atmospheric Science, Dalhousie University, Halifax, NS B3H 4R2, Canada}

\author{David Eden}
\affiliation{Astrophysics Research Institute, Liverpool John Moores University, IC2, Liverpool Science Park, 146 Brownlow Hill, Liverpool, L3 5RF, UK}

\author{Walter K. Gear}
\affiliation{College of Physical Sciences and Engineering, Cardiff University, Cardiff  CF24 3AA, UK}

\author{Qiu-Sheng Gu}
\affiliation{School of Astronomy and Space Science, Nanjing University, Nanjing 210093, China}

\author{Jin-Hua He}
\affiliation{Key Laboratory for the Structure and Evolution of Celestial Objects, Yunnan Observatories, Chinese Academy of Sciences, 396 Yangfangwang, Guandu District, Kunming, 650216, China}
\affiliation{Chinese Academy of Sciences South America Center for Astronomy, China-Chile Joint Center for Astronomy, Camino El Observatorio \#1515, Las Condes, Santiago, Chile}

\author{Qian Jiao}
\affiliation{Purple Mountain Observatory \& Key Laboratory for Radio Astronomy, Chinese Academy of Sciences, 8 Yuanhua Road, Nanjing 210034, China}

\author{Dai-Zhong Liu}
\affiliation{Max Planck Institute for Astronomy, K$\ddot{o}$nigstuhl 17, D-69117 Heidelberg, Germany}
\affiliation{Purple Mountain Observatory \& Key Laboratory for Radio Astronomy, Chinese Academy of Sciences, 8 Yuanhua Road, Nanjing 210034, China}

\author{Li-Jie Liu}
\affiliation{Department of Physics, University of Oxford, Oxford, OX1 3RH, UK}
\affiliation{Purple Mountain Observatory \& Key Laboratory for Radio Astronomy, Chinese Academy of Sciences, 8 Yuanhua Road, Nanjing 210034, China}

\author{Xiao-Hu Li}
\affiliation{National Astronomical Observatories, Chinese Academy of Sciences, Beijing 100012, China}

\author{Micha{\l}~J.~Micha{\l}owski}
\affiliation{Astronomical Observatory Institute, Faculty of Physics, Adam Mickiewicz University, ul.~S{\l}oneczna 36, 60-286 Pozna{\'n}, Poland}

\author{Quang Nguyen-Luong}
\affiliation{NAOJ Chile Observatory, National Astronomical Observatory of Japan, 2-21-1 Osawa, Mitaka, Tokyo 181-8588, Japan}
\affiliation{Korea Astronomy and Space Science Institute, 776 Daedeok daero, Yuseoung, Daejeon 34055, Korea}
\affiliation{CITA, University of Toronto, 60 St. George Street, Toronto, ON M5S 3H8, Canada}

\author{Jian-Jie Qiu}
\affiliation{School of Astronomy and Space Science, Nanjing University, Nanjing 210093, China}

\author{Matthew W. L. Smith}
\affiliation{School of Physics and Astronomy, Cardiff University, Queen's Buildings, The Parade, Cardiff CF24 3AA, UK}

\author{Giulio Violino}
\affiliation{Centre for Astrophysics Research, University of Hertfordshire, College Lane, Hatfield AL10 9AB, UK}

\author{Jian-Fa Wang}
\affil{Purple Mountain Observatory \& Key Laboratory for Radio Astronomy, Chinese Academy of Sciences, 8 Yuanhua Road, Nanjing 210034, China}

\author{Jun-Feng Wang}
\affiliation{Department of Astronomy, Xiamen University, 422 Siming South Road, Xiamen 361005, China }

\author{Jun-Zhi Wang}
\affiliation{Shanghai Astronomical Observatory, Chinese Academy of Sciences, 80 Nandan Road, Shanghai 200030, China}

\author{Sherry Yeh}
\affiliation{Subaru Telescope, NAOJ, 650 N. A'oh$\bar{o}$k$\bar{u}$ Place, Hilo, HI 96720, USA}
\affiliation{W. M. Keck Observatory, 65-1120 Mamalahoa Hwy, Kamuela, HI 96743, USA}

\author{Ying-He Zhao}
\affiliation{Yunnan Observatories, Chinese Academy of Sciences, Kunming 650011, China}

\author{Ming Zhu}
\affiliation{National Astronomical Observatories, Chinese Academy of Sciences, Beijing 100012, China}

%\author{et al.}
%\affiliation{}

%% Note that the \and command from previous versions of AASTeX is now
%% depreciated in this version as it is no longer necessary. AASTeX 
%% automatically takes care of all commas and "and"s between authors names.

%% AASTeX 6.1 has the new \collaboration and \nocollaboration commands to
%% provide the collaboration status of a group of authors. These commands 
%% can be used either before or after the list of corresponding authors. The
%% argument for \collaboration is the collaboration identifier. Authors are
%% encouraged to surround collaboration identifiers with ()s. The 
%% \nocollaboration command takes no argument and exists to indicate that
%% the nearby authors are not part of surrounding collaborations.

%% Mark off the abstract in the ``abstract'' environment. 
\begin{abstract}
We present \hcnfour\ and \hcopfour\ maps of six nearby star-forming galaxies, NGC 253, NGC 1068, IC 342, M82, M83, and NGC 6946, obtained with the James Clerk Maxwell Telescope as part of the MALATANG survey. All galaxies were mapped in the central $2\arcmin \times 2\arcmin$ region at 14$\arcsec$ (FWHM) resolution (corresponding to linear scales of $\sim0.2-1.0$ kpc). The $L_{\rm IR}-L'_{\rm dense}$ relation, where the dense gas is traced by the \hcnfour\ and the \hcopfour\ emission, measured in our sample of spatially-resolved galaxies is found to follow the linear correlation established globally in galaxies within the scatter. We find that the luminosity ratio, $L_{\rm IR}/L'_{\rm dense}$, shows systematic variations with $L_{\rm IR}$ within individual spatially resolved galaxies, whereas the galaxy-integrated ratios vary little. A rising trend is also found between $L_{\rm IR}/L'_{\rm dense}$ ratio and the warm-dust temperature gauged by the \mbox{70 $\mu$m}/\mbox{100 $\mu$m} flux ratio. We find the luminosity ratios of IR/HCN(4-3) and IR/HCO$^+$(4-3), which can be taken as a proxy for the efficiency of star formation in the dense molecular gas (SFE$_{\rm dense}$), appears to be nearly independent of the dense-gas fraction ($f_{\rm dense}$) for our sample of galaxies. The SFE of the total molecular gas (SFE$_{\rm mol}$) is found to increase substantially with $f_{\rm dense}$ when combining our data with that on local (ultra)luminous infrared galaxies and high-$z$ quasars. The mean $L'_{\rm HCN(4-3)}/L'_{\rm HCO^+(4-3)}$ line ratio measured for the six targeted galaxies is 0.9$\pm$0.6. No significant correlation is found for the $L'_{\rm HCN(4-3)}/L'_{\rm HCO^+(4-3)}$ ratio with the SFR as traced by $L_{\rm IR}$, nor with the warm-dust temperature, for the different populations of galaxies.
\end{abstract}

%% Keywords should appear after the \end{abstract} command. 
%% See the online documentation for the full list of available subject
%% keywords and the rules for their use.
%\keywords{editorials, notices --- 
%miscellaneous --- catalogs --- surveys}
\keywords{galaxies: ISM --- galaxies: star formation --- infrared: galaxies --- ISM: molecules --- radio lines: galaxies}

%% From the front matter, we move on to the body of the paper.
%% Sections are demarcated by \section and \subsection, respectively.
%% Observe the use of the LaTeX \label
%% command after the \subsection to give a symboli\includegraphics[]{resolved_SFL_v2.pdf}
%% subsection for cross-referencing in a \ref command.
%% You can use LaTeX's \ref and \label commands to keep track of
%% cross-references to sections, equations, tables, and figures.
%% That way, if you change the order of any elements, LaTeX will
%% automatically renumber them.

%% We recommend that authors also use the natbib \citep
%% and \citet commands to identify citations.  The citations are
%% tied to the reference list via symbolic KEYs. The KEY corresponds
%% to the KEY in the \bibitem in the reference list below. 

\section{Introduction}\label{sec:intro}

In the last two decades we have seen significant advances in our understanding of the relationship between star formation and the interstellar gas, from which stars form, in large part thanks to galaxy surveys at (sub)millimeter bands of multiple molecular species \citep[e.g.,][]{kennicutt98,gao04a,gao04b,bigiel08,leroy08,baan08,gracia08,daddi10,shi11,shi18,garcia12,zhang14,greve14,lu14,usero15}. It has become clear that the molecular gas, rather than the atomic gas, is the raw material for star formation. The Kennicutt-Schmidt (K-S) law that relates the global surface densities of star-formation rate (SFR) and that of total gas including atomic and molecular gas (traced by H{\sc i} 21 cm line and rotational lines of CO, respectively), is characterized by a power-law index of $n\approx 1.4$ ($\Sigma_{\rm SFR}\propto \Sigma^n_{\rm gas}$) \citep[see][and reference therein]{kennicutt98,kennicutt12}. The super linear slope derived in this empirical scaling relation suggests that the star-formation efficiency (SFE) indicated by the SFR per unit mass of molecular gas increases with SFR. 

\citet{bigiel08} examined the resolved K-S law on sub-kpc scale ($\sim$ 750 pc) in nearby spiral and dwarf galaxies, and did not find significant correlation between $\Sigma_{\rm H{\sc I}}$ and $\Sigma_{\rm SFR}$, as the H{\sc i} surface density is shown to saturate at about 9 $M_\odot\ {\rm pc^{-2}}$. In contrast, a slope of unity relates $\Sigma_{\rm SFR}$ and $\Sigma_{\rm H_2}$ for normal and dwarf galaxies \citep{schruba13}. However, the slope steepens for the $\Sigma_{\rm SFR}-\Sigma_{\rm H_2}$ relation if we expand the sample to include galaxies with extreme starbursts, such as luminous infrared galaxies (LIRGs, $10^{11} L_\odot\leqslant L_{\rm IR}<10^{12} L_\odot$) and ultraluminous infrared galaxies (ULIRGs, $L_{\rm IR}\geqslant10^{12}L_\odot$), which is evident both from observations \citep[e.g.,][]{gao04a,daddi10,lliu15} and theoretical predictions \citep[e.g.,][]{krumholz05,elmegreen15,elmegreen18}. In addition, a breakdown of the K-S law is found at giant molecular cloud (GMC) scales of a few tens of pc \citep[e.g.,][]{onodera10,nguyen16}, which is attributed to the dynamical evolution of GMCs and the drift of young clusters from their GMCs.

A large \hcnone\ survey in nearby spiral galaxies and (U)LIRGs performed by \citet{gao04a,gao04b} revealed a tight linear correlation between the infrared (IR) and the HCN luminosities for normal star-forming galaxies and starbursts. This linearity seemingly extends down to the scale of Galactic massive cores in the Milky Way and holds over a total range of luminosity of about eight orders of magnitude \citep{wu05}. These results imply that the dense molecular gas (i.e., $n({\rm H_2})\gtrsim 7\times 10^4\ {\rm cm^{-3}}$) as traced by the \hcnone\ line, rather than the total molecular gas, is the direct fuel for star formation. High-resolution simulation of dense clouds found that the HCN luminosity is related to mass of dense gas of $\gtrsim 10^4$ cm$^{-3}$ \citep{onus18}. In addition, {\it Spitzer} studies of Galactic molecular clouds also show evidence that star formation is restricted to the dense cores of GMCs \citep[e.g.,][]{evans08,lada10}. The critical density $n_{\rm crit}$\footnote{The critical density of rotational level $j$ is defined as $n_{\rm crit}(j)=\frac{\sum_{j>j'}A_{j\rightarrow j'}}{\sum_{j\neq j'}C_{j\rightarrow j'}(T_{\rm kin})}$, where $A_{j\rightarrow j'}$ is the Einstein coefficient for spontaneous emission, $A_{j\rightarrow j'}\propto \mu^2 \nu^3$ in units of s$^{-1}$, $C_{j\rightarrow j'}$ is the collision rate coefficient that depends on the gas temperature and in units of cm$^3$ s$^{-1}$. All critical densities in this work are calculated on the assumption of $T_{\rm kin}=100$ K and optically thin conditions. The critical densities will decrease if the lines are optically thick.} of rotational transitions is proportional to $\mu^2 \nu^3$ (for optically thin lines at frequency $\nu$; $\mu$ is the dipole moment of the molecule); therefore molecules with high dipole moment are expected to trace high-density molecular gas \citep[e.g., $\mu_{\rm HCN}\sim$ 2.98 D, $\mu_{\rm HCO^+}\sim$ 3.93 D, and  $\mu_{\rm CS}\sim$ 1.96 D versus $\mu_{\rm CO}\sim$ 0.11 D; see][]{schoier05}. Subsequently, a number of studies have explored the link between molecular lines of dense gas (e.g., HCN, HCO$^+$, and CS) and IR luminosities in different population of galaxies \citep[e.g.,][]{gao07,papadopoulos07,baan08,liu10,wu10,wang11,garcia12,zhang14,chen15,usero15,liu16}. 

While all of these studies generally agree that there is a close link between the dense gas and star formation, the exact nature of the relation is less clear. Specifically, it is not clear whether the star formation efficiency, as gauged by IR/HCN, is indeed universal from GMCs to distant starburst galaxies. While log-linear fits over eight decades in luminosity have been claimed as evidence of such a universality, claims to the contrary have also been made. For example, local (U)LIRGs have been observed to have 3-4 times higher IR/HCN(1-0) ratios than normal galaxies \citep{gracia08,garcia12}, which would suggest a slightly super-linear IR-HCN relation. Furthermore, resolved studies of dense-gas tracers in nearby galaxies have revealed systematic change in IR/HCN(1-0) with galactocentric radius \citep[e.g.,][]{chen15,bigiel16}.

The physical processes that can affect the observed $L_{\rm IR}-L'_{\rm dense}$ relation fall into two categories. One category contains the physical mechanisms that might compromise the ability of a given molecular transition(e.g., \hcnone) to trace the dense gas in a consistent manner that can be calibrated. For example, significant enhancements in the HCN abundance due to X-ray driven chemistry on large scales, such as might be found in AGN \citep[e.g.,][]{lepp96,kohno01}. The HCN abundance is also thought to be enhanced in hot cores and high-temperature chemistry regions driven by shock heating. Furthermore, it is predicted to be sensitive to the gas phase metallicity \citep[e.g.,][]{bayet12,davis13,braine17}. In the other category we find physical process that would affect the star formation efficiency. Recent observations of Orion A show that the \hcnone\ emission can trace gas with a characteristic H$_2$ density that is about two orders of magnitude below the value commonly adopted \citep{kauffmann17}, which argues that HCN may also can be excited through collisions with electrons\citep{goldsmith17}. It has been argued that the \hcnone\ line could be enhanced by infrared pumping through a vibrational transition at \mbox{14 $\mu$m} near strong mid-infrared sources \citep[e.g.,][]{aalto95,aalto12,gracia06}. In addition, self-absorption of HCN emission has been observed in the Galactic Center and in the compact obscured nuclei of nearby (U)LIRGs \citep{mills13,mills17,aalto15}, thereby rendering this line useless as a probe of the star-forming conditions in the center.

It is still not fully understood how the physical properties of molecular clouds affect the star-formation process in galaxies. In a theoretical study of the star-formation relation, \citet{krumholz05} and \citet{krumholz07} derive a turbulence-regulated model for understanding the K-S law. They propose that the star formation is controlled by the free-fall timescale of the gas, and that the slope of the IR--molecular line luminosity correlations, i.e., the relation between the SFR as probed by the IR luminosity and the molecular emission line, depends on the average gas density of the molecular clouds and the critical density of the molecular line. Similarly, non-local thermodynamic equilibrium radiative transfer calculations with hydrodynamical simulations of galaxies predict decreasing power-law indices of the SFR--molecular line luminosity relation with increasing $n_{\rm crit}$, due to the increase of subthermal emission of the gas tracers in galaxies \citep{narayanan08}. 

\citet{lada10,lada12} argue that the rate of star formation in a molecular cloud or galaxy does not depend on the overall average gas density, but only on the amount of molecular gas above a certain volume density or column density thresholds \citep[i.e., $n$(H$_2$)$\geqslant 10^4\ {\rm cm^{-3}}$; see also][]{heiderman10}. Observations of local molecular clouds by \citet{evans14} show evidence to support this density threshold model and find that the free-fall time {\bf ($t_{\rm ff}$)} is irrelevant to the SFR on small scales of a few pc within molecular clouds. \citet{zhang14} observed \hcnfour, \hcopfour, and \csseven\ in 20 nearby star-forming galaxies and found tight linear correlations of IR--molecular line luminosities for all three gas tracers that probe molecular gas with density higher than $10^6\ {\rm cm^{-3}}$, consistent with those found for \hcnone\ \citep[e.g.,][]{gao04a,gao04b,wu05} and \csfive\ \citep[e.g.,][]{wang11} observations, indicating that the free-fall time scale is likely irrelevant to the SFR on global scales for gas with densities $\gtrsim 10^4$ cm$^{-3}$. They argue that the shorter $t_{\rm ff}$ for the denser gas would not keep $L'_{\rm dense}-L_{\rm IR}$ linear if the $\Sigma_{\rm dense}/t_{\rm ff}-\Sigma_{\rm SFR}$ correlations are linear for all of the dense gas, since the gas content as traced by molecule at high-$J$ (e.g., \hcnfour) has a shorter $t_{\rm ff}$ than that at low-$J$ (e.g., \hcnone) with a lower critical density because $t_{\rm ff}\propto \rho^{-1/2}$. However, it is important to remember that the correlation of $\Sigma_{\rm dense}-\Sigma_{\rm SFR}$ is subject to uncertainties in the conversion from $L'_{\rm dense}$ to the gas mass and from $L_{\rm IR}$ to the SFR. In addition, the {\it Herschel} study of a large sample of nearby galaxies and Galactic clouds in mid-to-high-$J$ ($J=4\rightarrow3$ to $12\rightarrow11$) CO transitions by \citet{dliu15} found that all nine CO transitions are linearly correlated with IR luminosities over a luminosity range of about 14 orders of magnitude, from high-{\it z} star-forming galaxies down to Galactic young stellar objects. Recent observations of Galactic clouds also found that the dense-gas SFE is remarkably constant over a wide range of scales (i.e., from $\sim1-10$ pc to $>$10 kpc) and far-ultraviolet radiation environments \citep{shimajiri17}.

Up to now, observations of dense gas in galaxies have been mainly performed on the central nuclear regions of nearby galaxies with a single pointing or in local (U)LIRGs with global measures. Observations of dense-gas tracers toward the outer disks of galaxies that are more quiescent and relatively weaker in gas emission are still scarce.  \citet{chen15} presented a \hcnone\ map of M51 that covers a $4\arcmin \times 5\arcmin$ region. They found that the outer disk regions of M51 on a kpc-scale follow the IR-HCN relation established globally in galaxies within the scatter and these regions bridge the luminosity gap between GMCs and galaxies. Maps of M51 in HCN, HCO$^+$, and HNC \jone\ emission were also shown in \citet{bigiel16}. Both studies show that \hcnone\ is enhanced with respect to the IR emission in the nuclear region of M51 compared to the outer disk. These are consistent with the results reported by \citet{kohno96}, who found the HCN emission is enhanced compared to the CO emission in the central region ($<$200 pc). It has been suggested that the enhancement of the HCN abundance at the nucleus of M51 could be attributed to the shock produced by the interaction between AGN jets and molecular gas \citep{matsushita15}. \hcnone\ observations in several off-nuclear positions of nearby galaxies by \citet{usero15} show a systematic variation of the SFR per unit dense-gas mass with both the H$_2$ and stellar mass surface densities, which they argue is more consistent with models of turbulence-regulated star formation than with density threshold models.

In this work we present new mapping observations of six nearby star-forming galaxies, NGC~253, NGC~1068, IC~342, M82, M83, NGC~6946, in the \jfour\ lines of HCN and HCO$^+$. These observations were completed in the early stages of the MALATANG (Mapping the dense molecular gas in the strongest star-forming galaxies; Zhang et al. 2018, in prep.) survey with the James Clerk Maxwell Telescope (JCMT). In the MALATANG survey,  we select these six targets to be mapped in the central regions because of their broad distribution, strong emission lines of molecular gas, and concentrated star-formation activity in the nuclear regions. Table~\ref{tab:galaxies} outlines some of the basic physical properties of these six galaxies. These are the first \hcnfour\ and \hcopfour\ spatially resolved observations toward the central $2\arcmin \times 2\arcmin$ region (i.e., $\sim 2-9$ kpc) of these nearby galaxies to date. Three of our sample galaxies, i.e., NGC 253, NGC 1068, and M82, have previously been mapped in the \jfour\ lines of HCN and the HCO$^+$ \citep[e.g.,][]{seaquist00,knudsen07,garcia14,krips11} but over on smaller regions.  All six galaxies have been mapped in \coone\ by the 45 m telescope of the Nobeyama Radio Observatory (NRO) with almost the same angular resolution as the JCMT \citep[e.g.,][]{nakai87,sorai00,kuno07,salak13}, providing an excellent comparison with the total H$_2$ gas.

%%%%%%%%%%- Table-1 -%%%%%%%%%%
\begin{deluxetable*}{lrrccCccclc}[htbp]
\tablecaption{The basic properties of the galaxies in the MALATANG sample observed in jiggle-map mode\label{tab:galaxies}}
\addtolength{\tabcolsep}{-2.5pt}
\tablewidth{0pt}
\tablehead{
\colhead{Source} & \colhead{R.A.(J2000)} & \colhead{Decl.(J2000)} & \colhead{$V_{\rm hel}$\tablenotemark{a}} & \colhead{$D$\tablenotemark{b}} & \colhead{$D_{\rm 25}$\tablenotemark{c}} & \colhead{P.A.\tablenotemark{d}} & \colhead{Inclination} & \colhead{Spatial Scale} & \colhead{Type\tablenotemark{e}} & \colhead{Reference\tablenotemark{f}} \\
\colhead{} & \colhead{} & \colhead{} & \colhead{(km s$^{-1}$)} & \colhead{(Mpc)} & \colhead{(arcmin)} & \colhead{(deg)} &\colhead{(deg)} & \colhead{(1$\arcsec$)} & \colhead{} & \colhead{}
}
\startdata
NGC 253 & 00 47 33.1 & $-$25 17 19.7 & 243 & 3.5 & 27.5\times6.8 & 51 & 76 & 17 pc & SAB(s)c, SF & 1 \\
NGC 1068 & 02 42 40.8 & $-$00 00 47.8 & 1137 & 15.7 & 7.1\times6.0 & 90 & 43 & 76 pc & (R)SA(rs)b, AGN & \nodata \\
IC 342 & 03 46 48.5 & +68 05 46.0 & 31 & 3.4 & 21.4\times20.9 & 0 & 25 & 16 pc & SAB(rs)cd, SF & 2 \\
M82 & 09 55 52.4 & +69 40 46.9 & 203 & 3.5 & 11.2\times4.3 & 65 & 66 & 17 pc & \uppercase\expandafter{\romannumeral1}0 sp, SF & 3 \\
M83 & 13 37 00.9 & $-$29 51 56.0 & 513 & 4.8 & 12.9\times11.5 & 45 & 27 & 23 pc & SAB(s)c, SF & 1 \\
NGC 6946 & 20 34 52.3 & +60 09 13.2 & 40 & 4.7 & 11.5\times9.8 & 19 & 30 & 23 pc & SAB(rs)cd, SF & 4,5 \\
\enddata
\tablenotetext{a}{Heliocentric velocity drawn from NASA/IPAC Extragalactic Database (NED).}
\tablenotetext{b}{Source distance. For NGC 1068, the distance is calculated using $H_0=71\ {\rm km\ s^{-1}\ Mpc^{-1}}$ corrected for the Virgo infall motion. The distances of the remaining galaxies are recent values from the literature. See the last column for the reference.}
\tablenotetext{c}{Major and minor diameters of the galaxies, which is an optical size measured at the 25$^\mathrm{th}$-magnitude isophote in the blue band.}
\tablenotetext{d}{Position angle of the major axis of the galaxy, except for NGC 1068 and IC 342. This is used in mapping HCN and HCO$^+$ emission.}
\tablenotetext{e}{Galaxy types from NED. SF denotes galaxies that are star-forming without an AGN. NGC~1068 is a barred spiral galaxy hosting a Seyfert 2 type AGN.}
\tablenotetext{f}{Reference for the source distance. (1) \citet{radburn11}; (2) \citet{wu14}; (3) \citet{dalcanton09}; (4) \citet{poznanski09}; and (5) \citet{olivares10}.}
\end{deluxetable*}
%%%%%%%%%%%%%%%%%%%%%%%%%%

We describe our JCMT observations and the data reduction, and the processing of ancillary data in Section~\ref{sec:data}. Section~\ref{sec:results} presents the spectra and luminosity measurements. In Section~\ref{sec:correlation} the relationships between the dense-molecular-gas tracers and the star-formation properties are presented and discussed. In Section~\ref{sec:ratio}, we present the 
HCN to HCO$^+$ \jfour\ line ratio, and discuss possible explanations for the variations of line ratios in different populations of galaxies. Our main results are summarized in Section~\ref{sec:summary}. We adopt cosmological parameters of $H_0=71\ {\rm km\ s^{-1}\ Mpc^{-1}}$, $\Omega_{\rm M}=0.27$, $\Omega_\Lambda=0.73$ throughout this work \citep{spergel07}.

\section{Observations and data reduction} \label{sec:data}

\subsection{JCMT HCN(4$-$3) and HCO$^+$(4$-$3) Data}\label{subsec:jcmt}
Zhang et al. (2018, in prep.) describe the basic MALATANG observation and reduction strategy. In brief, observations of the \jfour\ lines of HCN and HCO$^+$ in the six galaxies that we are studying here, NGC~253, NGC~1068, IC~342, M82, M83, and NGC~6946, were obtained on the JCMT telescope with the 16-receptor array receiver Heterodyne Array Receiver Program \citep[HARP;][]{buckle09}, between 2015 December and 2016 November in the early stages of the MALATANG survey. The Auto-Correlation Spectral Imaging System (ACSIS) spectrometer was used as the receiver backend with a bandwidth of 1 GHz and a resolution of 0.488 MHz, which correspond to 840 \kms\ and 0.41 \kms\ at 354 GHz, respectively. We mapped the central $2\arcmin \times 2\arcmin$ region for all six targets using a 3$\times$3 jiggle mode with grid spacing of 10\arcsec. The full width at half maximum (FWHM) beamwidth of each receptor at 350 GHz is about 14\arcsec. To optimise the performance, we set up the rotator angle of the K-mirror to control the orientation of the HARP array so that there were four working receptors parallel to the major axis of the galaxy in each scan, since two receptors (H13, H14) at the edge of the array were not operational. The telescope pointing was checked before starting a new source, and every 1-1.5 hours by observing one or more calibrator sources in the \cothree\ line at 345.8 GHz. The uncertainty in the absolute flux calibration is estimated to be about 10\% for our sample of galaxies and is measured with the standard line calibrators. The velocity of each galaxy measured in our observations is radio defined with respect to the kinematical Local Standard of Rest (LSR). Details of the observations for the six galaxies are summarized in Table~\ref{tab:obs}.

%%%%%%%%%- Table-2 -%%%%%%%%%%%%
\begin{deluxetable*}{lcccCCCC}[htbp]
\tablecaption{A summary of observing parameters\label{tab:obs}}
\addtolength{\tabcolsep}{-2.5pt}
\tablewidth{0pt}
\tablehead{
\colhead{Source} & \colhead{Molecule} &\colhead{Dates of observations}  & \colhead{$f_{\rm obs}$} & \colhead{ROT\_PA} & \colhead{$\overline{T_{\rm sys}}$} & \colhead{$\overline{\tau}$}& \colhead{$t_{\rm int}$} \\ % & \colhead{$\Delta T$} \\
\colhead{} & \colhead{} & \colhead{} & \colhead{(GHz)} & \colhead{(deg)} & \colhead{(K)} & \colhead{(225 GHz)} & \colhead{(min)} \\ %& \colhead{(mK)}  \\
\colhead{(1)} & \colhead{(2)} & \colhead{(3)} & \colhead{(4)} & \colhead{(5)} & \colhead{(6)} & \colhead{(7)} & \colhead{(8)} % & \colhead{(9)} 
}
\startdata
NGC 253 & \hcnfour & 2015-(12-02,12-10,12-11) & 354.223 & 51,-39 & 231 & 0.024 & 142   \\
                & \hcopfour & 2015-12-12 & 356.447 & 51,-39 & 281 & 0.036 & 100  \\
NGC 1068 & \hcnfour & 2015-(12-13,12-30,12-31),2016-11-14 & 353.191 & 0 & 246 & 0.046 & 250  \\
                  & \hcopfour & 2015-(12-12,12-13),2016-(02-10,06-23,10-09) & 355.411 & 0 & 328 & 0.072 & 287  \\
IC 342 & \hcnfour & 2015-(12-02,12-12,12-16),2016-(10-08,10-09) & 354.474 & 90 & 458 & 0.076 & 300   \\
            & \hcopfour & 2015-(12-13,12-16,12-20,12-21,12-24),2016-10-07& 356.701 & 90 & 453 & 0.070 & 352   \\
M82 & \hcnfour & 2015-(12-10,12-12) & 354.265 & 65,155 & 270 & 0.031 & 150   \\
        & \hcopfour & 2015-12-13 & 356.494 & 65,155 & 338 & 0.051 & 100   \\
M83 & \hcnfour & 2016-(06-22,06-25,07-12,07-13,07-14) & 353.954  & -45 & 459 & 0.075 & 300   \\
        & \hcopfour & 2016-(06-26,07-11,07-15,07-16,07-17,07-18,07-31) & 356.132 & -45 & 619 & 0.097 & 350   \\
NGC 6946 & \hcnfour & 2016-(05-04,06-15,07-11,07-12) & 354.458 & 109 & 409 & 0.082 & 450   \\
                  & \hcopfour & 2016-(05-05,05-06,06-16,07-12,07-13,07-14,07-15) & 356.681 & 109 & 472 & 0.091 & 553   \\                                                       
\enddata
\tablecomments{Column 1: galaxy name. Column 2: observed spectral line. Column 3: the data obtained in the early stage of MALATANG survey that were used in this study. The date of the observations is listed in format of YYYY-MM-DD.  Column 4: observing frequency. Column 5: position of the K-mirror. In order to make sure there are four working receptors parallel to the major axis of the galaxies in each scan, we set up the rotator angle of the K-mirror to control the orientation of the HARP array (see Sect.~\ref{subsec:jcmt}). Column 6: median system temperature over all observations. Column 7: median atmospheric opacity at 225 GHz over the observations that was recorded at the start and end of each scan. Column 8: total integration time including time spent integrating on the source and the reference position.}
\end{deluxetable*}
%%%%%%%%%%%%%%%%%%%%%%%%%%

\begin{figure*}[t!]
\plottwo{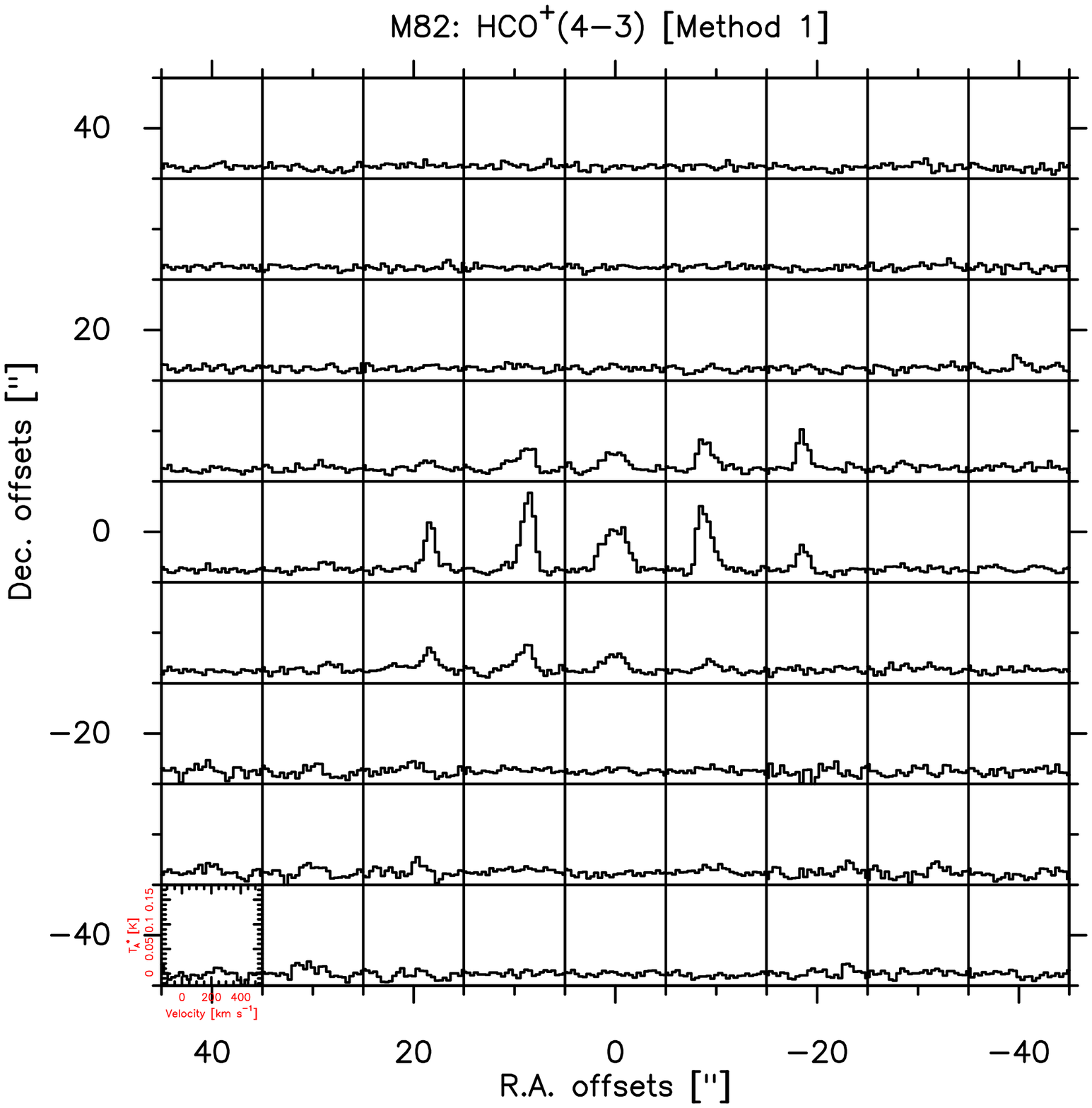}{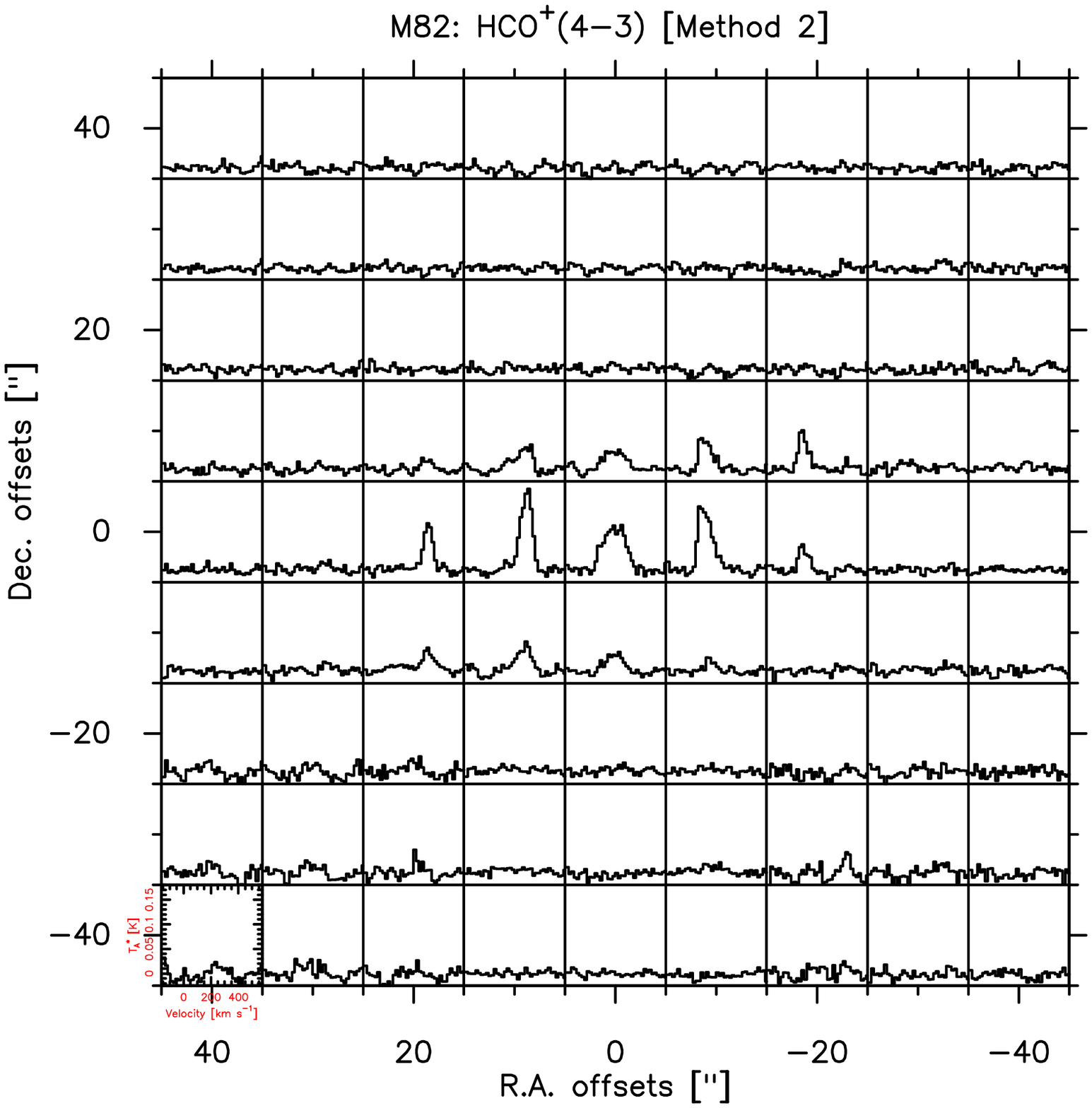}
\caption{JCMT HCO$^+\ J=4-3$ spectra map in the central $90\arcsec \times 90\arcsec$ region (grid spacing of 10$\arcsec$) of M82 processed with the ORAC-DR pipeline (left panel) and the method described in \citet{wilson12} (right panel). All spectra are on the $T_{\rm A}^*$ scale for the same range from -0.025 K to 0.18 K, and smoothed to 26 \kms\ and 20 \kms\ for the two methods, respectively. \label{fig:f1}}
\end{figure*}

We reduce the data using the Starlink software package ORAC-DR \citep{jenness15} to obtain pipeline-processed data and then convert the spectra to GILDAS/CLASS\footnote{\url{http://www.iram.fr/IRAMFR/GILDAS/}} format for further data processing. A recipe of REDUCE$\_$SCIENCE$\_$BROADLINE with default parameters was adopted for the ORAC-DR pipeline. We further assess the qualify of the data by inspecting the flatness of the baseline and the deviation of the rms noise level between the measured and expected values based on the radiometer equation and then attribute a quality tag to each spectrum. After flagging the data with a bad quality grade (i.e., spectra with distorted baselines or abnormal rms noise levels, which is defined as three times higher than the value calculated with the radiometer equation), the data for each spectral line were gridded into a cube. For positions observed with the central four receptors, on average about 15\% of the spectra were flagged due to bad baselines, while about $30-40$\% were discarded for positions observed with receptors on the edge of the array which are less stable. We fitted and subtracted a first-order baseline from the data cube using  channels outside of the velocity range of the line emission. The final cubes were converted from antenna temperature $T^\ast_{\rm A}$ to main beam temperature $T_{\rm mb}$ adopting a main beam efficiency of $\eta_{\rm mb}=0.64$ ($T_{\rm mb}\equiv T^\ast_{\rm A}/\eta_{\rm mb}$). To check the validity of the data processing results with the ORAC-DR pipeline, we used a different reduction method which combines the raw data into a cube using the task \texttt{makecube} in SMURF package \citep{jenness13}, after first flagging poor data \citep[see][]{wilson12,warren10}. Similarly, the spectra were converted to CLASS format for further analysis. Figure~\ref{fig:f1} is a comparison of the spectra in the central $90\arcsec \times 90\arcsec$ region of M82 obtained from the reduction method with the ORAC-DR pipeline with those processed using the method described in \citet{wilson12}. It is clear that both the profile and the intensity of the spectra derived from different reduction methods are in good agreement for each position where significant signal is detected. A refined data reduction method, which aims primarily at the processing of weak emission lines by converting the raw data to GILDAS/CLASS format and qualifying the data automatically, is under development (Zhang et al. 2018, in prep.). 

%%%%%%%%-Fig-1-%%%%%%%
%%%%%%%%%%%%%%%%%%

\subsection{Ancillary Data}\label{subsec:ancillary}
The six galaxies in this study have a large number of ancillary data available at multiple wavelengths. In this work, we will focus on  molecular line and infrared data for a comprehensive analysis.

\subsubsection{Infrared Data}\label{subsec:ir}
We retrieved the calibrated IR image data obtained using the {\it Spitzer} MIPS and {\it Herschel} PACS instruments from the NASA/IPAC Infrared Science Archive (IRSA). The data have been processed to level 2 for MIPS \mbox{24 $\mu$m} and level 2.5 for PACS \mbox{70 $\mu$m}, 100 $\mu$m, and \mbox{160 $\mu$m} bands in the pipeline. For galaxies that were observed as part of the KINGFISH (Key Insights on Nearby Galaxies: A Far-Infrared Survey with Herschel; \citet{kennicutt11}) programme, IC 342 and NGC 6946, we use the PACS images made with the Scanamorphos code version 16.9. The FWHM angular resolutions are approximately 6$\arcsec$.0, 5$\arcsec$.8, 7$\arcsec$.8, and 12$\arcsec$ at \mbox{24 $\mu$m}, \mbox{70 $\mu$m}, \mbox{100 $\mu$m}, and \mbox{160 $\mu$m}, respectively. The {\it Herschel} SPIRE data were not used in our study because of their lower angular resolution ($\gtrsim 18 \arcsec$) compared with our JCMT line observations.

To estimate the infrared luminosity of each position in our target galaxies, we measure the infrared flux densities from \mbox{24 $\mu$m} to \mbox{160 $\mu$m}. In a first step, we use the convolution kernels provided by \citet{aniano11} to convolve the {\it Spitzer} and {\it Herschel} maps to match the 14$''$ beam of our line data. For the PACS data, we scale the image by a factor of 1.133$\times$(14/{\it pixel size})$^2$ where {\it pixel size} is the length of a pixel in arcseconds, to convert the units from Jy into Jy\,beam$^{-1}$, while for MIPS images, we first convert the pixel value in MJy\,sr$^{-1}$ into Jy and then scale the image to Jy\,beam$^{-1}$. We estimate an average of the pixel values within the sky area to subtract the mean sky background for each galaxy. We then measure the central pixel flux for each position in the convolved image to obtain the flux of each infrared band in units of Jy beam$^{-1}$.

\subsubsection{NRO 45m CO \jone\ Data}\label{sec:codata}
We obtain the CO \jone\ data from the Nobeyama CO-mapping survey \citep{kuno07} at the NRO website\footnote{\url{http://www.nro.nao.ac.jp/~nro45mrt/html/COatlas/}}. The beam size (FWHM=15$''$) of the CO \jone\ mapping is comparable to our JCMT observations. We align the CO data with our JCMT data by gridding the cube, and then extract the spectra from each matched position. Except for M82, for which we adopt the \cothree\ data from the JCMT NGLS survey \citep{wilson12}, the CO data for the remaining five galaxies are the \jone\ transition and were observed with the NRO 45m. With the CO data, we determine for each position the velocity range over which line emission is to be integrated. The CO line is detected at high signal-to-noise ratio, at all positions that we observed with the JCMT. Integrating over the CO emitting velocity ranges guarantees that we have an integrated-intensity measurement along each line of sight, which is particularly important for positions with weak emission of HCN or HCO$^+$ \jfour\ line.

\section{Molecular line and Infrared measurements}\label{sec:results}
\subsection{Spectra}\label{subsec:spectra}
Fig.~\ref{fig:spe} shows mosaics of \hcnfour\ and \hcopfour\ spectra of the central $\sim 50\arcsec \times 50\arcsec$ regions of the six galaxies, which have been mapped in their central $2\arcmin\times 2\arcmin$ regions. Our observations show that the dense molecular-line emission is mainly concentrated within the central $\sim 1\arcmin$ region. We will present a further analysis of the data in the outer disks (i.e., $\gtrsim 1\arcmin$ region) using the refined data-reduction method mentioned in Sect.~\ref{subsec:jcmt} in a follow-up paper. All the six galaxies have been detected in both \hcnfour\ and \hcopfour\  lines in off-central positions, except M83 where only the central position was detected in HCN and HCO$^+$ emission with significance of \mbox{4.9 $\sigma$} and \mbox{8.5 $\sigma$} respectively. For M83, we only show the spectral line toward the central position and a spectrum averaged by stacking all the positions observed, excluding the center for each line. For spectrum from each position, we shift the velocity relative to the line center, which is derived from \coone\ spectra at the same position with a Gaussian fitting, then stack the HCN and HCO$^+$ spectra from all off-center positions. As expected, the line profiles are very similar between HCN and HCO$^+$ since both trace the dense molecular gas in galaxies. For the five galaxies with detections in off-central positions, the rotation of circumnuclear gas is apparent in both lines based on the line profiles and the shifts in centroid velocity.

\subsection{HCN and HCO$^+$ Line Luminosities}\label{subsec:line}

%%%%%%%%%- Fig-2 -%%%%%%%%

\begin{figure*}[htbp]
\centering
\includegraphics[scale=0.45]{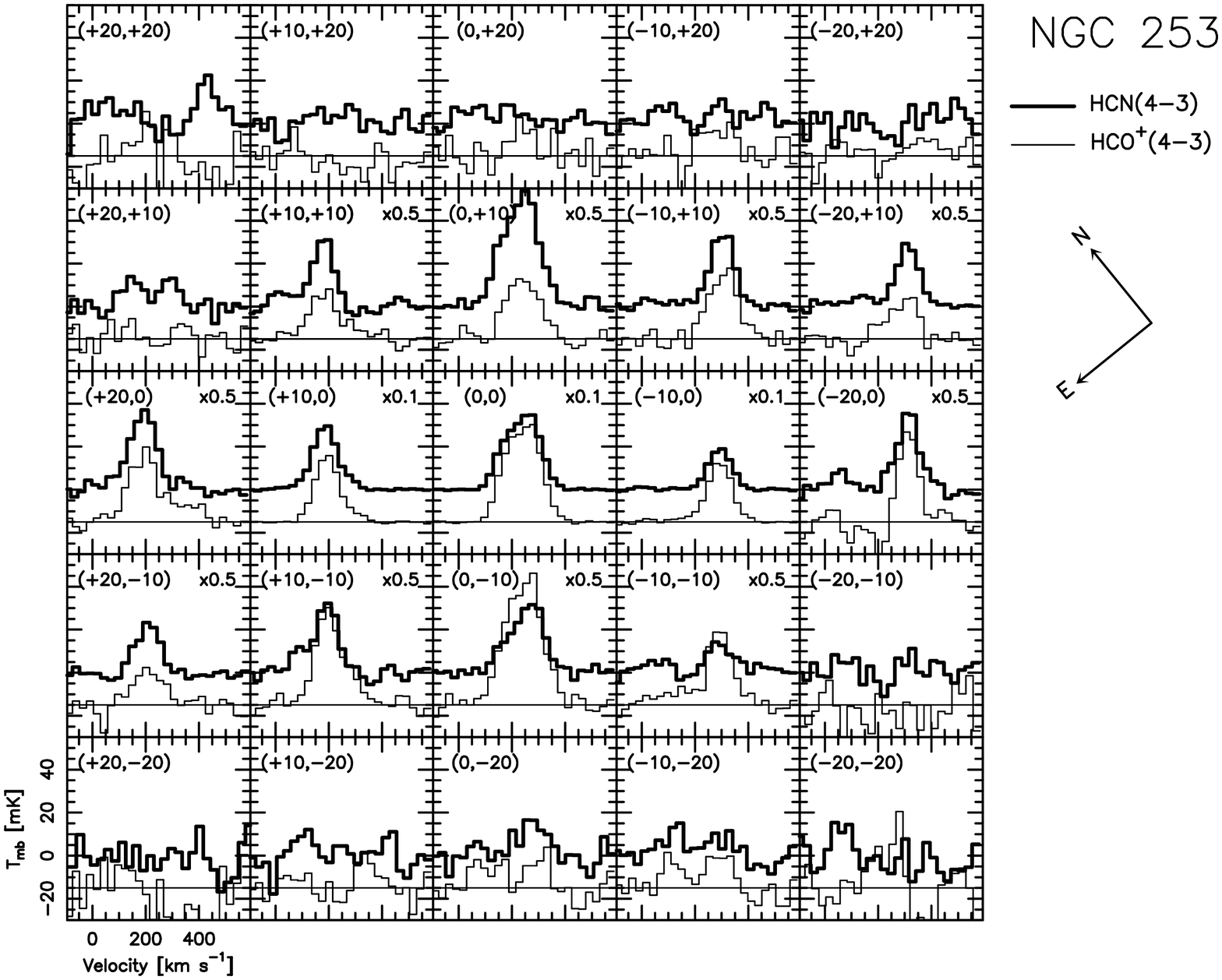}\\
\vspace{4pt}
\includegraphics[scale=0.45]{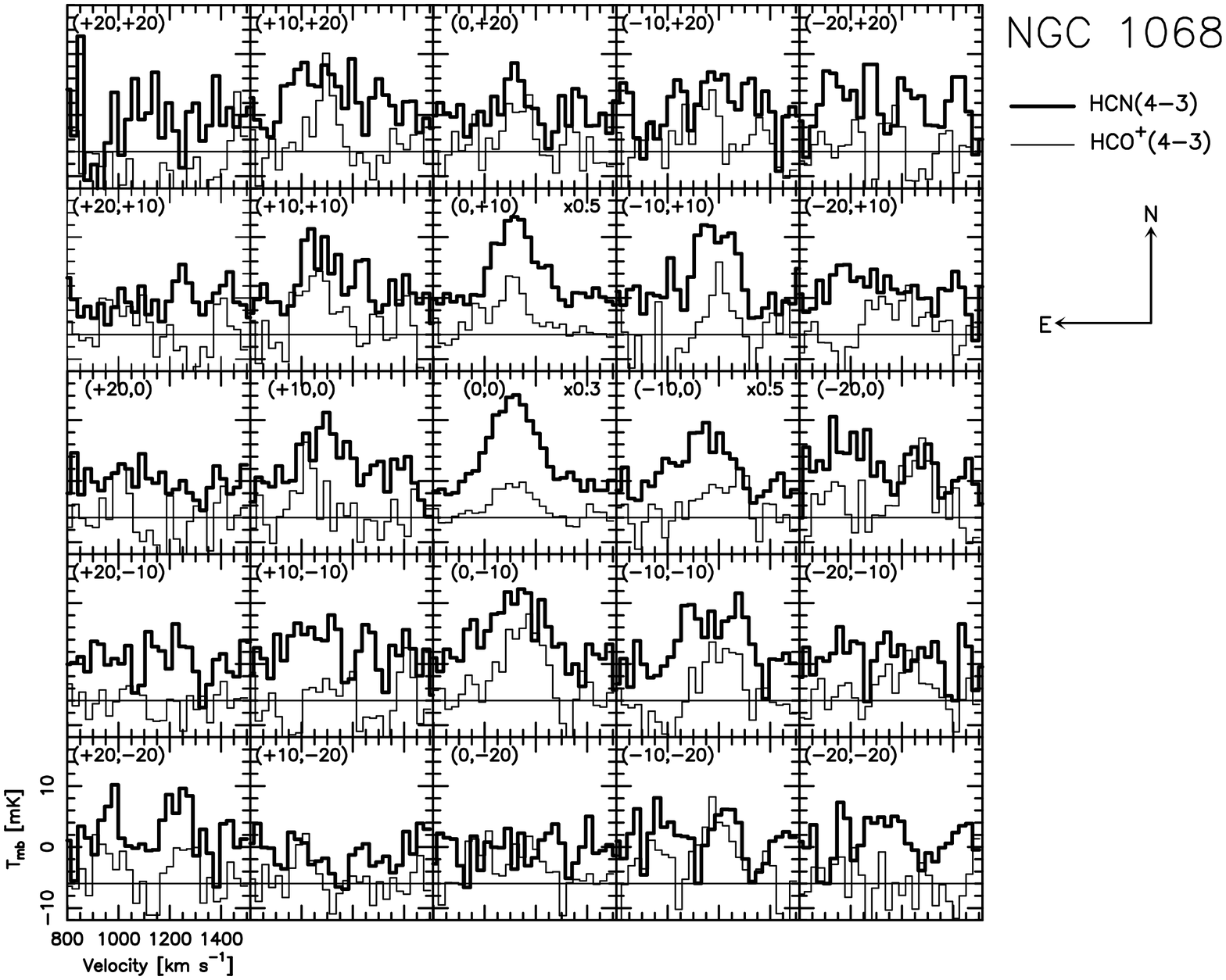}
\caption{\hcnfour \ ({\it thick lines}) and \hcopfour \ ({\it thin lines}) spectra map at the central $\sim 50\arcsec \times 50\arcsec$ region of galaxies that were observed in jiggle-mapping mode (3$\times$3 pattern, 10$''$ spacing) with the JCMT. To facilitate comparison with the spectra of positions with weak emission, we scaled down the spectra for those positions with relatively stronger emission of both \hcnfour \ and \hcopfour \ by the same multiplying factor which is listed in the upper right of each grid. For each line and position, we averaged all the spectra taken at that position excluding those with poor baseline or abnormal noise levels to produce a final spectrum.  All spectra are on the $T_{\rm mb}$ scale and smoothed to a velocity resolution of $\sim$ 26 \kms\ unless otherwise noted. The offset from the center position in units of arcsecond is indicated in the upper left of each grid.  The \hcopfour\ lines were shifted downwards with zero intensity level indicated by the horizontal lines. The directions North and East are shown to the right of each spectra grid. \label{fig:spe}}
\end{figure*}

\setcounter{figure}{1}
\begin{figure*}
\centering
\includegraphics[scale=0.45]{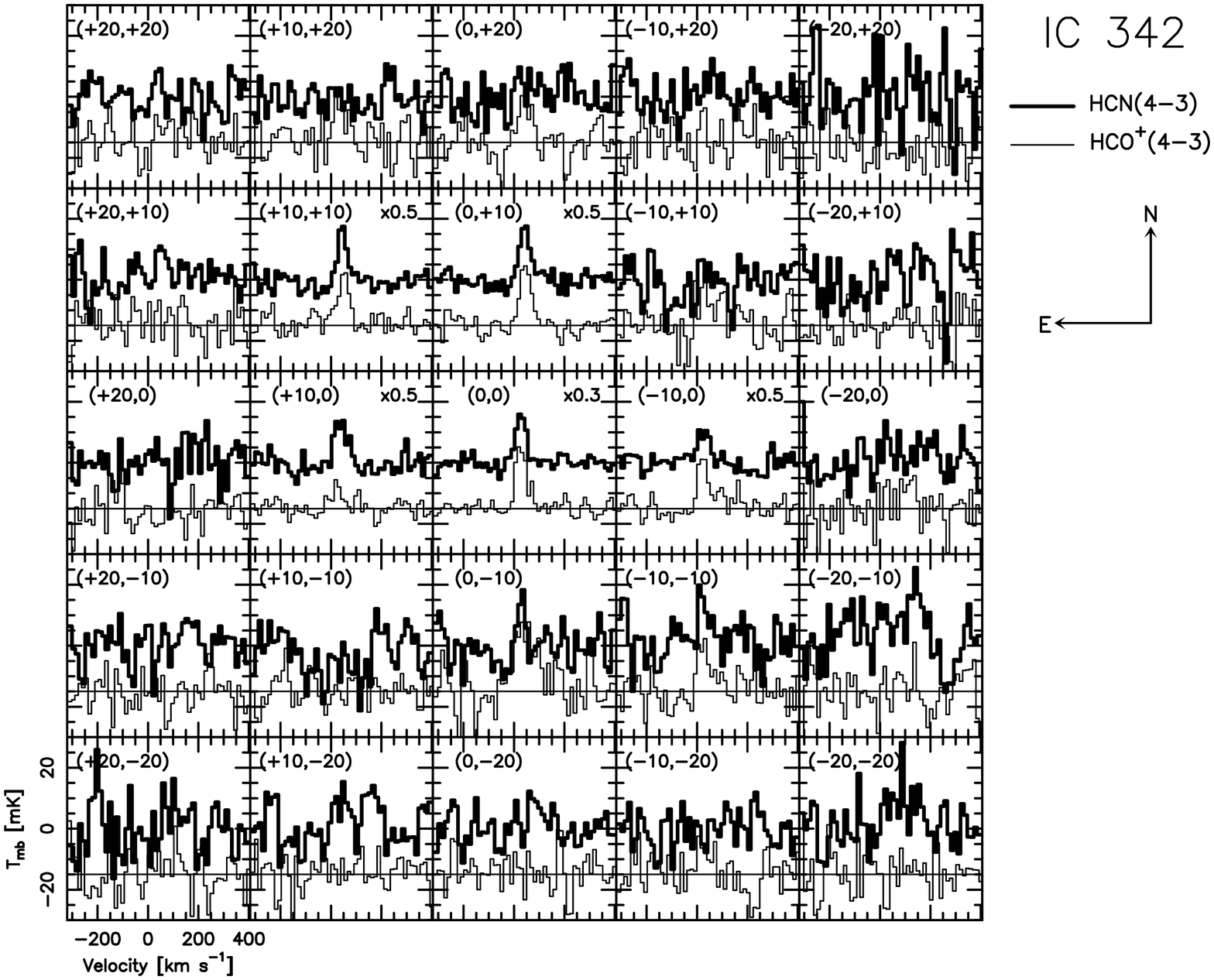}\\
\vspace{4pt}
\includegraphics[scale=0.46]{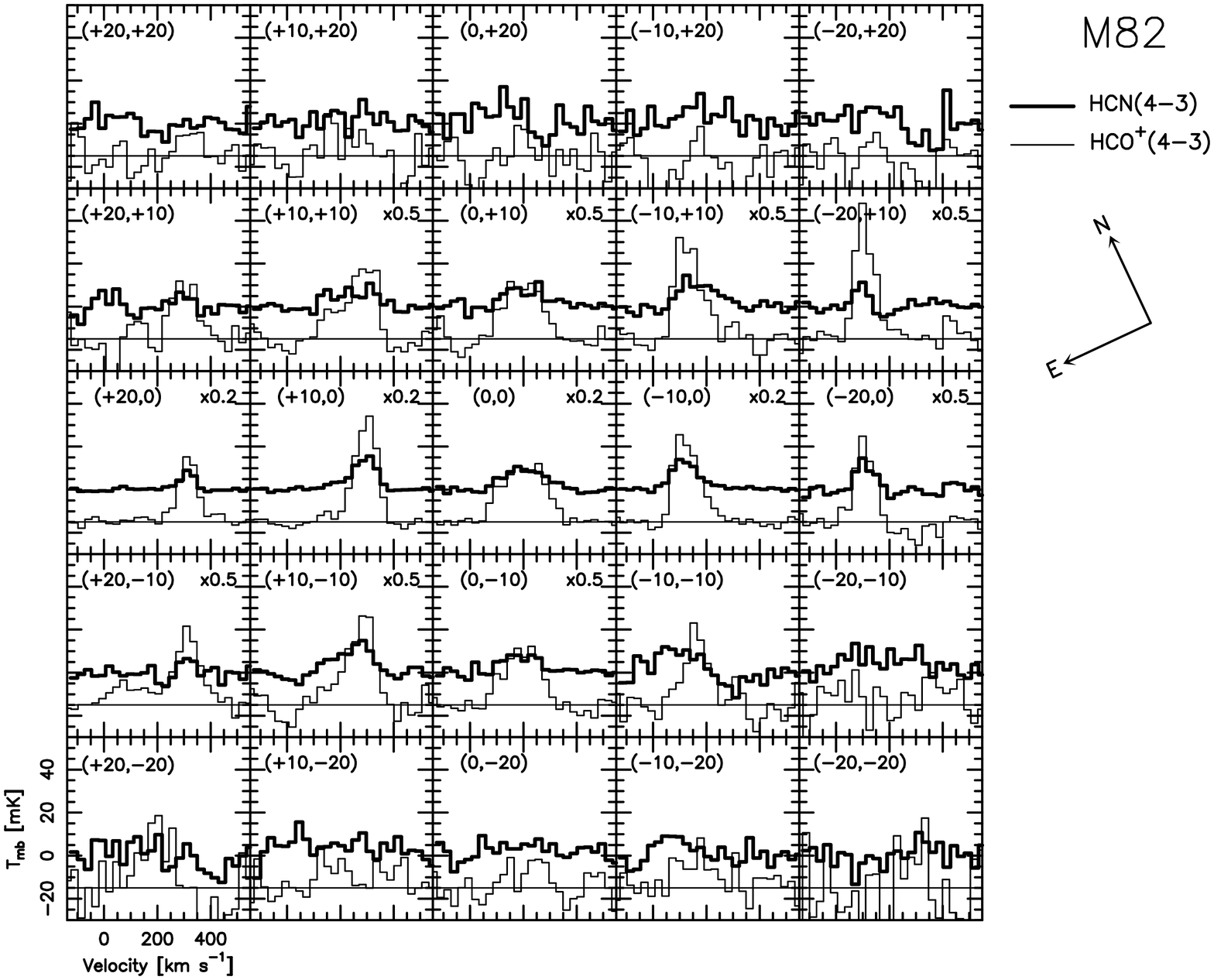}
\caption{{\it Continued}. The spectra of IC 342 were smoothed to a velocity resolution of 13 \kms.}
\end{figure*}

\setcounter{figure}{1}
\begin{figure*}
\centering
\includegraphics[scale=0.45]{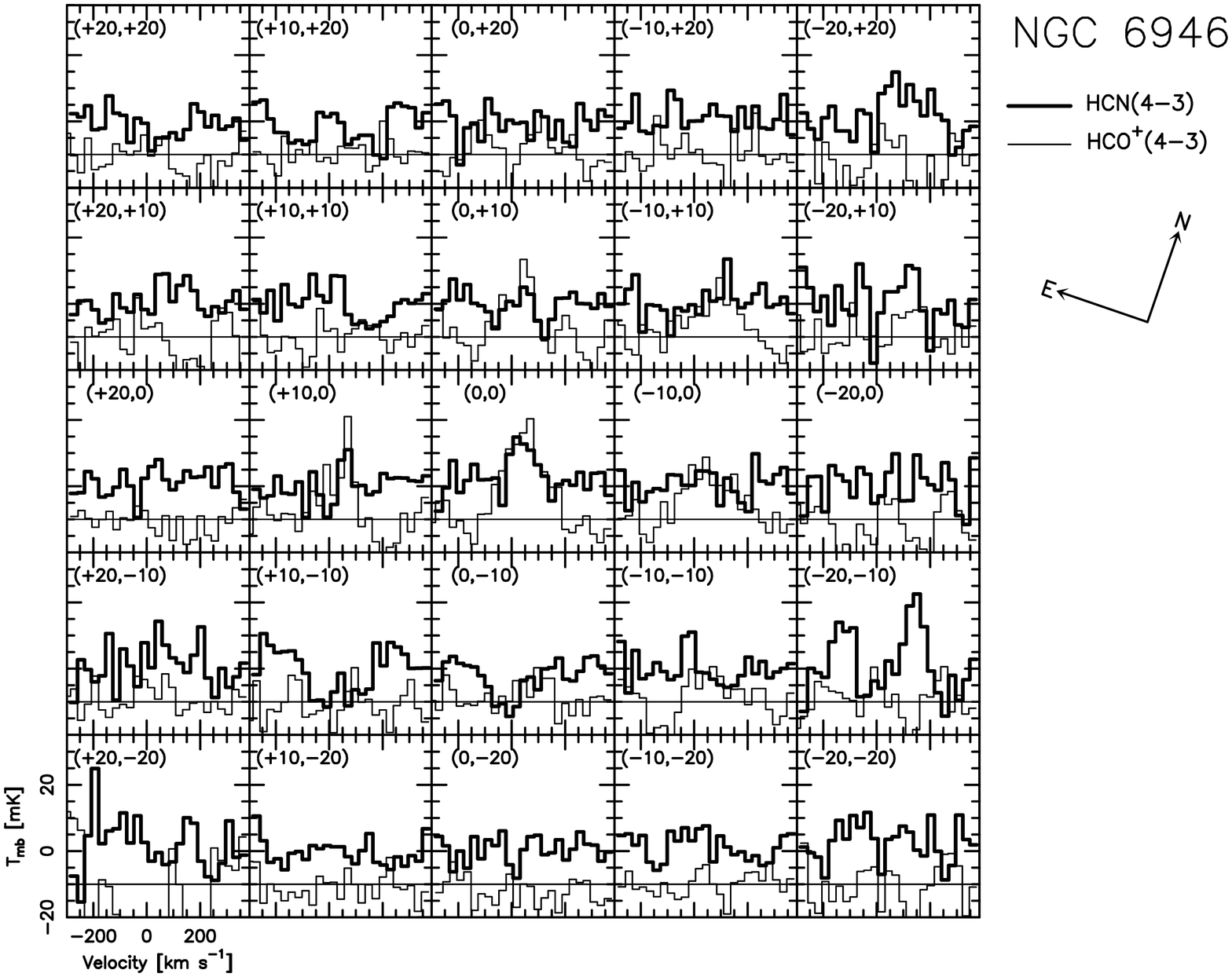}\\
\vspace{4pt}
\includegraphics[scale=0.43]{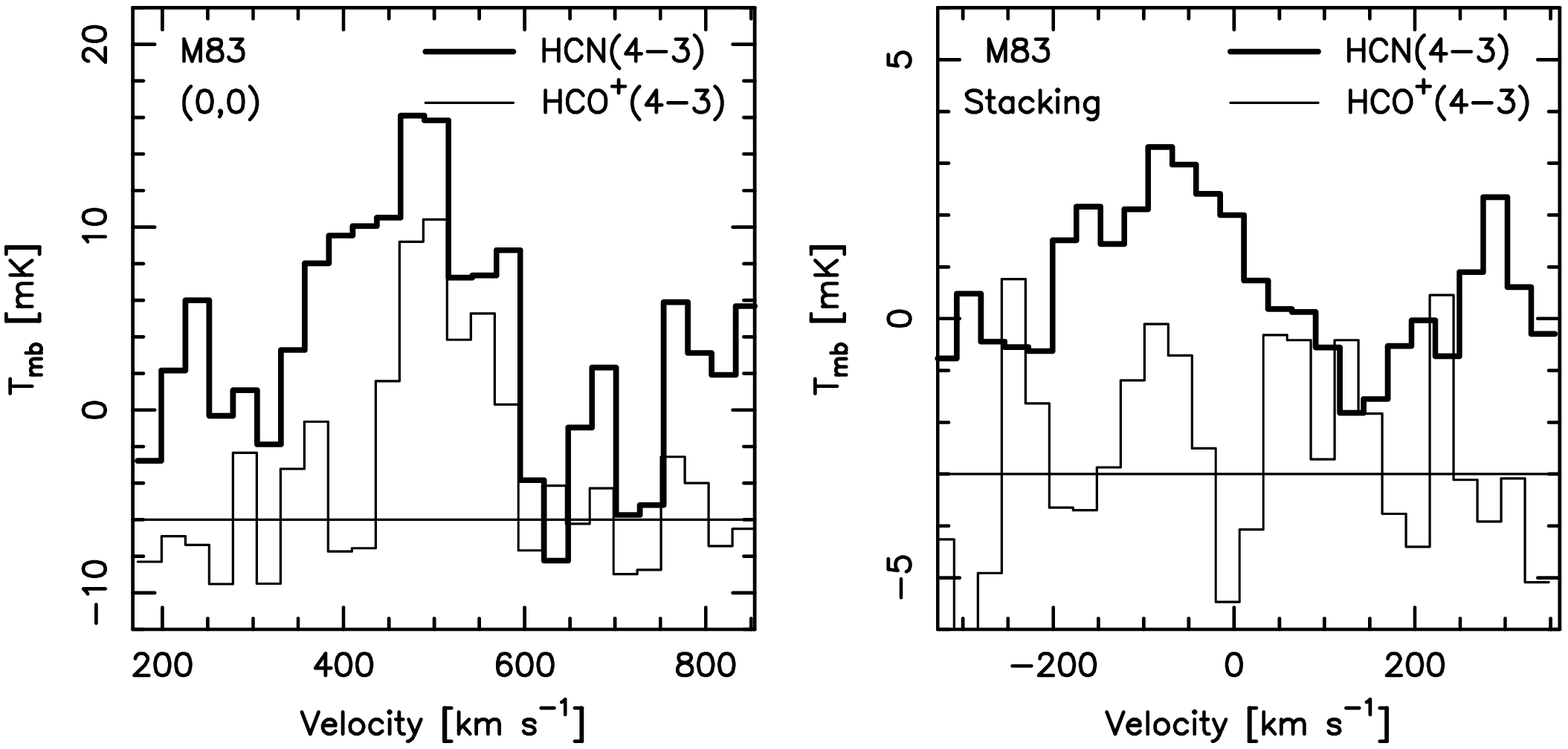}
\caption{{\it Continued}. For M83, we show the spectra toward the central position ({\it left}) and the spectra stacked for all the observed positions excluding the center ({\it right}). We stack the spectra by averaging the off-center positions with velocities shifted to the line center of the \coone\ data, which was derived by a single-velocity-component Gaussian fitting. }
\end{figure*}
%%%%%%%%%%%%%%%%%%%%%%
The observed line intensities, $I \tbond \int T_{\rm mb}dv$, for the positions with \mbox{$\geqslant 3\ \sigma$} detection in the \jfour\ lines of HCN and HCO$^+$ are listed in Table~\ref{tab:measurements}, along with the line luminosities. We define a detection if the velocity-integrated line intensity is higher than or equal to \mbox{3 $\sigma$}. The uncertainties ($\sigma$) in the integrated intensities were derived via 
\begin{equation}\label{eq1}
\sigma_I=T_{\rm rms} \sqrt{\Delta v_{\rm line} \Delta v_{\rm res}} \sqrt{1+\Delta v_{\rm line}/\Delta v_{\rm base}},
\end{equation}
where $T_{\rm rms}$ is the rms main-beam temperature of the line data for a spectral velocity resolution of $\Delta v_{\rm res}$, $\Delta v_{\rm line}$ is the velocity range of the emission line, and $\Delta v_{\rm base}$ is the velocity range used to fit the baseline \citep{gao96}. The velocity range is determined based on the \coone\ data with a Gaussian fit to the line profile, on the assumption that the velocity range of dense gas is covered by the CO line emitting range (see Sect.~\ref{sec:codata}). For the positions without significant detections, we estimated a \mbox{3 $\sigma$} upper limit to the line integrated intensities. The \coone\ luminosities of M82 were estimated based on the JCMT \cothree\ data by assuming a line brightness temperature ratio of $r_{31}=0.8\pm0.2$ for all the positions mapped in M82 \citep[e.g.,][]{weiss05,mao10}.

The line luminosities $L^\prime_{\rm dense}$\footnote{The line luminosity $L^\prime_{\rm dense}$ is often expressed in units of K km s$^{-1}$ pc$^2$. The line luminosity $L_{\rm dense}$ measured in $L_\odot$ can be converted from $L^\prime_{\rm dense}$ by multiplying a factor of 8$\pi k \nu^3_{\rm rest}/c^3$ \citep{solomon92}.} for each position were calculated following \citet{solomon97}:
\begin{equation}\label{eq2}
\begin{split}
L^\prime_{\rm dense}= & 3.25\times10^7 \left(\frac{S\Delta v}{{\rm 1\ Jy\ km\ s^{-1}}}\right)\left(\frac{\nu_{\rm obs}}{{\rm 1\ GHz}}\right)^{-2}\\
& \times\left(\frac{D_{\rm L}}{{\rm 1\ Mpc}}\right)^2 \left(1+z\right)^{-3}\ {\rm K\ km\ s^{-1}\ pc^2},
\end{split}
\end{equation}
where $S\Delta v$ is the velocity-integrated flux density, $\nu_{\rm obs}$ is the observed line frequency, and $D_{\rm L}$ is the luminosity distance. We convert the line intensity to flux density using a conversion factor of $S/T_{\rm mb}=15.6/\eta_{\rm mb}=24.4\ {\rm Jy\ K^{-1}}$ for the JCMT telescope by assuming that the line emission from each individual region fills the main beam, given that the gas emission is rather clumpy in these nearby galaxies.

Three of our sample galaxies have published fluxes of HCN and HCO$^+$ \jfour\ emission towards the galaxy center in the literature \citep[NGC 253, NGC 1068, M82;][]{zhang14,knudsen07,seaquist00}. We compared our fluxes with those previous efforts and found good agreement for these sources (i.e., agree to within $\sim20\%$).

\subsection{Infrared Luminosities}\label{subsec:lir}

We estimate the total infrared luminosities $L_{\rm TIR}$ from \mbox{3 $\mu$m} to \mbox{1100 $\mu$m} using the prescription of \citet{galametz13} based on a combination of $Spitzer$/MIPS \mbox{24 $\mu$m} and $Herschel$/PACS luminosities:
\begin{equation}
L_{\rm TIR}=\Sigma\ c_i \nu L_\nu(i)\ L_\odot,
\end{equation}
where $\nu L_\nu(i)$ is the resolved luminosity in a given band $i$ in units of $L_\odot$ and measured as $4\pi {\rm D}^2_\mathrm{L}(\nu f_\nu)_i$, and $c_i$ are the calibration coefficients for various combinations of $Spitzer$ and $Herschel$ bands. For galaxies without a MIPS \mbox{24 $\mu$m} image or which are saturated in the \mbox{24 $\mu$m} image cores, we use PACS bands alone to estimate $L_{\rm TIR}$. With the exception of NGC~1068 and M82, for which only PACS \mbox{70 $\mu$m} and \mbox{160 $\mu$m} data are available, we have photometry data in at least three bands for the remaining four galaxies. The total uncertainties estimated for $L_{\rm TIR}$ comprise the photometric uncertainty, the flux calibration uncertainty \citep[assumed to be 5\%;][]{balog14}, and the uncertainty of the TIR calibration from combined luminosities \citep[$\sim$20\% for galaxies with data in four IR bands available and $\sim$25\% for those have fewer IR images;][]{galametz13}. The IR luminosity derived for each position with significant (\mbox{$\geqslant 3\ \sigma$}) \hcnfour\ or \hcopfour\ detections is listed in Table~\ref{tab:measurements}.

\section{The relationships between dense molecular gas tracers and dust/star formation properties}\label{sec:correlation}

\subsection{Correlation Between Molecular Lines of Dense Gas and Infrared Luminosities}\label{sec:ir2dense}

\begin{figure*}[t!]
\plottwo{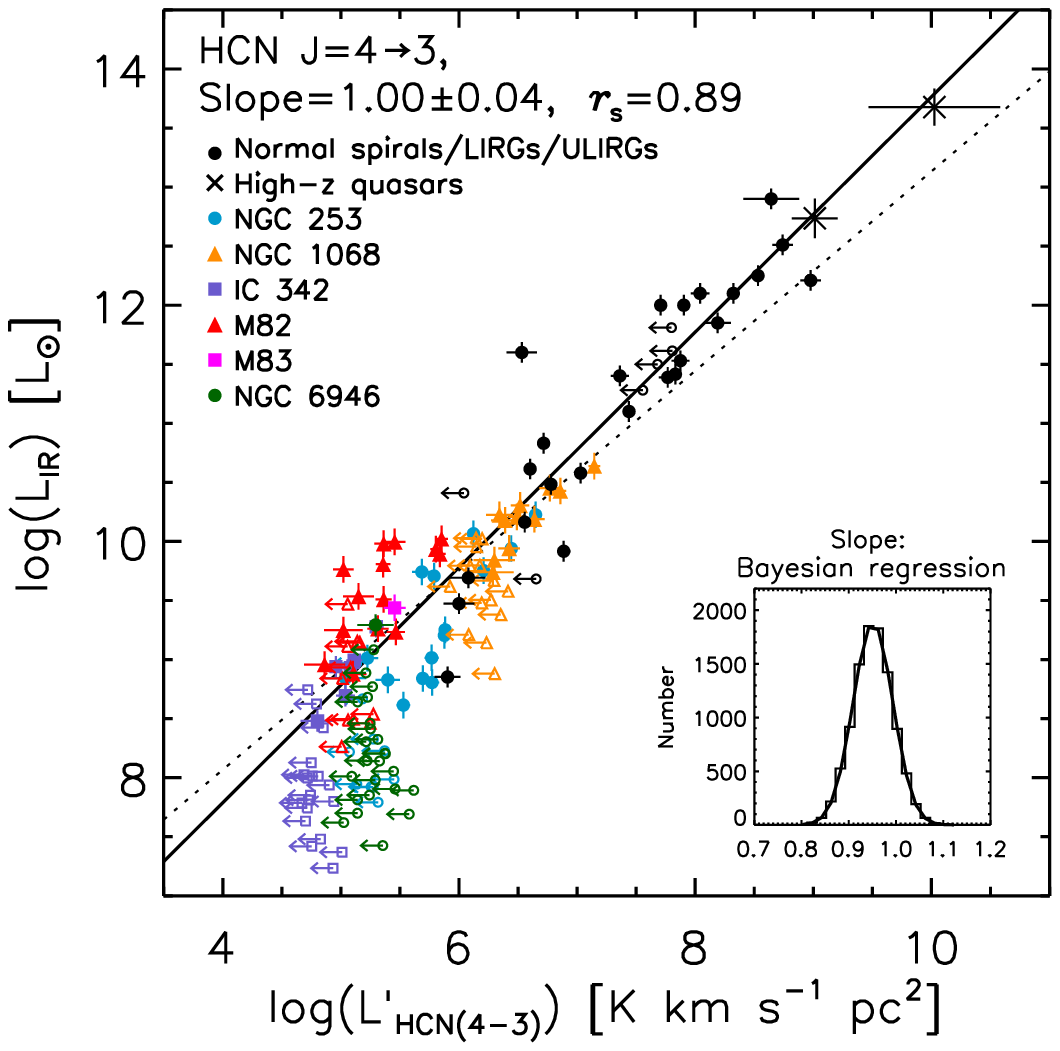}{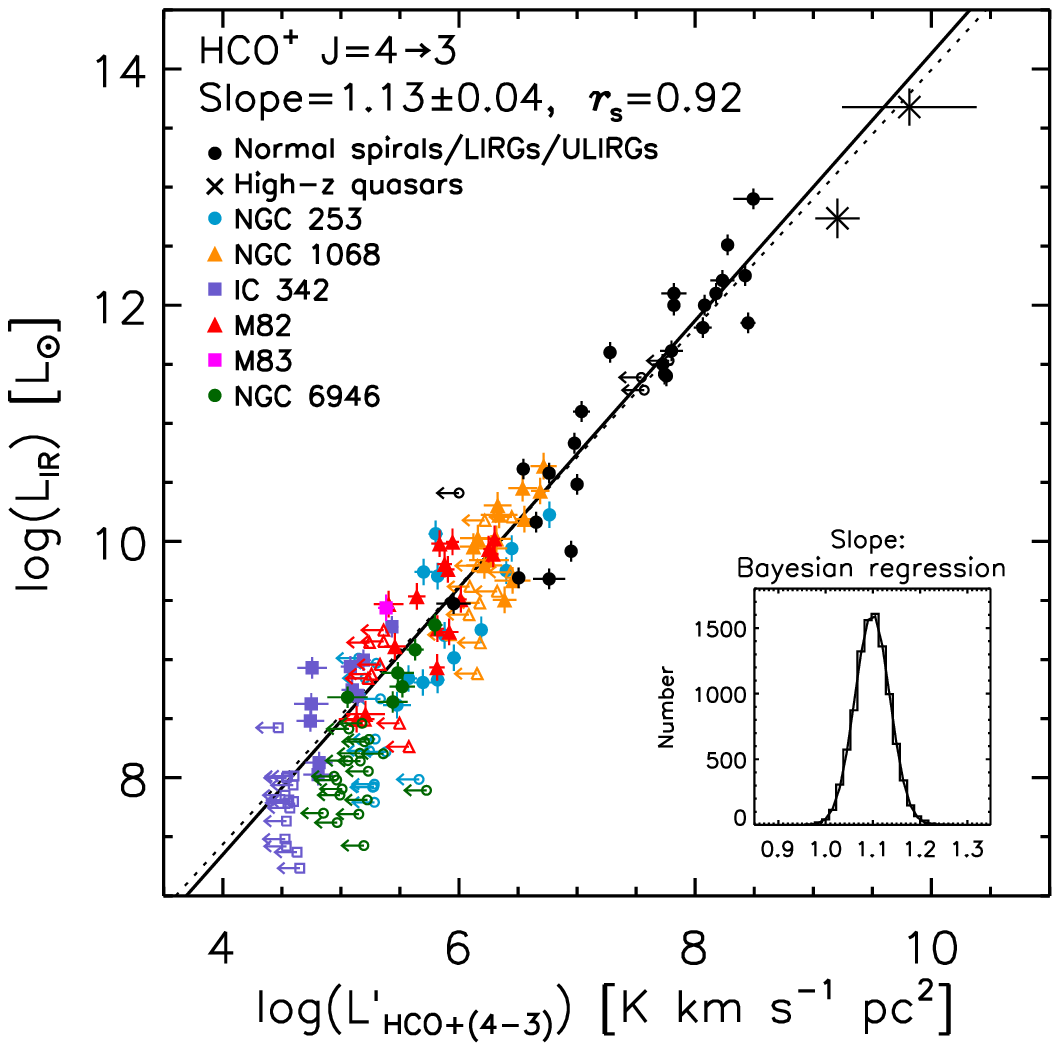}
\caption{Correlations between the molecular line luminosities of dense-gas tracers log($L'_{\rm dense}$) and the IR luminosity log($L_{\rm IR}$) for galaxies spatially resolved on sub-kpc scales ({\it colored symbols}) and galaxies with integrated measurements ({\it black symbols}). {\it Left}: \hcnfour. {\it Right}: \hcopfour. The colored symbols represent the spatially resolved sub-kpc structures in the central $\sim 50\arcsec \times 50\arcsec$ region of our sample galaxies, and the black symbols indicate the data from the literature (see legend in the top left of each panel). The solid lines in the left and right panels indicate the best-fit relations of Equations (4) and (5) respectively, while the black dotted lines show the relation considering the new JCMT data alone. The upper limits are marked with open symbols with leftward arrows and are not included in the fitting. The total uncertainties on the individual data points, including the statistical measurement uncertainties and the systematic uncertainties, are indicated by error bars (see Table~\ref{tab:measurements}). The best-fit power-law index and the Spearman rank correlation coefficient for the $L_{\rm IR}-L'_{\rm HCN(4-3)}$ and the $L_{\rm IR}-L'_{\rm HCO^+(4-3)}$ relation are listed in the top left of each panel. The inset shows the probability density distribution of the slope derived from the Bayesian fitting. \label{fig:correlation}}
\end{figure*}

In Fig.~\ref{fig:correlation}, we show the $L_{\rm IR}-L'_{\rm dense}$ relation for the different populations of galaxies compiled for this work using our new data (Table~\ref{tab:measurements}) and the data from the literature, including \hcnfour\ and \hcopfour\ detections in the center of nearby normal galaxies and in (U)LIRGs \citep{zhang14}, and six local (U)LIRGs observed with ALMA \citep{imanishi13b,imanishi13a,imanishi14}. We also included two high-redshift quasars, the Cloverleaf at {\it z} = 2.56 and APM 08279+5255 at {\it z} = 3.91. The Cloverleaf quasar is the only high-{\it z} galaxy that is detected in both \hcnfour\ and \hcopfour\ emission \citep{riechers11a,barvainis97}\footnote{Updated measurements of \hcnfour\ and \hcopfour\ emission with the IRAM PdBI for the Cloverleaf were reported by Michel Gu\'elin (2010), and published in the IRAM Newsletter at \url{http://www.iram-institute.org/medias/uploads/NewsletterAug2010.pdf}}. 

%%%%%%%%%%%- Table-3 -%%%%%%%%%%%%
\startlongtable

%\begin{deluxetable*}{lhRRRCCC}
\begin{deluxetable*}{lRhRRCCC}
\centering
\tablecaption{Derived properties for sampled positions of the six galaxies in our sample\label{tab:measurements}}
\tablewidth{0pt}
\tablehead{
\colhead{Source} & \colhead{Offsets\tablenotemark{a}} & \nocolhead{Offsets(real)} & \colhead{$I_{\rm HCN(4-3)}$} & \colhead{$I_{\rm HCO^{+}(4-3)}$} &  \colhead{$L'_{\rm HCN(4-3)}$} & \colhead{$L'_{\rm HCO^+(4-3)}$} & \colhead{$L_{\rm TIR}$}\\
\colhead{} & \colhead{(arcsec)} & \nocolhead{(arcsec)} & \colhead{(K km s$^{-1}$)} & \colhead{(K km s$^{-1}$)} & \colhead{($10^4$ K \kms \ pc$^2$)} & \colhead{($10^4$ K \kms \ pc$^2$)} & \colhead{($10^{7}\ L_{\odot}$)}
}
%\colnumbers
\startdata
NGC 253 & (0,0) & (0.0,0.0) & 57.7\pm0.7 & 76.6\pm0.8 & 445\pm5 & 583\pm6 & 1681\pm102 \\
                & (10,0) & (7.8,6.3)  & 35.9\pm0.7 & 36.7\pm0.5 & 277\pm6 & 280\pm4 & 870\pm53 \\
                & (20,0) & (15.5,12.6)  & 9.8\pm0.6 & 9.9\pm0.9 & 75\pm4 & 76\pm7 & 160\pm9  \\
                & (-10,0) & (-7.8,-6.3)  & 20.9\pm0.7 & 33.0\pm0.9 & 162\pm5 & 251\pm7 & 566\pm35  \\
                & (-20,0) & (-15.5,-12.6)  & 7.7\pm0.7 & 6.5\pm0.9 & 59\pm6 & 50\pm7 & 64\pm4  \\
                & (0,10) & (-6.3,7.8)  & 17.2\pm0.6 & 8.3\pm0.8 & 133\pm5 & 64\pm6 & 1162\pm71 \\
                & (10,10) & (1.5,14.1)  & 6.3\pm1.0 & 6.6\pm1.0 & 48\pm8 & 50\pm7 & 552\pm33 \\
                & (20,10) & (9.2,20.4)  & 2.2\pm0.4 & <1.9 & 17\pm3 & <14 & 103\pm6  \\
                & (-10,10) & (-14.1,1.5)  & 8.0\pm0.4 & 8.7\pm0.6 & 61\pm3 & 66\pm5 & 509\pm30  \\
                & (-20,10) & (-21.8,-4.8)  & 6.4\pm0.5 & 4.9\pm0.6 & 49\pm4 & 37\pm5 & 69\pm4  \\
                & (0,-10) & (6.3,-7.8)  & 9.9\pm0.5 & 20.3\pm0.8 & 76\pm4 & 154\pm6 & 178\pm11  \\
                & (10,-10) & (14.1,-1.5)  & 7.6\pm0.7 & 11.9\pm1.0 & 59\pm5 & 91\pm8 & 103\pm6  \\
                & (20,-10) & (21.8,4.8)  & 4.4\pm0.4 & 3.9\pm1.0 & 34\pm3 & 30\pm7 & 41\pm2 \\
                & (-10,-10) & (-1.5,-14.1)  & 3.2\pm0.7 & 8.6\pm0.8 & 25\pm5 & 66\pm6 & 67\pm4   \\
NGC 1068 & (0,0) & (0.0,0.0) & 9.0\pm0.5 & 3.4\pm0.8 & 1398\pm72 & 521\pm121 & 4347\pm280  \\
                  & (10,0) & (10.0,0.0) & 1.6\pm0.4 & <1.1 & 245\pm62 & <165 & 1511\pm96  \\                  
                  & (-10,0) & (-10.0,0.0) & 3.8\pm0.7 & 2.3\pm0.6 & 589\pm106 & 346\pm91 & 2832\pm179  \\
                  & (-20,0) & (-20.0,0.0) & <1.0 & 1.4\pm0.4 & <157 & 208\pm54 & 1052\pm65 \\
                  & (0,10) & (0.0,10.0) & 4.7\pm0.3 & 3.2\pm0.5 & 721\pm41 & 487\pm71 & 2675\pm169  \\
                  & (10,10) & (10.0,10.0) & 1.4\pm0.3 & 1.4\pm0.5 & 220\pm53 & 219\pm74 & 1678\pm105  \\
                  & (-10,10) & (-10.0,10.0) & 2.0\pm0.3 & <1.8 & 310\pm41 & <280 & 1614\pm101  \\
                  & (-20,10) & (-20.0,10.0) & <1.0 & 1.1\pm0.4 & <150 & 164\pm55 & 627\pm38 \\
                  & (0,-10) & (0.0,-10.0) & 2.8\pm0.5 &2.3\pm0.3 & 435\pm74 & 358\pm50 & 1545\pm96   \\
                  & (-10,-10) & (-10.0,-10.0) & 2.1\pm0.3 & 1.4\pm0.3 & 329\pm48 & 213\pm53 & 2018\pm125   \\
                  & (-20,-10) & (-20.0,-10.0) & <0.9 & 0.9\pm0.2 & <141 & 133\pm33 & 903\pm55   \\     
                  & (0,20) & (0.0,20.0) & <0.9 & 0.9\pm0.3 & <139 & 144\pm47 & 1066\pm66 \\           
                  & (10,20) & (10.0,20.0) & 1.7\pm0.3 & 1.2\pm0.4 & 266\pm49 & 188\pm55 & 874\pm54   \\
                  & (0,-20) & (0.0,-20.0) & <1.2 & 1.6\pm0.3 & <187 & 243\pm50 & 320\pm20 \\
                  & (-10,-20) & (-10.0,-20.0) & <1.3 & 1.9\pm0.6 & <197 & 284\pm95 & 467\pm30 \\
 IC 342 & (0,0) & (0.0,0.0) & 2.7\pm0.3 & 3.8\pm0.3 & 20\pm2 & 27\pm2 & 190\pm11  \\
            & (10,0) & (10.0,0.0) & 1.7\pm0.3 & 0.8\pm0.2 & 12\pm2 & 6\pm2 & 85\pm5  \\
            & (-10,0) & (-10.0,0.0) & 1.2\pm0.2 & 1.7\pm0.3 & 9\pm2 & 12\pm2 & 87\pm5  \\
            & (0,10) & (0.0,10.0) & 1.8\pm0.2 & 2.2\pm0.2 & 13\pm1 & 16\pm2 & 99\pm6   \\
            & (10,10) & (10.0,10.0) & 1.5\pm0.2 & 1.9\pm0.2 & 11\pm2 & 14\pm2 & 49\pm3  \\
            & (-10,10) & (-10.0,10.0) & <0.9 & 0.8\pm0.2 & <6 & 6\pm2 & 42\pm2  \\
            & (0,-10) & (0.0,-10.0) & <0.7 & 1.7\pm0.3 & <5 & 12\pm2 & 55\pm3   \\
            & (-10,-10) & (-10.0,-10.0) & 0.9\pm0.3 & 0.8\pm0.2 & 6\pm2 & 6\pm1 & 30\pm2  \\            
            & (0,20) & (0.0,20.0) & <0.8 & 0.9\pm0.2 & <6 & 7\pm2 & 13\pm1  \\
            & (10,20) & (10.0,20.0) & <0.7 & 0.9\pm0.2 & <5 & 6\pm2 & 11\pm1  \\
M82 & (0,0) & (0.0,0.0) & 9.2\pm0.7 & 26.3\pm0.7 & 71\pm6 & 200\pm5 &  1052\pm68 \\
        & (10,0) & (9.1,4.2) & 8.9\pm0.4 & 25.5\pm0.8 & 69\pm3 & 194\pm6 & 785\pm51   \\
        & (20,0) & (18.1,8.5) & 3.0\pm0.4 & 13.6\pm0.8 & 23\pm3 & 103\pm6 & 323\pm21  \\
        & (-10,0) & (-9.1,-4.2) & 8.3\pm0.4 & 23.4\pm0.9 & 64\pm3 & 179\pm7 & 860\pm57   \\
        & (-20,0) & (-18.1,-8.5) & 1.8\pm0.5 & 5.8\pm0.8 & 14\pm4 & 44\pm6 & 343\pm22   \\
        & (0,10) & (-4.2,9.1) & 3.0\pm0.5 & 9.0\pm1.1 & 23\pm4 & 69\pm9 & 957\pm61   \\
        & (10,10) & (4.8,13.3) & 3.0\pm0.3 & 9.9\pm1.0 & 23\pm2 & 75\pm8 & 643\pm41   \\
        & (20,10) & (13.9,17.5) & <1.5 & 3.3\pm0.9 & <11 & 25\pm7 & 295\pm19 \\
        & (-10,10) & (-13.3,4.8) & 3.7\pm0.7 & 11.6\pm0.9 & 29\pm5 & 88\pm7 & 991\pm63   \\
        & (-20,10) & (-22.4,0.6) & 1.3\pm0.2 & 10.6\pm0.8 & 10\pm2 & 81\pm6 & 581\pm37   \\
        & (0,-10) & (4.2,-9.1) & 2.7\pm0.5 & 8.6\pm0.6 & 21\pm4 & 66\pm5 & 182\pm12  \\
        & (10,-10) & (13.3,-4.8) & 3.8\pm0.5 & 10.8\pm1.6 & 29\pm4 & 83\pm12 & 171\pm11   \\
        & (20,-10) & (22.4,-0.6) & <1.6 & 8.5\pm0.9 & <12 & 65\pm7 & 86\pm6   \\
        & (-10,-10) & (-4.8,-13.3) & <1.5 & 3.8\pm0.7 & <11 & 29\pm5 & 129\pm9  \\
        & (-20,-10) & (-13.9,-17.5) & 1.6\pm0.4 & <2.4 & 13\pm3 & <18 & 75\pm5   \\
        & (-10,-20) & (-0.6,-22.4) & <1.7 & 1.8\pm0.6 & <13 & 14\pm5 & 31\pm2 \\
M83 & (0,0) & (0.0,0.0) & 2.0\pm0.4 & 1.7\pm0.2 & 29\pm5 & 24\pm3 & 274\pm16  \\
NGC 6946 & (0,0) & (0.0,0.0) &  1.4\pm0.5 & 4.5\pm0.5 & 20\pm7 & 63\pm7 & 196\pm11 \\
                  & (10,0) & (9.5,-3.3) & <1.0 & 2.0\pm0.5 & <14 & 27\pm8 & 44\pm2   \\   
                  & (-10,0) & (-9.5,3.3) & <1.4 & 3.1\pm0.4 & <19 & 43\pm6 & 121\pm7   \\
                  & (0,10) & (3.3,9.5) & <1.2 & 2.2\pm0.7 & <16 & 30\pm9 & 77\pm4  \\
                  & (-10,10) & (-6.2,12.7) & <1.3 & 2.4\pm0.5 & <19 & 33\pm7 & 59\pm3  \\    %3 sigma upper limits
\enddata
\tablecomments{All uncertainties are estimated statistically from the measurements. For the line velocity-integrated intensity and luminosity, an additional 10\% uncertainty should be added to account for the systematic uncertainties in absolute flux calibration. For the IR luminosity, we need to take into account the additional uncertainties from the flux calibration ($\sim5\%$) and the TIR calibration ($\sim20-25\%$) (see Sect.~\ref{subsec:lir}). We report a \mbox{3$\sigma$} upper limit for non-detections.}
\tablenotetext{a}{Offsets without correcting for the rotation of the receiver array. See the observing settings in Sect.~\ref{subsec:jcmt} and the directions North and East for each galaxy in Fig.~\ref{fig:spe}.}
\end{deluxetable*}

%%%%%%%- Fig-3 -%%%%%%%%%%
%%%%%%%%%%%%%%%%%%%%%

%%%%%%%%%%%%%%%%%%%%%%%
\noindent
For APM 08279+5255, we estimate the \hcnfour\ and the \hcopfour\ line luminosities based on the \jsix\ lines of HCN and HCO$^+$ measured by \citet{riechers10}, by assuming a \jsix/\jfour\ line luminosity ratio of 1.1$\pm$0.6, for both HCN and HCO$^+$. This line ratio is roughly estimated by taking the average of the HCN \jsix/\jfive\ luminosity ratio \citep[$r_{65{\rm (HCN)}}=1.36\pm0.31$;][]{riechers10} and the CO line ratio \citep[$r_{64{\rm (CO)}}=0.86\pm0.29$;][]{weiss07}. Note that it is likely that the uncertainties in the \jsix/\jfour\ line ratios are underestimated due to the presumably non-uniform physical conditions in the molecular gas as traced by CO and the dense gas as traced by HCN and HCO$^+$ in this galaxy \citep{weiss07}. The SFR are calibrated based on the total IR luminosity \citep[e.g.,][]{kennicutt98,murphy11}. For the high-{\it z} quasars, however, we used the far-IR luminosity \citep[i.e., integrated from \mbox{40 $\mu$m} to \mbox{120 $\mu$m} rest-wavelength,][]{helou85} as a measure of SFR due to the powerful AGN heating of dust in the mid-IR band. The IR luminosity of these two quasars shown in Fig.~\ref{fig:correlation} thus corresponds to the far-IR luminosity plus an additional uncertainty of 30\% from converting the FIR luminosity to the total IR luminosity \citep{weiss03,weiss07,sanders03}. We note that a tentative detection of \hcopfour\ and upper limit of \hcnfour\ emission in a {\it z} = 2.64 lensed star-forming galaxy was reported recently by \citet{roberts17}, and stacked detections of these two lines are reported in high-{\it z} dusty galaxies by \citet{spilker14}. These data were not included in our analysis as no IR measurements are yet available.

We adopt the IDL routine \texttt{linfitex.pro} of the MPFIT package \citep{markwardt09} for the linear least-squares fit and LINMIX\_ERR of \citet{kelly07} which uses the Markov Chain Monte Carlo approach to account for measurement uncertainties for the Bayesian regression. The uncertainties in $L_{\rm IR}$ and $L'_{\rm dense}$ accounted for in the fitting include the statistical measurement uncertainties and the systematic uncertainties which mainly originate from calibration (see Sect.~\ref{subsec:jcmt} and Sect.~\ref{subsec:lir} for details). The best linear least-squares fit (logarithmic) to our new data (the upper limits are not included in the fitting, marked in open symbols with leftward arrows in Fig.~\ref{fig:correlation}) combined with the literature data yields 
\begin{equation}
{\rm log}L_{\rm IR}=1.00(\pm0.04)\,{\rm log}L'_{\rm HCN(4-3)}+3.80(\pm0.27).
\end{equation}
This fit is shown as the solid line in Fig.~\ref{fig:correlation}\,(left). A Spearman rank correlation test yields a correlation coefficient of $r_{\rm s}=0.89$, with a probability ($p$-value) of $1.1\times10^{-26}$ for the null hypothesis. The Bayesian regression fits give a slope of 0.95$\pm$0.04, consistent with the linear least-squares fit, and the posterior distribution of possible slopes is shown in the inset of Fig.~\ref{fig:correlation}\,(left). A linear least-squares fit to our JCMT data ({\it colored symbols}) alone yields
${\rm log}L_{\rm IR} = 0.84(\pm0.09)\,{\rm log}L'_{\rm HCN(4-3)} + 4.69(\pm0.53)$
and with a Spearman rank correlation coefficient of 0.75, which is shown as a black dotted line in Fig.~\ref{fig:correlation}\,(left).

In Fig.~\ref{fig:correlation}\,(right) we plot the relation between $L_{\rm IR}$ and $L'_{\rm HCO^+(4-3)}$ and perform the same comparison. A linear least-squares fit to the data points excluding the upper limits gives a correlation close to linear,
\begin{equation}
{\rm log}L_{\rm IR}=1.13(\pm0.04)\,{\rm log}L'_{\rm HCO^+(4-3)}+2.83(\pm0.24),
\end{equation}
with a Spearman rank correlation coefficient of 0.92. The Bayesian regression fits give a similar slope of 1.10$\pm$0.04 and the fit using measurements in Table~\ref{tab:measurements} alone gives a consistent relation with 
${\rm log}L_{\rm IR} = 1.09(\pm0.08)\,{\rm log}L'_{\rm HCO^+(4-3)} + 3.06(\pm0.49)$
and with a Spearman rank correlation coefficient of 0.84. 

%%%%%%- Fig-4 -%%%%%%%
\begin{figure}[htbp]
\includegraphics[scale=0.7]{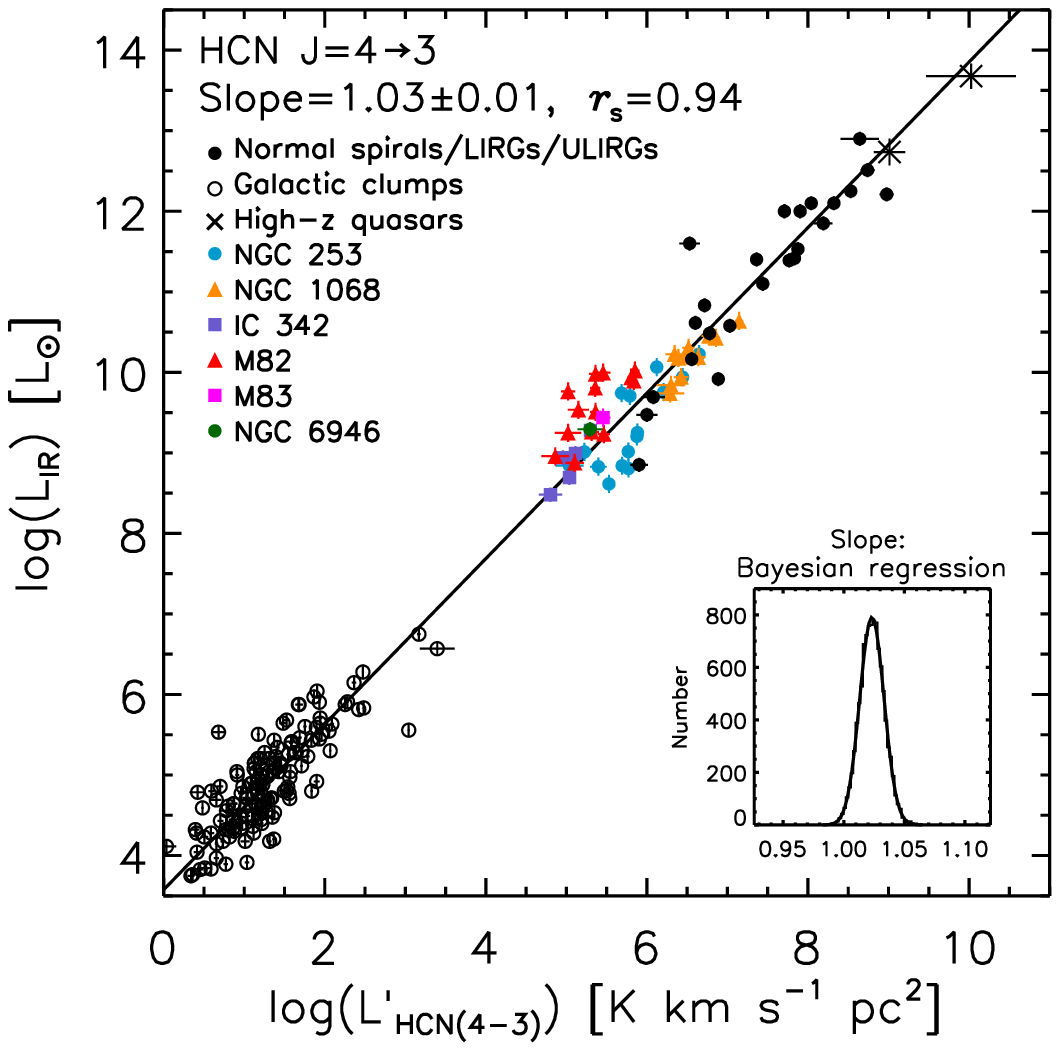}
\caption{Correlation between \hcnfour\ and IR luminosities for Galactic clumps ({\it circles}), our sample of galaxies resolved at sub-kpc scales ({\it colored symbols}), normal galaxies and local (U)LIRGs ({\it solid circles}), and high-$z$ quasars ({\it crosses}). The upper limits of \hcnfour\ are not included in the fitting and are not shown in this plot. The solid line represents the best-fit relation of  ${\rm log}L_{\rm IR} = 1.03(\pm0.01)\,{\rm log}L'_{\rm HCN(4-3)} + 3.58$. The probability density distribution of the slope derived from the Bayesian fitting is shown in the inset panel. A Spearman rank correlation analysis yields a correlation coefficient of 0.94. \label{fig:ir2hcn}}
\end{figure}
%%%%%%%%%%%%%%%%%

\citet{liu16} observed \hcnfour\ and \csseven\ lines in Galactic clumps and found that the $L_{\rm IR}$ are tightly correlated with both HCN and CS luminosities down to clumps with $L_{\rm IR}\sim 10^3\ L_\odot$. We compiled the \hcnfour\ data of Galactic clumps to compare with the data shown in Fig.~\ref{fig:correlation}\,(left). A linear least-squares fit to all data yields a slope of 1.03$\pm$0.01 (see Fig.~\ref{fig:ir2hcn}), in good agreement with the fit for the sample of galaxies measured globally.

To check the reliability of the best-fit relations obtained above, we adopted a Monte Carlo (MC) approach to fit the data. This approach, which is based on \citet{blanc09} and \citet{leroy13}, includes observational uncertainties, upper limits, and intrinsic scatter in the fits. Following \citet{blanc09} we fitted the following relation with three parameters 
\begin{equation}\label{eq:mc}
\left(\frac{L_{\rm IR}}{L_{\odot}}\right) = A \left(\frac{L^\prime_{\rm gas}}{{\rm K\ km\ s^{-1}\ pc^2}}\right)^N\ \times 10^{\mathcal{N}(0,\epsilon)}
\end{equation}
where $A$ is the normalization factor, $N$ is the power-law index, and $\mathcal{N}(0,\epsilon)$ is the intrinsic, log-normally distributed scatter on the relation with zero mean and standard deviation $\epsilon$. Our data are mainly limited by the sensitivity of the dense-gas observations as can be seen in Fig.~\ref{fig:correlation}. For the non-detections of \hcnfour\ and \hcopfour\ emission, we use the measurements in our fits and exclude data with velocity-integrated intensity $I_{\rm gas}\leq 0$. Similarly to \citet{leroy13}, we grid our data in ${\rm log}_{\rm 10} (L_{\rm IR})-{\rm log}_{\rm 10} (L^\prime_{\rm gas})$ space using cells 0.75 dex wide in both dimensions. 

%%%%%%%%%%- Table-4 -%%%%%%%%%%

\begin{deluxetable}{lccc}[t!]
\tablecaption{Results of Monte Carlo Fitting to Equation(\ref{eq:mc}) \label{tab:mc}}
\addtolength{\tabcolsep}{-1.0pt}
\tablewidth{0pt}
\tablehead{
\colhead{Molecule} & \colhead{log$_{10}\ A$} & \colhead{$N$} & \colhead{$\epsilon$}\\
\colhead{} & \colhead{($L_\odot$)} & & \colhead{(dex)} 
}
\startdata
\hcnfour\ & 3.58$\pm$0.25 & 1.00$\pm$0.04 & 0.53$\pm$0.04 \\
\hcopfour\ & 2.95$\pm$0.34 & 1.10$\pm$0.04 & 0.32$\pm$0.11 \\
\enddata
\tablecomments{The Best-fit values for parameters in Equation(\ref{eq:mc}) for the $L_{\rm IR}-L^\prime_{\rm HCN(4-3)}$ and the $L_{\rm IR}-L^\prime_{\rm HCO^+(4-3)}$ relations. The uncertainties are estimated from a bootstrapping approach described in Appendix~\ref{appendix}.}
\end{deluxetable}

%%%%%%%%%%%%%%%%%%%%%%%%%%

The detailed fitting procedure to our data using a MC approach is described in Appendix A. Table~\ref{tab:mc} reports the results of the MC fits for different dense-gas tracers. For the $L_{\rm IR}-L^\prime_{\rm HCN(4-3)}$ relation we measure a power-law index $N=1.00\pm0.04$, an amplitude $A=10^{3.58\pm0.25}$, and an intrinsic scatter $\epsilon=0.53\pm0.04$ dex, while for the $L_{\rm IR}-L^\prime_{\rm HCO^+(4-3)}$ relation we obtain a power-law index $N=1.10\pm0.04$, an amplitude $A=10^{2.95\pm0.34}$, and an intrinsic scatter $\epsilon=0.32\pm0.11$ dex. The best-fit slope and amplitude are in good agreement with the results obtained based on the bivariate linear fit using clipped data in Fig.~\ref{fig:correlation}. The intrinsic scatter of $0.53\pm0.04$ dex and $0.32\pm0.11$ dex derived based on the MC fits is significant, implying that the IR luminosity can vary by a factor of $\sim 2-4$ for regions having the same dense molecular line luminosity.

%%%%%%%%%- Fig-5 -%%%%%%%%
\begin{figure*}[htbp]
\centering
\includegraphics[scale=0.7]{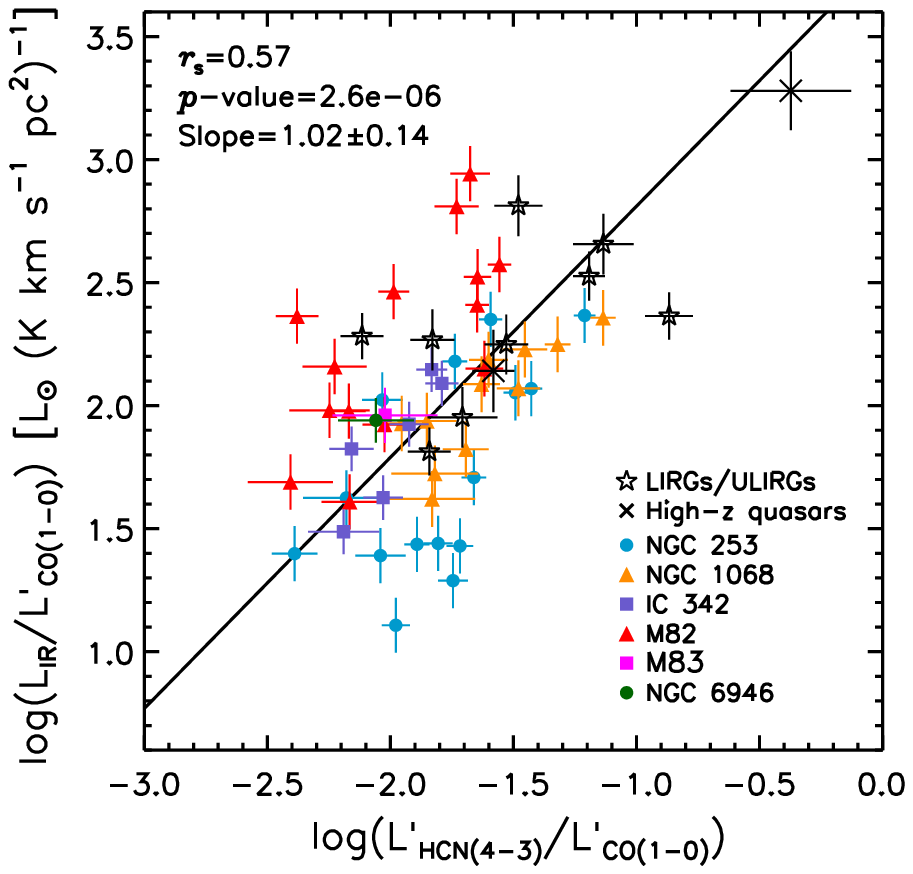}
\includegraphics[scale=0.7]{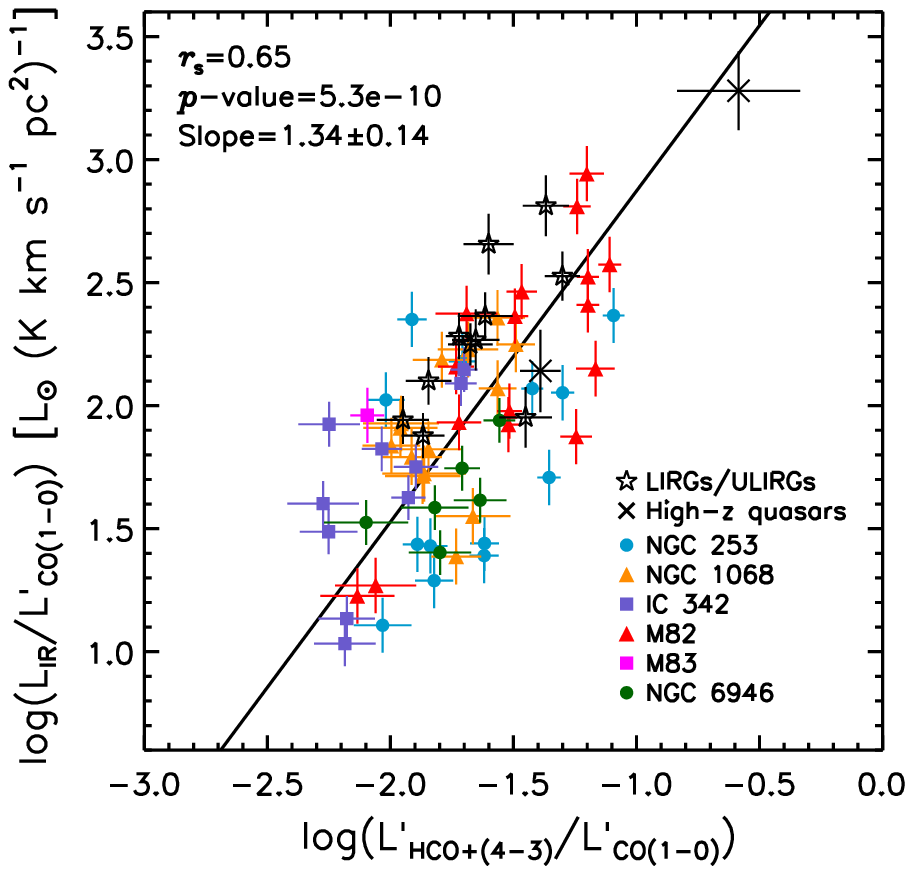}
\includegraphics[scale=0.7]{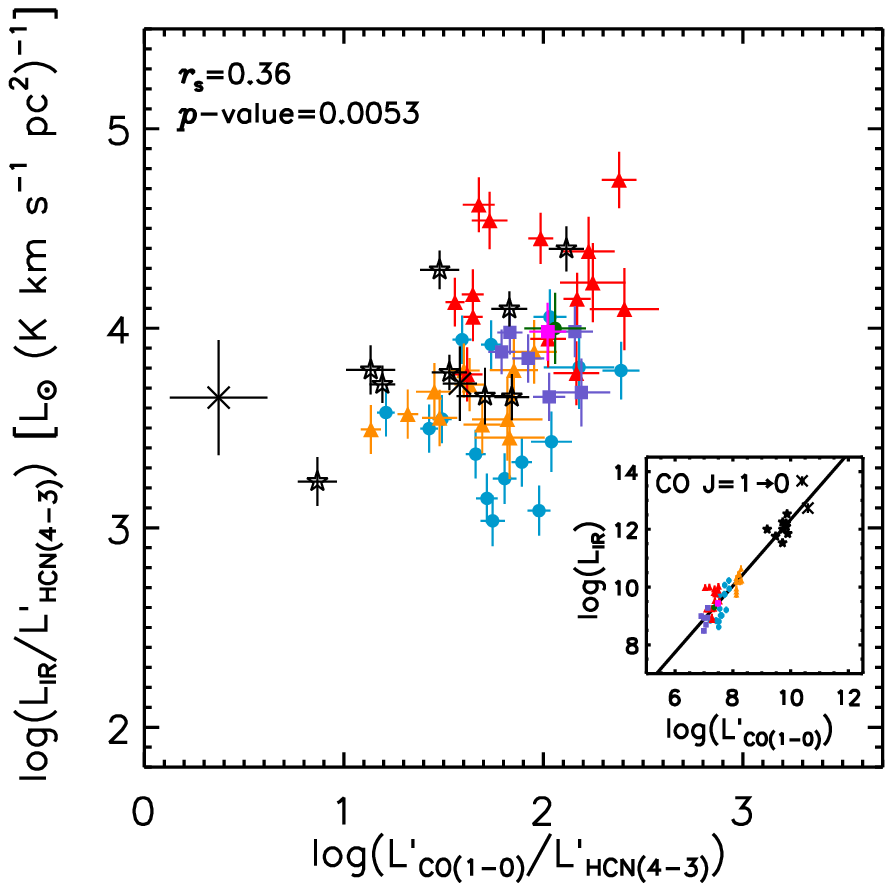}
\includegraphics[scale=0.7]{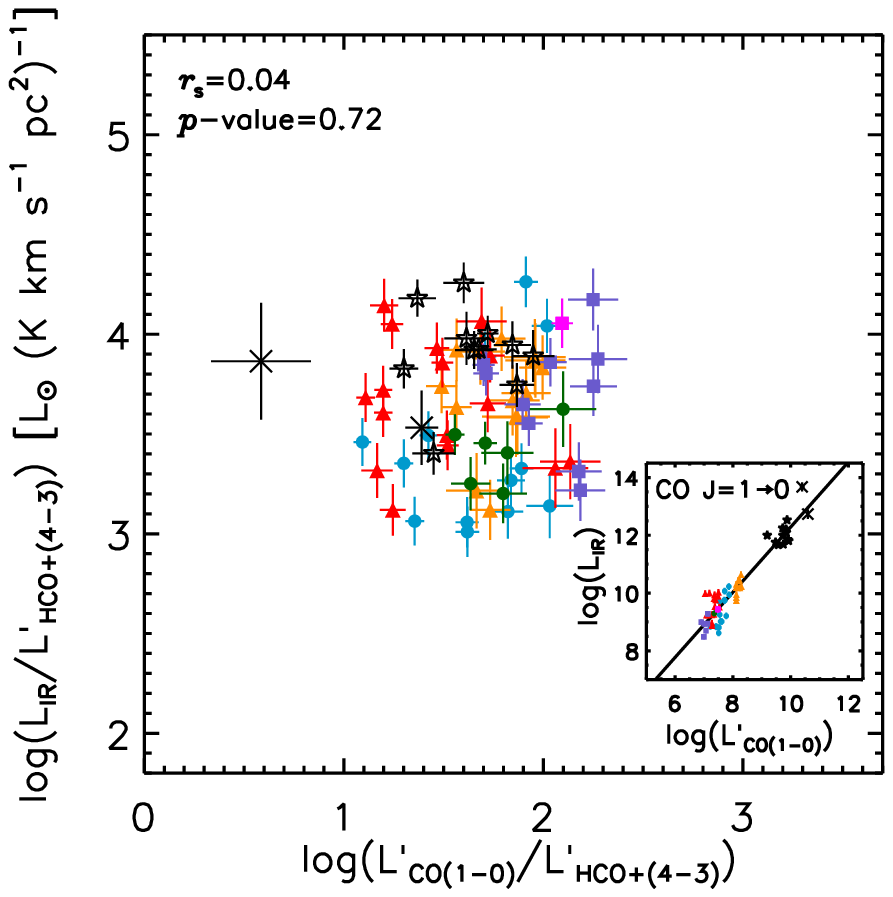}
\caption{{\it Top row}: $L_{\rm IR}/L'_{\rm CO(1-0)}$ as a function of $L'_{\rm HCN(4-3)}/L'_{\rm CO(1-0)}$ ({\it top left}) and $L'_{\rm HCO^+(4-3)}/L'_{\rm CO(1-0)}$ ({\it top right}) for nearby star-forming galaxies ({\it colored symbols}), local (U)LIRGs ({\it open stars}), and high-$z$ quasars ({\it crosses}). The IR and dense molecular line luminosities are normalized by $L'_{\rm CO(1-0)}$ to remove the galaxy distance and size dependencies. {\it Bottom row}: similar to the top panels, but instead normalized by \hcnfour\ (bottom-left panel) and \hcopfour\ (bottom-right panel) instead. The inset shows the correlation between $L_{\rm IR}$ and $L'_{\rm CO(1-0)}$. The Spearman rank correlation coefficient for each panel is listed at the top left.\label{fig:fraction}}
\end{figure*}
%%%%%%%%%%%%%%%%%%%%%

\subsection{Comparison of Correlations Between the Ratios}

To eliminate the distance and the galaxy size dependencies that could introduce a potentially strong correlation of $L_{\rm IR}$ with $L'_{\rm dense}$, where the dense gas is traced by the \hcnfour\ and the \hcopfour\ emission, we follow the same approach as that adopted by \citet{gao04b} to examine the correlation between the luminosity ratios $L_{\rm IR}$/$L'_{\rm CO}$ and $L'_{\rm dense}$/$L'_{\rm CO}$. A linear least-squares fit to our data including global measurements of local (U)LIRGs and high-z quasars gives slopes of 1.27$\pm$0.15 and 1.45$\pm$0.15, respectively. A Spearman test yields a correlation coefficient of $r_{\rm s}=0.57$ with the significance of its deviation from the zero of a $p$-value of 2.6$\times 10^{-6}$ for \hcnfour\ , and $r_{\rm s}=0.65$ with a $p$-value of 5.3$\times 10^{-10}$ for \hcopfour\ , suggesting a moderately significant correlation between $L_{\rm IR}/L'_{\rm CO}$ and $L'_{\rm dense}/L'_{\rm CO}$ (see Fig.~\ref{fig:fraction}\, (top)). 

Similarly, in Fig.~\ref{fig:fraction}\,(bottom) we plot the correlation between $L_{\rm IR}$ and $L'_{\rm CO}$ divided by $L'_{\rm dense}$ for normalization. The correlation between $L_{\rm IR}/L'_{\rm HCN(4-3)}$ and $L'_{\rm CO}/L'_{\rm HCN(4-3)}$ is found to be weaker ($r_{\rm s}=0.36$) than the correlation between $L_{\rm IR}$ and $L'_{\rm HCN(4-3)}$ normalized by $L'_{\rm CO}$, and with a higher $p$-value of 0.0053. For the correlation between $L_{\rm IR}/L'_{\rm HCO^+(4-3)}$ and $L'_{\rm CO}/L'_{\rm HCO^+(4-3)}$, the Spearman test gives a correlation coefficient of 0.04 with a $p$-value of 0.72, suggesting that the significance of the correlation between the luminosity ratios is very low, although a strong correlation is seen between IR and CO (see the insets in Fig.~\ref{fig:fraction}\,(bottom)). 

The results of this work are limited by the dynamical range of the $L'_{\rm dense}/L'_{\rm CO}$ ratio (about 2 dex) and the large scatter, as well as the effect of correlated axes (both normalized with a same variable), it remains unclear how strong the physical correlation between SFR and dense gas is. It is beyond the scope of this paper to analyze in detail the origin of the possible physical correlation. Our results are consistent with the correlation between the IR and the HCN(1-0) luminosities shown in \citet{gao04b}. Moreover, a tight linear correlation between the surface densities of the dense molecular gas and the SF rates, as well as between the HCN luminosity and the radio continuum luminosity, has been established for a large sample of galaxies \citep[e.g.,][]{liu10,lliu15,chen15,chen17}. All of these results indicate that the star formation is very much likely physically related to the dense molecular gas.

To statistically quantify the detailed physical relationship between dense molecular gas and star formation with models, analysis of the entire dataset of MALATANG and the combination of all datasets from available dense gas surveys \citep[e.g.,][]{gao04a,gao04b,zhang14,usero15,bigiel16}, and the investigation of the dependence on different parameters are required. We will address this subject in future work.

\subsection{Comparison with Literature Data}\label{subsec:compare}

The nearly unity power-law slopes derived for the $L_{\rm IR}-L'_{\rm HCN(4-3)}$ and the $L_{\rm IR}-L'_{\rm HCO^+(4-3)}$ correlations from our fits are in good agreement with \citet{zhang14}. The slightly super-linear slope of the $L_{\rm IR}-L'_{\rm HCO^+(4-3)}$ correlation derived from our fit also agrees with that obtained by \citet{zhang14}, who speculate the super-linear slope is likely to be a result of a decrease of the HCO$^+$ abundance in extreme physical conditions. For example, in extreme IR luminous galaxies, an increase of free electrons created by cosmic-ray ionization would accelerate the destruction of HCO$^+$ by dissociative recombination \citep{seaquist00}. The self-absorption feature of HCO$^+$ emission line is often observed in the Galactic dense clumps \citep[e.g.,][]{reiter11}. However, it is not easy to investigate thoroughly the physical origin, since HCO$^+$ is an ion and follows a more complex chemistry \citep{omont07,papadopoulos07}. Observations of other molecular ions that probe dense gas, such as N$_2$H$^+$, would provide clues to test the hypothesis.

Nevertheless, the linear correlation between $L_{\rm IR}$ and $L'_{\rm HCN(4-3)}$ is similar to that derived for the \jone\ lines of HCN and HCO$^+$ \citep[e.g.,][]{gao04b,wu05,baan08,bigiel15,bigiel16,usero15,chen17} and the \csseven\ line \citep{zhang14}. All of these correlations hold over a wide IR luminosity range covering nearly 10 orders of magnitude, providing evidence to support the argument that the SFR is directly proportional to the total mass of dense gas, and does not depend on the exact value of the gas density once the gas is denser than a threshold density of $\sim 10^4$ cm$^{-3}$ \citep{lada12}. All of these dense-gas tracers have a critical density higher than this threshold and the critical densities ($n_{\rm crit} \sim (3-6)\times10^6$ cm$^{-3}$) for \hcnfour\ and \csseven\  are about two orders of magnitude higher than \hcnone. These results are inconsistent with the sub-linear relations (e.g., power-law slope of 0.6$\pm$0.1 and 0.7$\pm$0.1 for the $L_{\rm IR}-L'_{\rm HCN(4-3)}$ and the $L_{\rm IR}-L'_{\rm HCO^+(4-3)}$ relations, respectively) predicted by numerical simulations, which concluded that the SFR$-L'_{\rm gas}$ slope tends to decrease with increasing $n_{\rm crit}$ \citep{narayanan08,juneau09}. Our mapping observations show direct evidence that a portion of dense gas as traced by the \hcnfour\ and the \hcopfour\ emission is distributed in the off-nuclear regions. It thus cannot be ruled out that the sub-linear slope ($\sim0.8\pm0.1$) obtained by \citet{bussmann08} is a result of underestimating of the total \hcnthree\ emission for nearby galaxies, as the line intensities are measured from a single point toward the galaxy center with beam size of $\sim 30\arcsec$, while the IR luminosities are derived from IRAS flux densities that measured with a larger beam size. 

\subsection{Variation in the Infrared-to-Molecular-Line Luminosity Ratio}\label{subsec:ir2lgas}

\begin{figure*}[htbp]
\plottwo{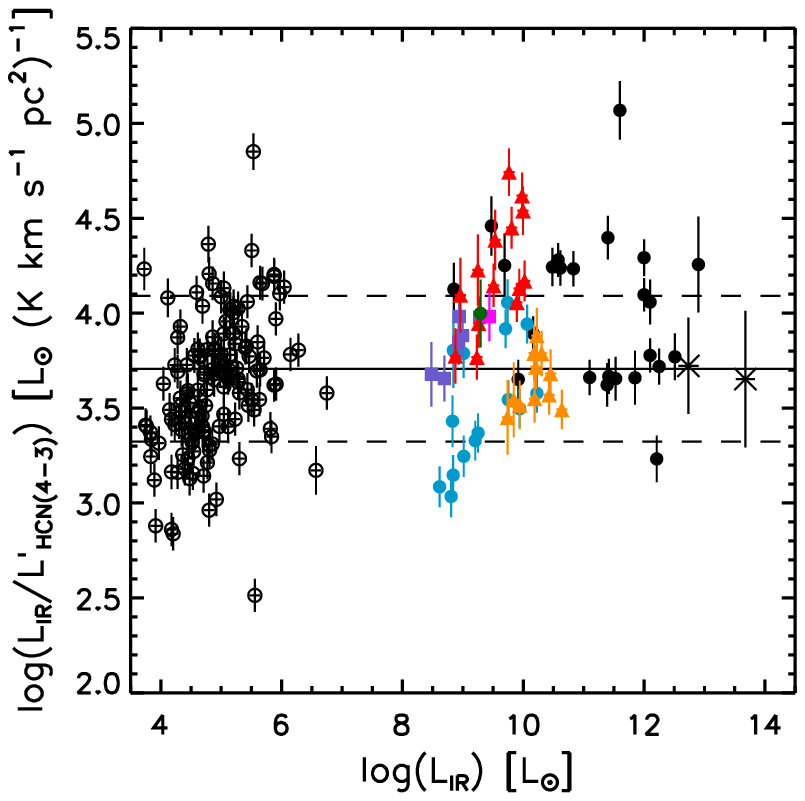}{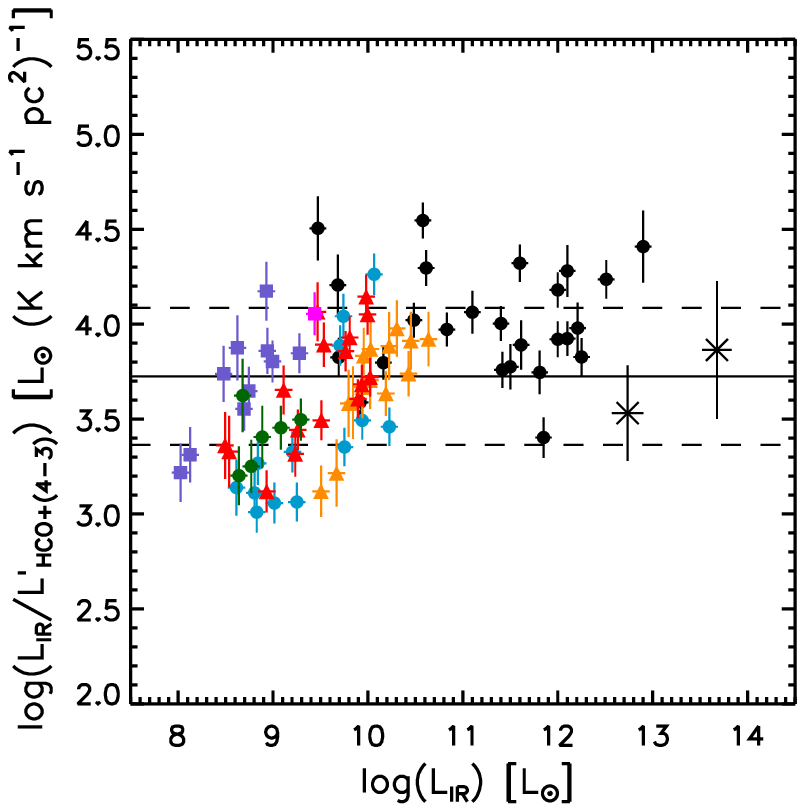}
\caption{{\it Left}: the luminosity ratio of IR to HCN(4-3) as a function of $L_{\rm IR}$ for Galactic clumps ({\it circles}), our sample of galaxies resolved to sub-kpc scales ({\it colored symbols}), normal galaxies and local (U)LIRGs ({\it solid circles}), and high-$z$ quasars ({\it crosses}). {\it Right}: the luminosity ratio of IR to HCO$^+$(4-3) as a function of $L_{\rm IR}$ for the same set of galaxies as in the left panel, but without the sample of Galactic clumps. Symbols are as in Fig.~\ref{fig:ir2hcn}. The mean values of log($L_{\rm IR}/L'_{\rm HCN(4-3)}$) and log($L_{\rm IR}/L'_{\rm HCO^+(4-3)}$) for the full sample of galaxies are 3.71$\pm$0.03 and 3.73$\pm$0.04 ({\it solid lines}), with an rms scatter of 0.38 dex and 0.36 dex ({\it dashed lines}), respectively. \label{fig:scatter}}
\end{figure*}

In Fig.~\ref{fig:scatter}\,(left) we show the ratio of $L_{\rm IR}/L'_{\rm HCN(4-3)}$ as a function of $L_{\rm IR}$. The mean value of log($L_{\rm IR}/L'_{\rm HCN(4-3)}$) for Galactic clumps, normal star-forming galaxies, and (U)LIRGs/high-$z$ quasars are 3.62$\pm$0.03, 3.89$\pm$0.06, and 3.86$\pm$0.10, with an rms scatter of 0.35 dex, 0.35 dex, and 0.32 dex, respectively. The mean value of log($L_{\rm IR}/L'_{\rm HCN(4-3)}$) for the full sample of galaxies is 3.71$\pm$0.03, with an rms scatter of 0.38 dex. The ratio of $L_{\rm IR}/L'_{\rm HCO^+(4-3)}$ shows a similar scatter (see Fig.~\ref{fig:scatter}\,(right)). The mean log($L_{\rm IR}/L'_{\rm HCO^+(4-3)}$) for normal galaxies and (U)LIRGs/high-$z$ quasars are 3.77$\pm$0.05 and 4.02$\pm$0.08 with an rms scatter of 0.33 dex and 0.26 dex respectively, while the mean value measured for the full sample is 3.73$\pm$0.04 with an rms scatter of 0.36 dex. The mean $L_{\rm IR}/L'_{\rm dense}$ ratio measured across the whole population of galaxies in our sample appears to vary little. This is similar to the IR/HCN(1-0) and the IR/HCO$^+$(1-0) data, which are found to be independent of $L_{\rm IR}$ extending from galaxy scales to individual GMCs \citep[e.g.,][]{gao04b,wu05,chen17}. Note that there is significant scatter measured within our sample of galaxies, which is in good agreement with the intrinsic scatter derived from MC fitting (see Sect.~\ref{sec:ir2dense}). A plausible explanation for the large scatter could be a wide range of physical conditions for the molecular gas in the dense phase and/or abundance variations \citep[e.g.,][]{jackson95,papadopoulos07}. However, it is worth noting that both the IR/HCN(4-3) and the IR/HCO$^+$(4-3) ratios show systematic variations with IR luminosity within individual spatially resolved galaxies, and also the Galactic clumps, though with significant scatter. We discuss possible explanations for these trends in the next subsection.

%%%%%%%%%- Fig-6 -%%%%%%%%
%%%%%%%%%%%%%%%%%%%%%

Compared with other galaxies in our sample, M82 appear weakened in HCN(4-3) relative to IR with ratio of IR/HCN(4-3) mostly above \mbox{1 $\sigma$} scatter (see Fig.~\ref{fig:scatter}\,(left)), while the IR/HCO$^+$(4-3) ratio shown in Fig.~\ref{fig:scatter}\,(right) is well within the \mbox{1 $\sigma$} scatter. A plausible explanation for the decrease of $L'_{\rm HCN(4-3)}/L'_{\rm HCO^+(4-3)}$ could be a low HCN abundance in M82. \citet{braine17} observed various molecular lines in low-metallicity Local Group galaxies and found that both HCN and HNC lines are weak with respect to the IR emission, while HCO$^+$ follows the trends observed in galaxies with solar metallicity. They attributed the weakness of the nitrogen bearing molecules to the low nitrogen abundance in these galaxies, based on the observed trend in the HCN/HCO$^+$ ratio with metallicity. The weak \hcnfour\ emission observed in M82 may be a similar effect, as there is some evidence of sub-solar metallicity for this galaxy \citep[e.g.,][]{origlia04,nagao11}.

Another possible explanation is related to the relatively low gas density observed in M82. In a study of HCN and HCO$^+$ in transitions up to \jfour, \citet{jackson95} found that the HCN \jfour/\jone\ line ratio is significantly smaller for M82 than for NGC 253, both of which are starburst galaxies with intense star formation in galactic nuclei and have comparable IR luminosities. A single-component gas excitation model indicates that the average gas density $n({\rm H_2})$ is at least 10 times lower in M82 ($\sim 10^4$ cm$^{-3}$) than in NGC 253 ($\sim 5\times 10^5$ cm$^{-3}$)  \citep{jackson95,knudsen07,naylor10}. Compared with the extremely low line ratio of HCN \jfour/\jone\ ($<0.1$) observed in M82, a factor of more than 10 times lower than in NGC 253, the HCO$^+$ \jfour/\jone\ line ratio of M82 also shows a lower value ($\sim0.3$) than that of NGC 253, but only by a factor of 2-3 \citep{jackson95}. Other observations of M82 show that the HCN/HCO$^+$ \jthree/\jone\ line ratio is larger than \jfour/\jone\ ratio \citep[e.g.,][]{nguyen89,wild92,seaquist00}. This may imply a lack of molecular gas with high density, in which case HCO$^+$ is more easily collisionally excited to \jfour\ than HCN, since the critical density of \hcnfour\ ($n_{\rm crit}\sim 5.6\times10^6$ cm$^{-3}$) is higher than that of \hcopfour\ ($n_{\rm crit}\sim 1.3\times10^6$ cm$^{-3}$) \citep{meijerink07,yamada07,greve09}. In addition, studies of chemical complexity toward the nuclear regions of M82 and NGC 253 by molecular line surveys reveal different chemical compositions for these two galaxies \citep[e.g.,][]{martin06,aladro11}. It is found that the nuclear starburst in M82 represents an evolved state where the heating of molecular clouds is driven by photo-dominated regions (PDRs), while the heating of NGC 253 is dominated by large-scale shocks \citep{martin06}. This could be a plausible explanation for the systematic difference of the IR/HCN(4-3) ratio between M82 and NGC 253 that is shown in Fig.~\ref{fig:scatter}\,(left).

%%%%%%%%%- Fig-7 -%%%%%%%%%
\begin{figure*}[htbp]
\plottwo{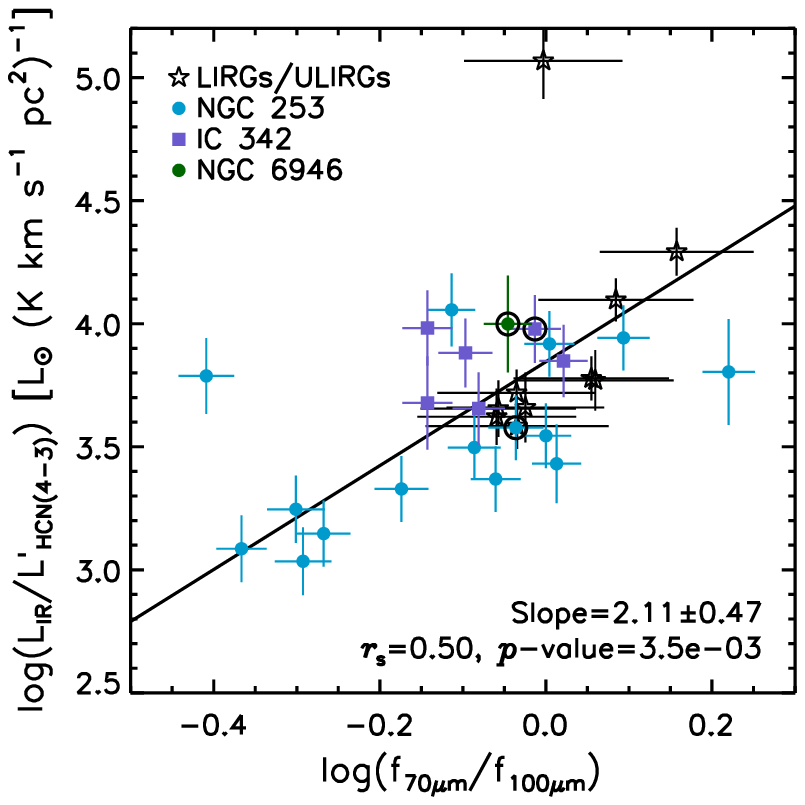}{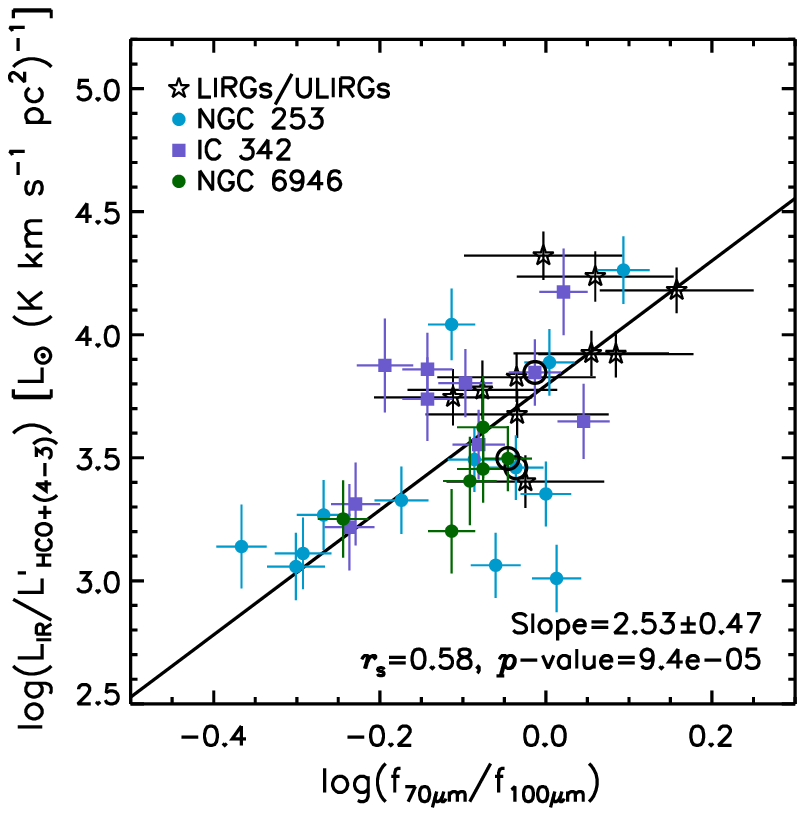}
\caption{$L_{\rm IR}/L'_{\rm HCN(4-3)}$ (left panel) and $L_{\rm IR}/L'_{\rm HCO^+(4-3)}$ (right panel) as a function of \mbox{70/100 $\mu$m} flux ratio for the galaxies in our sample where we have both PACS \mbox{70 $\mu$m} and \mbox{100 $\mu$m} data. The galaxy centers are highlighted with a black circle. The best-fit power-law index and the Spearman rank correlation coefficient for each panel are listed in the bottom right. \label{fig:dust}}
\end{figure*}
%%%%%%%%%%%%%%%%%%%%%%

\subsection{Correlations with Warm Dust Temperature}\label{subsec:tdust}

Fig.~\ref{fig:dust} shows the $L_{\rm IR}/L'_{\rm HCN(4-3)}$ ratio (left) and the $L_{\rm IR}/L'_{\rm HCO^+(4-3)}$ ratio (right) as a function of $f_{70\mu m}/f_{100\mu m}$ for NGC~253, IC~342, and NGC~6946 where we have both PACS \mbox{70 $\mu$m} and \mbox{100 $\mu$m} data. We adopt the PACS \mbox{70/100 $\mu$m} flux ratio as a proxy for warm-dust temperature, similar to the IRAS \mbox{60/100 $\mu$m} color which is often used to estimate the temperature of the warm-dust component \citep[$T_{\rm d}\sim 25-60$ K; e.g.,][]{solomon97,chanial07}. We also include a sample of local (U)LIRGs with PACS data from \citet{chu17} for comparison. A least-squares fit and a Spearman test yield log($L_{\rm IR}/L'_{\rm HCN(4-3)}$) = 2.1($\pm$0.5) log($f_{70}/f_{100}$) + 3.8 ($r_{\rm s}=0.50$, {\it p}-value=$3.5\times10^{-3}$) and log($L_{\rm IR}/L'_{\rm HCO^+(4-3)}$) = 2.5($\pm$0.5) log($f_{70}/f_{100}$) + 3.8 ($r_{\rm s}=0.58$, {\it p}-value=$9.4\times10^{-5}$) respectively, indicating that there is a statistically significant correlation between $L_{\rm IR}/L'_{\rm dense}$ and $T_{\rm d}$. These correlations are slightly stronger than the correlation between $L_{\rm IR}/L'_{\rm HCN(1-0)}$ and $f_{60\mu m}/f_{100\mu m}$, but not as strong as $L_{\rm IR}/L'_{\rm CO(1-0)}$ versus $f_{60\mu m}/f_{100\mu m}$ correlation \citep[correlation coefficient of 0.85; see Fig. 9 in][]{gao04b,lliu15}.

Comparing the $L_{\rm IR}/L'_{\rm dense}-f_{70\mu m}/f_{100\mu m}$ relation with the $L_{\rm IR}/L'_{\rm dense}-L_{\rm IR}$ relation for the individual galaxies, spatially resolved at sub-kpc scales shown in Fig.~\ref{fig:scatter}, we find that the ratio of $L_{\rm IR}/L'_{\rm dense}$ correlates with both the dust temperature indicated by the observed \mbox{70/100 $\mu$m} flux ratio and the IR luminosity. We speculate that the rising trend of $L_{\rm IR}/L'_{\rm dense}$ with $L_{\rm IR}$ observed is likely driven primarily by or related to the correlation of $L_{\rm IR}/L'_{\rm dense}$ with $T_{\rm d}$, since the IR emission from dust grains depends closely on the dust temperature.

\subsection{Correlations Between SFE and Dense Gas Fraction}

\begin{figure*}[htbp]
\centering
\includegraphics[scale=1.0]{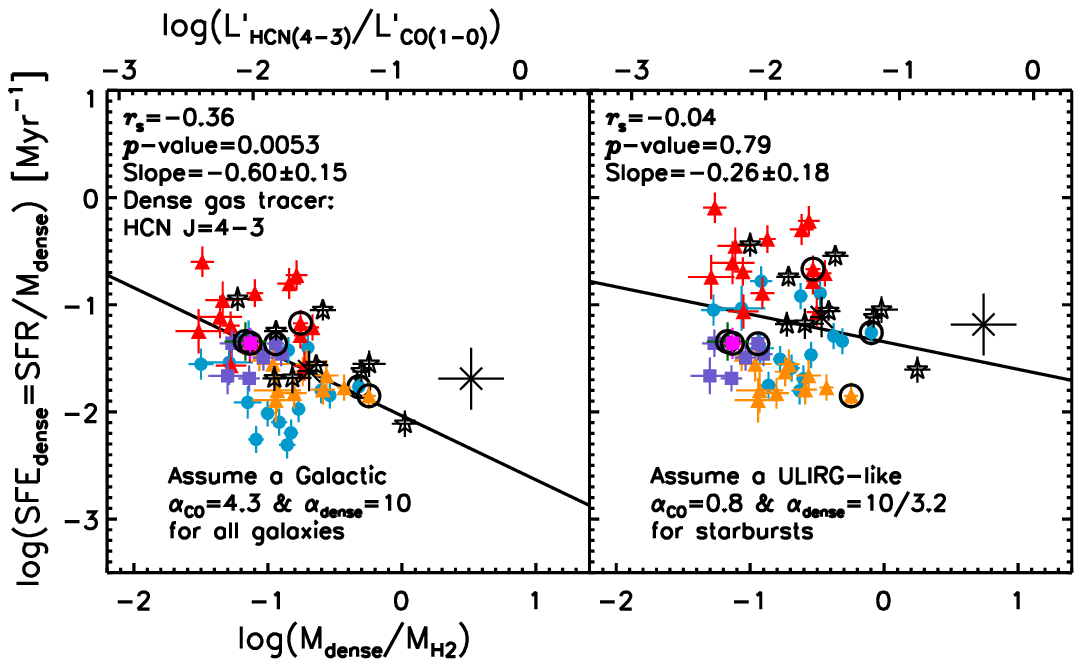}
\includegraphics[scale=1.0]{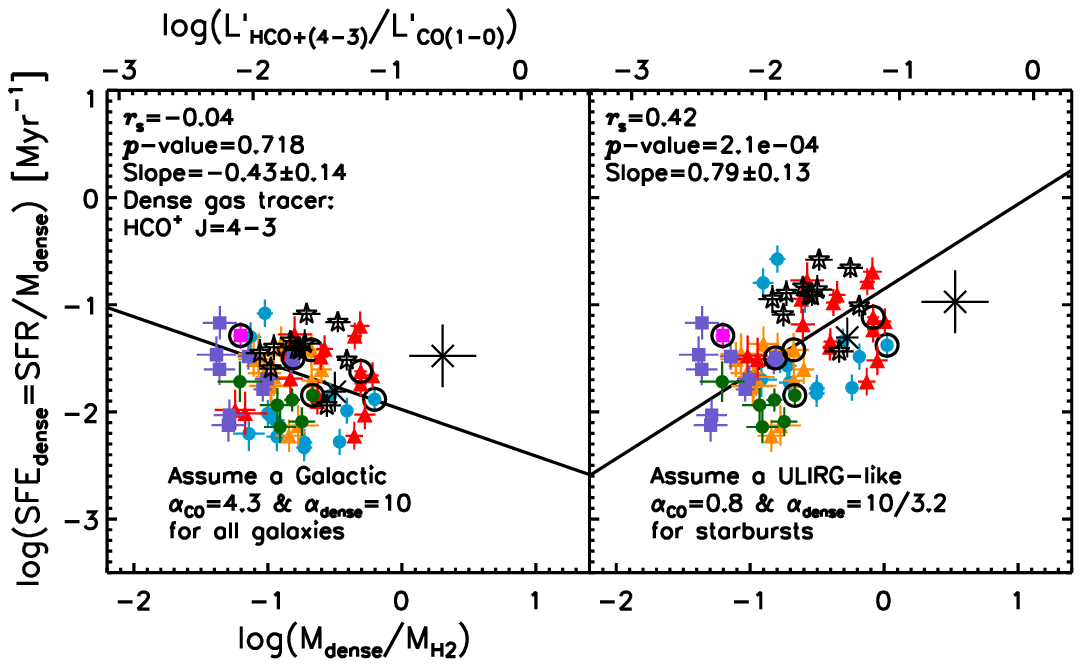}
\caption{SFE of the dense molecular gas as a function of the dense-gas fraction, with dense gas as traced by the \hcnfour\ (top row) and the \hcopfour\ (bottom row) lines for the sample of galaxies compiled in this work. The left panels show the data that assume a Galactic $\alpha_{\rm CO}$ of 4.3 and $\alpha_{\rm dense}$ of 10 for the full sample of galaxies, while the right panels represent the data that adopt a ULIRG-like $\alpha_{\rm CO}$ of 0.8 and $\alpha_{\rm dense}$ of 10/3.2 for starbursts (NGC 253, M82, (U)LIRGs, and high-$z$ quasars). We assume a fixed $\alpha_{\rm dense}$ and line brightness temperature ratio $r_{41}$ to estimate the mass of molecular gas in the dense phase for the full sample of galaxies. Symbols are as in Fig.~\ref{fig:fraction}. The data points highlighted with a black circle denote the central position of each galaxy. The Spearman rank correlation coefficient for each panel is listed in the top left.\label{fig:sfedense}}
\end{figure*}

Fig.~\ref{fig:sfedense} shows the SFE of the dense molecular gas (SFE$_{\rm dense} \tbond$ SFR/$M_{\rm dense}$) as a function of the dense-molecular-gas fraction ($f_{\rm dense}$). The dense-gas content is traced by the \hcnfour\ (top panel) and the \hcopfour\ (bottom panel) lines, respectively. The SFR is estimated from the total IR luminosity based on the calibrations of \citet{kennicutt98} and \citet{murphy11}:
\begin{equation}
\left(\frac{{\rm SFR}}{M_\odot\ {\rm yr}^{-1}}\right) = 1.50\times 10^{-10} \left(\frac{L_{\rm IR}}{L_\odot}\right).
\end{equation}
The SFR is calculated based on a \citet{kroupa01} IMF. To better compare with previous work which focuses mostly on the luminosity ratio of HCN/CO at \jone\ as a measure of the dense-gas fraction \citep[e.g.,][]{gao04b,usero15}, we convert the \jfour\ line luminosity of HCN and HCO$^+$ to the dense-gas mass, and the CO \jone\ luminosity to the total molecular-gas mass, by assuming  conversion factors\footnote{The units of the luminosity-to-mass conversion factor, $M_\odot$ (K km s$^{-1}$ pc$^2$)$^{-1}$, are omitted from the text for brevity.} of $\alpha_{\rm dense}$ and $\alpha_{\rm CO}$, respectively. We initially assume a Galactic $\alpha_{\rm CO}$ of 4.3 for the full sample of galaxies \citep{bolatto13} (left column).  

The mass of dense molecular gas can be estimated from $L'_{\rm HCN(4-3)}$ and $L'_{\rm HCO^+(4-3)}$,
\begin{equation}
M_{\rm dense}=\alpha_{\rm dense} \left(\frac{L'_{{\rm gas}\ J=4-3}}{r_{41}}\right),
\end{equation}
where $\alpha_{\rm dense}$ is the HCN(HCO$^+$) \jone -to-dense-gas-mass conversion factor and $r_{41}$ is the HCN(HCO$^+$) \jfour / \jone\ line ratio. For simplicity, we assume a fixed $\alpha_{\rm dense}$ = 10 for both HCN \jone\ and HCO$^+$ \jone\ emission, which was estimated by \citet{gao04b} for normal SF galaxies with a brightness temperature of $T_{\rm b}$ = 35 K. We adopt $r_{41}$ = 0.3, which is an average ratio estimated by comparing the \hcnfour\ data of \citet{zhang14} (including the data presented in this study) with the \hcnone\ data of \citet{gao04a}. Note that the $r_{41}$ we used is a rough estimate with large uncertainty partly due to the slightly different angular resolution of the \hcnfour\ and the \hcnone\ observations. Also note that the dense-gas mass of extreme systems, i.e., galaxies or regions that are more excited in molecular gas emission with higher gas temperature, is likely overestimated under the assumption of a fixed $r_{41}$. The \hcnfour\ observations of a few (U)LIRGs indeed show higher $r_{41}$ ranging from $\sim$0.3 to 1.0 \citep{papadopoulos07}. 

We adopt here the assumption that the $L_{\rm IR}/L'_{\rm HCN(4-3)}$ ratio is a proxy for the SFE of the dense gas (SFE$_{\rm dense}\propto L_{\rm IR}/L'_{\rm HCN(4-3)}$), assuming that both the $\alpha_{\rm dense}$ and the line ratio $r_{41}$ are constant for the full sample of galaxies. Similar assumptions are applied to the dense gas as traced by the \hcopfour\ line. A Spearman test yields a statistically insignificant correlation with $r_{\rm s}$ = -0.36 and a $p$-value of 0.0053 for \hcnfour\ and with $r_{\rm s}$ = -0.04 and a $p$-value of 0.718 for \hcopfour\ , indicating a very weak dependence, if any, between SFE$_{\rm dense}$ and $f_{\rm dense}$ within our sample. It is clearly illustrated in Fig.~\ref{fig:sfedense} that the fraction of dense gas is higher in starbursts and galactic centers ({\it circled point}) than in the outer regions of our sample galaxies, although there are some off-nuclear positions that have similar $f_{\rm dense}$ as the central region (e.g., M82). These are in good agreement with previous \hcnone\ studies and confirm the findings by \citet{gao04a,gao04b} that the starburst strength can be better indicated by the fraction of molecular gas in dense phase. For the nearby normal, star-forming galaxies, the mean log($f_{\rm dense}$) are -0.89$\pm$0.07 and -0.98$\pm$0.04, with an rms scatter of 0.28 dex and 0.22 dex, for the dense gas as traced by \hcnfour\ and \hcopfour\ , respectively. The small scatter in these ratios indicates that the dense-gas fraction varies little for galaxies with normal star formation activity, which seems to be a plausible explanation for the linear relation found between $\Sigma_{\rm SFR}$ and $\Sigma_{\rm gas}$ in nearby normal galaxies \citep[e.g.,][]{bigiel08,lada12}.

%%%%%%%%- Fig-8 -%%%%%%%%
%%%%%%%%%%%%%%%%%%%%

It is evident from Fig.~\ref{fig:sfedense}(left) that one of high-$z$ quasars in our sample exhibits excess dense-gas content with an unphysical dense-gas fraction ($f_{\rm dense}>1$), if we adopt a Galactic $\alpha_{\rm CO}$ and $\alpha_{\rm dense}$. Observations of luminous IR galaxies show that molecular clouds in starbursts are highly concentrated with large velocity dispersions and have higher average gas volume densities than a typical GMC in the Milky Way \citep[][and references therein]{bolatto13}, implying that a smaller $\alpha_{\rm CO}$ is more appropriate for these galaxies \citep[e.g.,][]{leroy15b,leroy15}. Note that the molecular-gas mass measured for the central nuclear regions of star-forming galaxies may be overestimated for a Galactic $\alpha_{\rm CO}$ \citep{sandstrom13}. It is expected that the CO-to-H$_2$ conversion factor has a dependence on the physical conditions in the molecular clouds, $\alpha_{\rm CO}\varpropto \rho^{0.5}/T_{\rm b}$, if we assume the emission orginates in the gravitationally bound and virialized cloud cores \citep[e.g.,][]{bolatto13}. The multiline analysis of HCN and HCO$^+$ by \citet{gracia08} presents evidence that $\alpha_{\rm HCN}$ is probably about three times lower in IR luminous galaxies. The potential variation of the dense-gas excitation (e.g., HCN \jfour/\jone\ line ratio) in different physical conditions could also play an important role in estimating the dense-gas content. 

With these results we assume a ULIRG-like $\alpha_{\rm CO}$=0.8, and $\alpha_{\rm dense}=\alpha^{\rm MW}_{\rm dense}$/3.2 for the dense gas as traced by the HCN and the HCO$^+$ emission in extreme starbursts (NGC 253, M82, (U)LIRGs, and high-z quasars), similar to the value adopted for LIRGs/ULIRGs in \citet{garcia12}. For comparison, in Fig.~\ref{fig:sfedense}\,(right) we also plot the data points for starbursts by assuming a revised $\alpha_{\rm CO}$ and $\alpha_{\rm dense}$, while keep using Galactic conversion factors for the remaining normal disk galaxies. A Spearman test to the data of extreme starbursts with gas content calculated with the revised $\alpha_{\rm CO}$ and $\alpha_{\rm dense}$ combined with normal disk galaxies, gives similar correlation coefficients ($r_{\rm s}$ = -0.04 and $p$-value = 0.79 for \hcnfour, $r_{\rm s}$ = 0.42 and $p$-value = 2.1$\times$10$^{-4}$ for \hcopfour) as the results derived based on the assumption of fixed conversion factors. The weak correlation revealed suggests that the efficiency of star formation in the dense gas is likely to be independent of dense-gas fraction. Keeping in mind the dependency revealed for $L_{\rm IR}/L'_{\rm dense}$ ratio with warm dust temperature shown in Sect.~\ref{subsec:tdust}, we note that the uncertainties of conversion factors ($\alpha_{\rm CO}$ and $\alpha_{\rm dense}$) may introduce some biases in interpreting the correlations between the SFE of dense gas and the dense gas fraction.

%%%%%%%%- Fig-9 -%%%%%%%%
\begin{figure*}[htbp]
\centering
\includegraphics[scale=1.0]{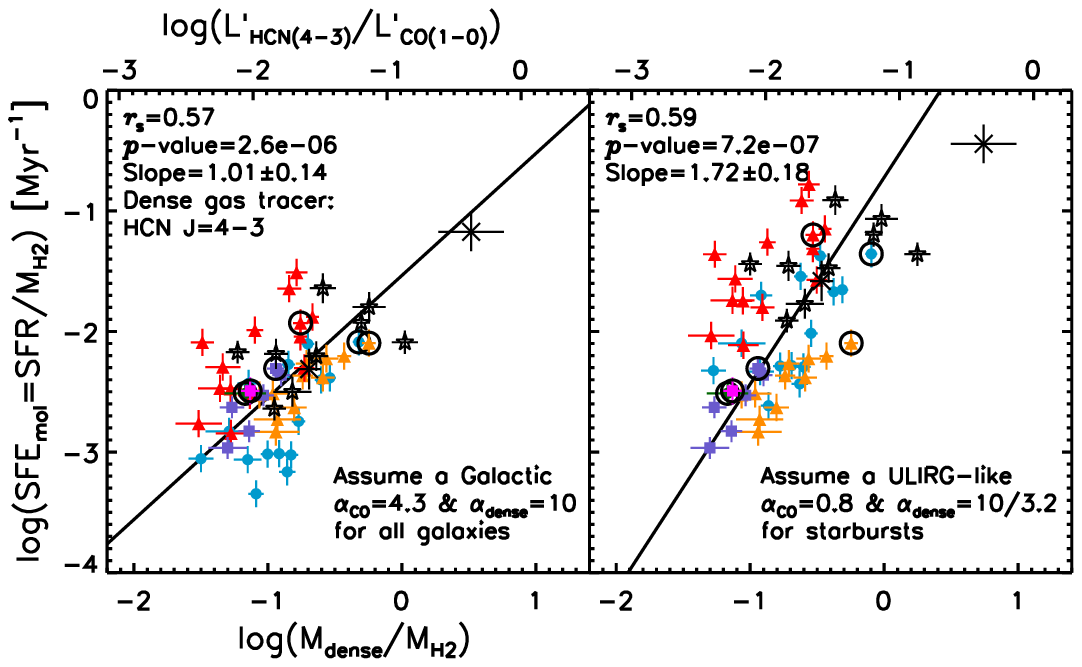}
\includegraphics[scale=1.0]{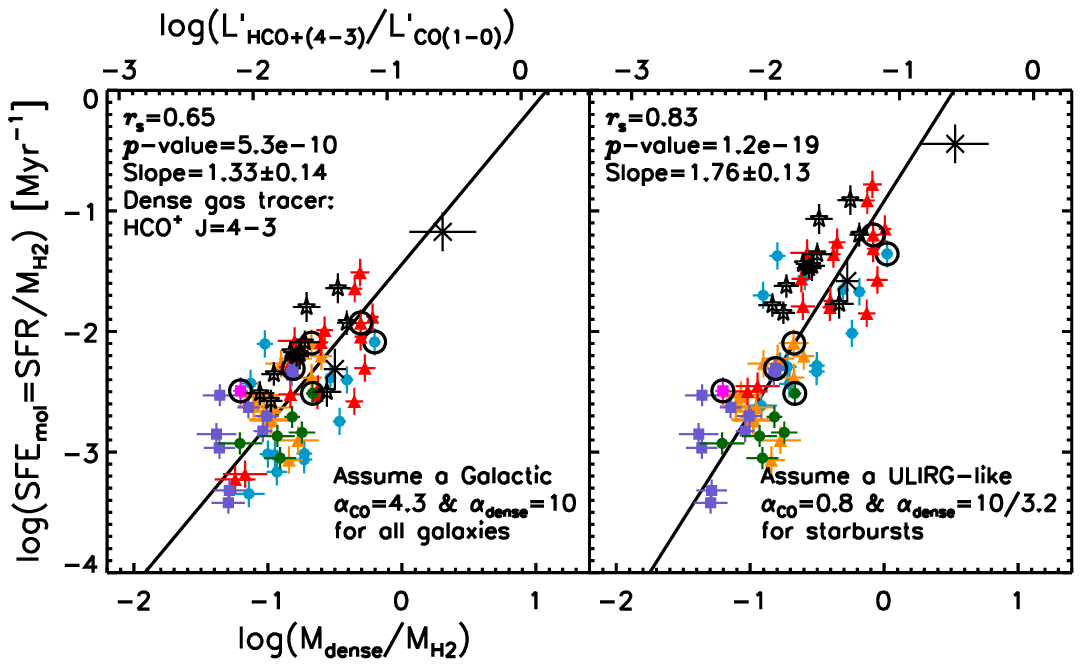}
\caption{Similar to Fig.~\ref{fig:sfedense}, but we plot the SFE of the total molecular gas as a function of the dense-gas fraction. \label{fig:sfemol}}
\end{figure*}
%%%%%%%%%%%%%%%%%%%%

We also plot the SFE of the total molecular gas, i.e., the inverse of the molecular-gas depletion time ($\tau_{\rm gas}$), as a function of $f_{\rm dense}$ (see Fig.~\ref{fig:sfemol}). These are similar to the relations shown in Fig.~\ref{fig:fraction}\,(top), but we convert the IR and line luminosities to SFR and gas mass, respectively. Similar as in Fig.~\ref{fig:sfedense}, we assume Galactic and ULIRG-like conversion factors for the starbursts in our sample for comparison, respectively. It is clearly that the SFE$_{\rm mol}$ increases with $f_{\rm dense}$ with a strong correlation coefficient ($r_{\rm s}\sim 0.6$ with $p$-value $<$10$^{-6}$ for \hcnfour\, and $r_{\rm s}\sim 0.7-0.8$ with $p$-value $<$10$^{-10}$ for \hcopfour). While a nearly constant SFE$_{\rm mol}$ is found for normal star-forming galaxy disks by \citet{usero15}, our data show that the SFE$_{\rm mol}$ is strongly correlated with $f_{\rm dense}$ when combining normal disks with more extreme luminous IR galaxies. Here we also note the large uncertainties involved in the derivation of correlations, due to the limited data points and the effect of correlated axes, as well as the assumption of conversion factors.

Compared with the normal galaxies, the higher SFE found in (U)LIRGs and the high-{\it z} quasars indicates that the latter will consume their total gas reservoir more quickly. This is consistent with the trend seen between $\tau_{\rm gas}$ and $L_{\rm FIR}$ where the depletion timescale is typically one order of magnitude shorter for ULIRGs and high-$z$ quasars with higher $L_{\rm FIR}$  than for normal spiral galaxies \citep[e.g.,][]{solomon05,daddi10,combes13,carilli13}. As expected, the galaxy centers tend to show higher SFE$_{\rm mol}$ than the outer regions as a result of the starburst environment in the galactic nuclear region.

%%%%%%%%%- Fig-10 -%%%%%%%%%
\begin{figure*}[htbp]
\centering
\includegraphics[scale=0.75]{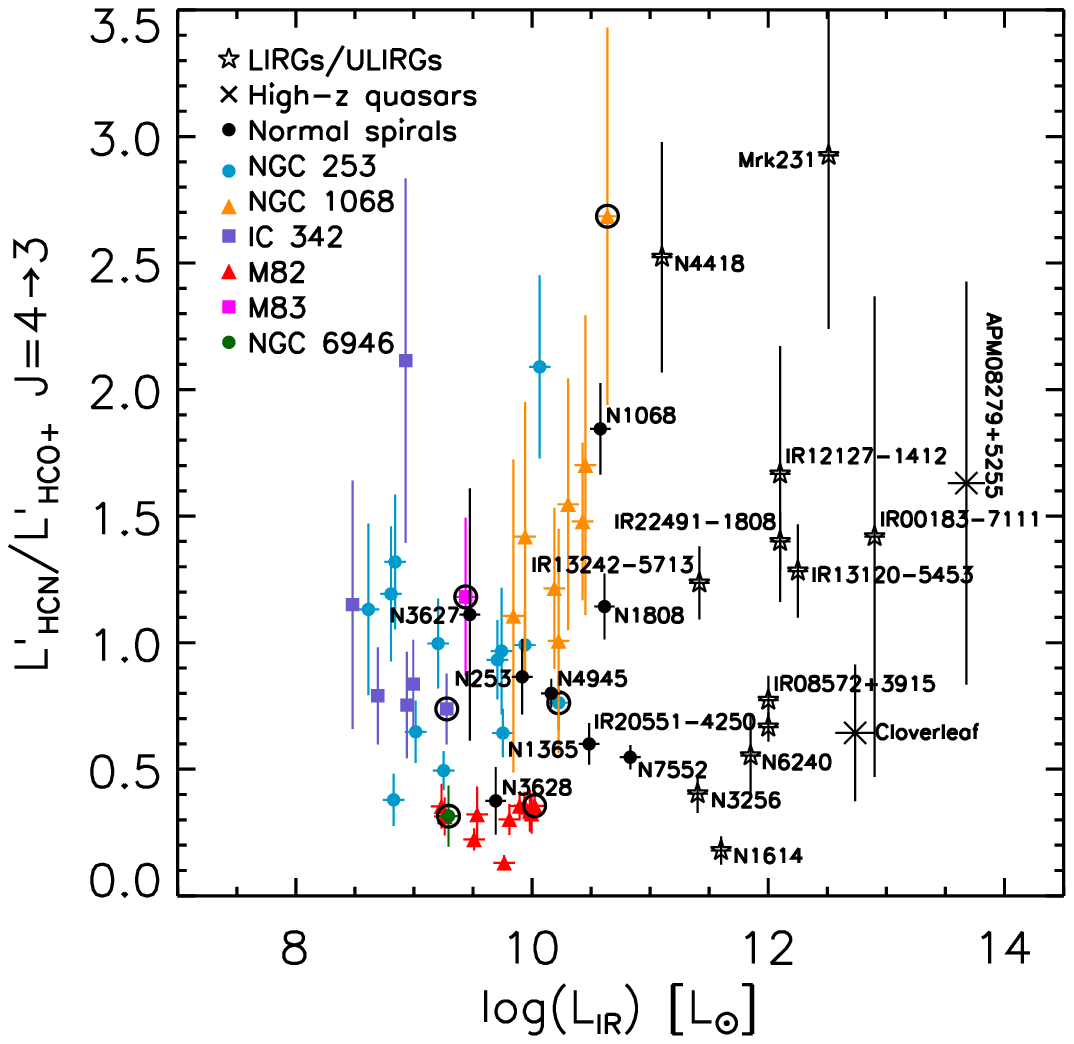}
\includegraphics[scale=0.75]{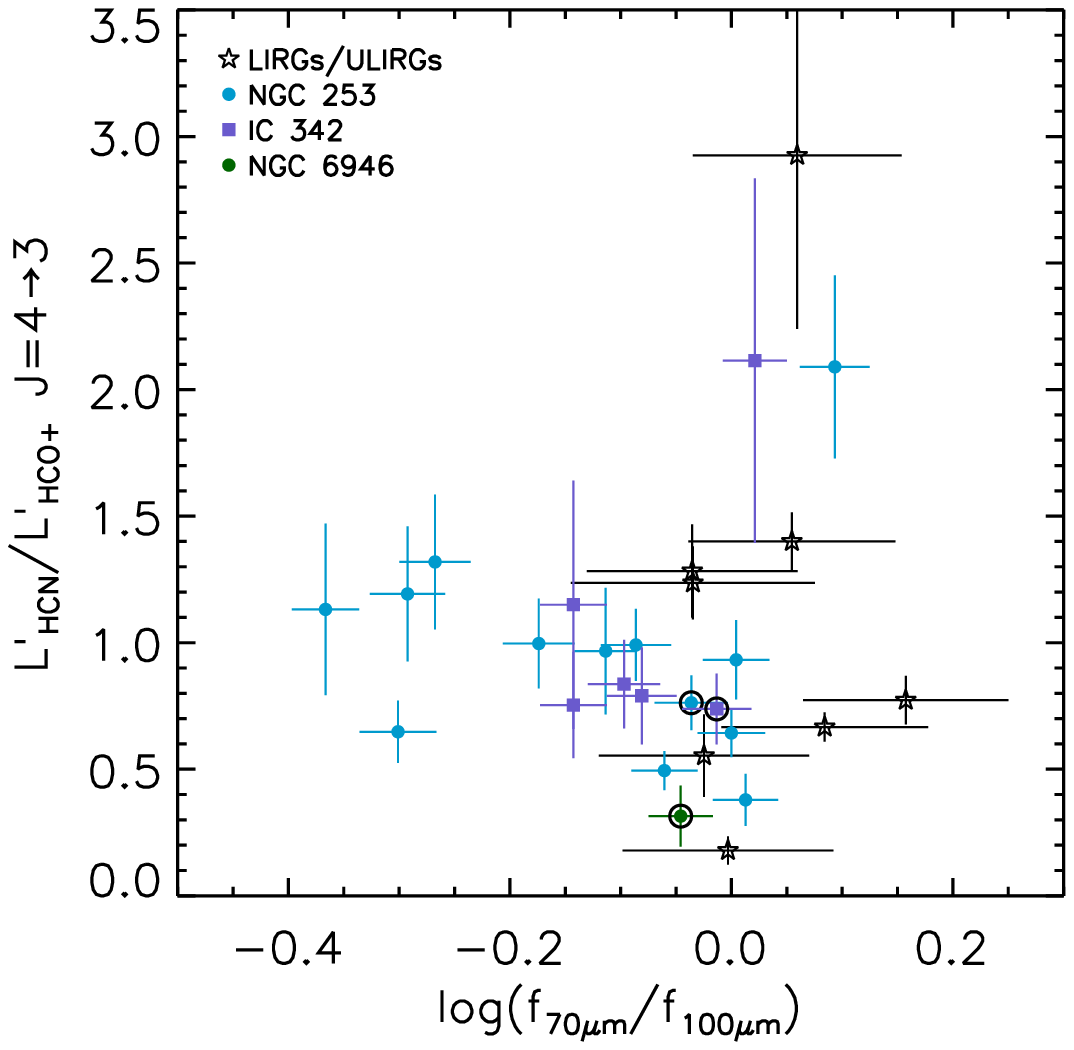}
\caption{ HCN/HCO$^+$ \jfour\ luminosity ratio as a function of IR luminosity (left panel) and \mbox{70 $\mu$m/100 $\mu$m} flux ratio (right panel) for our sample of galaxies that are spatially resolved ({\it colored symbols}), and the normal galaxies ({\it solid circles}), local (U)LIRGs ({\it open stars}), and high-$z$ quasars ({\it crosses}) from the literature. The galaxy centers of the six targeted galaxies are highlighted with black circles. \label{fig:ratio}}
\end{figure*}
%%%%%%%%%%%%%%%%%%%%%%%

\section{The HCN(4$-$3) to HCO$^+$(4$-$3) Line Ratio}\label{sec:ratio}

In Fig.~\ref{fig:ratio}\,(left), we plot the HCN to HCO$^+$ \jfour\ luminosity ratio as a function of IR luminosity for our target galaxies combining measurements of normal spirals, (U)LIRGs, and quasars from  the literature to inspect the variation of the HCN/HCO$^+$ line ratio. It is apparent that no systematic trend is found between $L'_{\rm HCN}/L'_{\rm HCO^+}$ and $L_{\rm IR}$. The $L'_{\rm HCN}/L'_{\rm HCO^+}$ \jfour\ ratio varies from 0.1 to 2.7 with a mean value of 0.9 and an rms scatter of 0.6 for the targeted six galaxies, while an average ratio of 0.8$\pm$0.5 is found for normal star-forming galaxies without AGN embedded. The $L'_{\rm HCN}/L'_{\rm HCO^+}$ \jfour\ ratio varies from 0.1 to 3.0 for the full sample of galaxies, in agreement with the HCN/HCO$^+$ luminosity ratios observed at \jone\ and \jthree\ transitions \citep[e.g.,][]{imanishi16,privon15,knudsen07,krips08}. 

The lowest HCN/HCO$^+$ line ratios in our sample are found in M82 and appear to be constant across the starburst disk with a mean value of $\sim0.3$, which is consistent with previous JCMT observations of M82 \citep{seaquist00}. As discussed in Sect.~\ref{subsec:ir2lgas}, we consider that the low HCN/HCO$^+$ ratio observed in M82 is more likely due to a deficit of HCN, rather than an increase of HCO$^+$, given that HCN is much weaker than HCO$^+$ with respect to the IR emission (see Fig.~\ref{fig:scatter}). Moreover, we speculate that the weakness of HCN in M82 could be attributed to the decrease of nitrogen abundance in the sub-solar metallicity environment and/or the relatively low gas density condition of this galaxy. For NGC~3628 and NGC~6946, which show comparably low HCN/HCO$^+$ ratios in Fig.~\ref{fig:ratio}\,(left), it has been found that their metallicities are sub-solar \citep{engelbracht08,gazak14}. In addition, NGC 3256 and NGC 1614 also show relatively low HCN/HCO$^+$ ratios ($\leq0.4$, see Fig.~\ref{fig:ratio}\,(left)). Weak \hcnone\ emission and a relatively low HCN/HCO$^+$ \jone\ ratio for NGC 1614 have also been reported by \citet{garcia12}. We speculate that the weakness of HCN in these two galaxies may be related to the deficiency of high-density gas, since both galaxies are merger remnants at an advanced merger stage that probably have dispersed their molecular gas by shocks from supernova explosions \citep[e.g.,][]{jackson95,costagliola11}. 

An enhancement of the HCN/HCO$^+$ abundance ratio in $X$-ray-dominated regions with modest densities ($n<10^4-10^5$ cm$^{-3}$) is predicted by theoretical models \citep[e.g.,][]{lepp96,meijerink07}. Observations show evidence for HCN enhancement in nearby galaxies hosting AGNs \citep[e.g.,][]{kohno01,krips08,izumi16}. We see from Fig.~\ref{fig:ratio}\,(left) that the Seyfert 2 galaxy NGC 1068 shows a high HCN/HCO$^+$ ratio with the highest value in the center, in contrast to the low ratio observed in the pure starburst, such as in M82. High HCN/HCO$^+$ line ratios are also found in NGC 4418 and Mrk 231, which could be associated with the enhancement of HCN by X-ray radiation from the AGN. The HCN and HCO$^+$ \jone\ observations also show relatively high line ratios for these galaxies \citep[e.g.,][]{imanishi04,costagliola11}. However, for the Cloverleaf, which is a high-$z$ quasar hosting AGN, a similar enhancement of HCN is not found. Instead, a relatively low HCN/HCO$^+$ ratio that is comparable to starburst-dominated systems is obtained for this galaxy. It has been argued that the variation of the HCN/HCO$^+$ ratio is likely determined by multiple processes including the interplay of radiation field and gas density \citep[e.g.,][]{papadopoulos07,harada10,harada13,privon15}. 

We also examine the relationship between the HCN/HCO$^+$ \jfour\ line ratio and the $f_{70\mu m}/f_{100\mu m}$ flux ratio for the galaxies where we have both PACS \mbox{70 $\mu$m} and \mbox{100 $\mu$m} data (see Fig.~\ref{fig:ratio}\,(right)). No significant correlation is found between HCN/HCO$^+$ and \mbox{70/100 $\mu$m} color temperature. A study of the excitation mechanisms for HCN and HCO$^+$ emission which includes low-$J$ observations will be presented in a future paper. 

\section{Summary} \label{sec:summary}
We have presented observations of the \hcnfour\ and the \hcopfour\ lines in the central $\sim50\arcsec \times 50\arcsec$ regions of six nearby star-forming galaxies from the JCMT program MALATANG. We combined these new data with previous multi-wavelength observations to study the relationships between the dense molecular gas as traced by the \jfour\ lines of HCN and HCO$^+$, the IR luminosity, and the dust and star-formation properties. Finally, we discussed the variation of the HCN/HCO$^+$ \jfour\ line ratio in different populations of galaxies. We summarize below the main results and conclusions of this work.

1. We detect HCN and HCO$^+$ \jfour\ emission in all six targeted galaxies at multiple positions except for M83 where only weak detections at the central position were obtained. Both the line profiles and line widths are found to be very similar for HCN and HCO$^+$, indicating that these two molecules are arising from the same region. 

2. All galaxies observed in our sample are spatially resolved at sub-kpc scales and follow the linear relation of $L_{\rm IR}-L'_{\rm dense}$ (dense gas as traced by HCN and HCO$^+$ \jfour) established globally in galaxies within the scatter. Our new data extend the relation to an intermediate luminosity regime to bridge the gap between Galactic clumps and integrated galaxies. The nearly linear slopes obtained for the $L_{\rm IR}-L'_{\rm HCN(4-3)}$ and $L_{\rm IR}-L'_{\rm HCO^+(4-3)}$ relations are inconsistent with the sub-linear relations predicted by some theoretical models.

3. We find that the $L_{\rm IR}/L'_{\rm dense}$ ratio shows a systematic trend with $L_{\rm IR}$ within individual galaxies, whereas the galaxy-integrated ratios vary little. Similar trends are also found between the $L_{\rm IR}/L_{\rm gas}$ ratio and the warm-dust temperature gauged by the \mbox{70/100 $\mu$m} flux ratio. 

4. Using appropriate conversion factors of $\alpha_{\rm CO}$ and $\alpha_{\rm dense}$ for normal star-forming galaxies, local (U)LIRGs, and high-$z$ quasars, we find the fraction of dense gas is higher in (U)LIRGs, high-$z$ quasars, and galactic centers than in the outer regions of our sample galaxies where a small variation of dense-gas fraction is found. The SFE of the dense molecular gas appears to be nearly independent of dense-gas fraction for our sample of galaxies, while the SFE of the total molecular gas increases substantially with dense-gas fraction when combining our data with local (U)LIRGs and high-$z$ quasars.

5. The HCN/HCO$^+$ \jfour\ ratio varies from 0.1 to 2.7 with a mean value of 0.9 and an rms scatter of 0.6 for the targeted six galaxies. No obvious correlation is found between HCN/HCO$^+$ line ratio and either IR luminosity or warm-dust temperature. We speculate that the low HCN/HCO$^+$ \jfour\ ratio found in M82 could be attributed to either a low HCN abundance and/or lack of gas with high enough density to excite the \hcnfour\ emission in this galaxy. The highest ratios are found in AGN-dominated systems, consistent with a scenario in which the presence of an AGN could cause an enhancement of the HCN abundance. 
\medskip

\acknowledgements

We are very grateful to the EAO/JCMT staff for their help during the observations and data reduction, and we thank the anonymous referee for useful comments and suggestions which greatly improved this manuscript. The James Clerk Maxwell Telescope is operated by the East Asian Observatory on behalf of The National Astronomical Observatory of Japan, Academia Sinica Institute of Astronomy and Astrophysics, the Korea Astronomy and Space Science Institute, the National Astronomical Observatories of China and the Chinese Academy of Sciences, with additional funding support from the Science and Technology Facilities Council of the United Kingdom and participating universities in the United Kingdom and Canada. MALATANG is a JCMT Large Program with project code of M16AL007. We are grateful to P. P. Papadopoulos for his warm help and support with the JCMT observations. We acknowledge the ORAC-DR, Starlink, and GILDAS software for the data reduction and analysis.

This work was supported by National Key R\&D Program of China grant \#2017YFA0402704, NSFC grants \#11420101002 and \#11603075, and Chinese Academy of Sciences Key Research Program of Frontier Sciences grant \#QYZDJ-SSW-SLH008. This work has been also supported by National Research Foundation of Korea grants No. 2015R1D1A1A01060516, and the Center for Galaxy Evolution Research (No. 2010-0027910). CDW acknowledges support from the Natural Sciences and Engineering Research Council of Canada. Z-Y.Z. acknowledges support from ERC in the form of the Advanced Investigator Programme, 321302, COSMICISM. LH acknowledges the support of National Key R\&D Program of China grant \#2016YFA0400702 and NSFC grant \#11721303. EB acknowledges support from the UK Science and Technology Facilities Council [Grant No. ST/M001008/1]. MJM acknowledges the support of the National Science Centre, Poland through the POLONEZ grant 2015/19/P/ST9/04010. This project has received funding from the European Union's Horizon 2020 research and innovation programme under the Marie Sk{\l}odowska-Curie grant agreement No. 665778. SM acknowledge the support of the Ministry of Science and Technology (MoST) of Taiwan, MoST 103-2112-M-001-032-MY3 and MoST 106-2112-M-001-011.
%\clearpage

\appendix

%%%%%%%%%- Fig-11 -%%%%%%%%%
\begin{figure*}[htbp]
\centering
\includegraphics[scale=0.75]{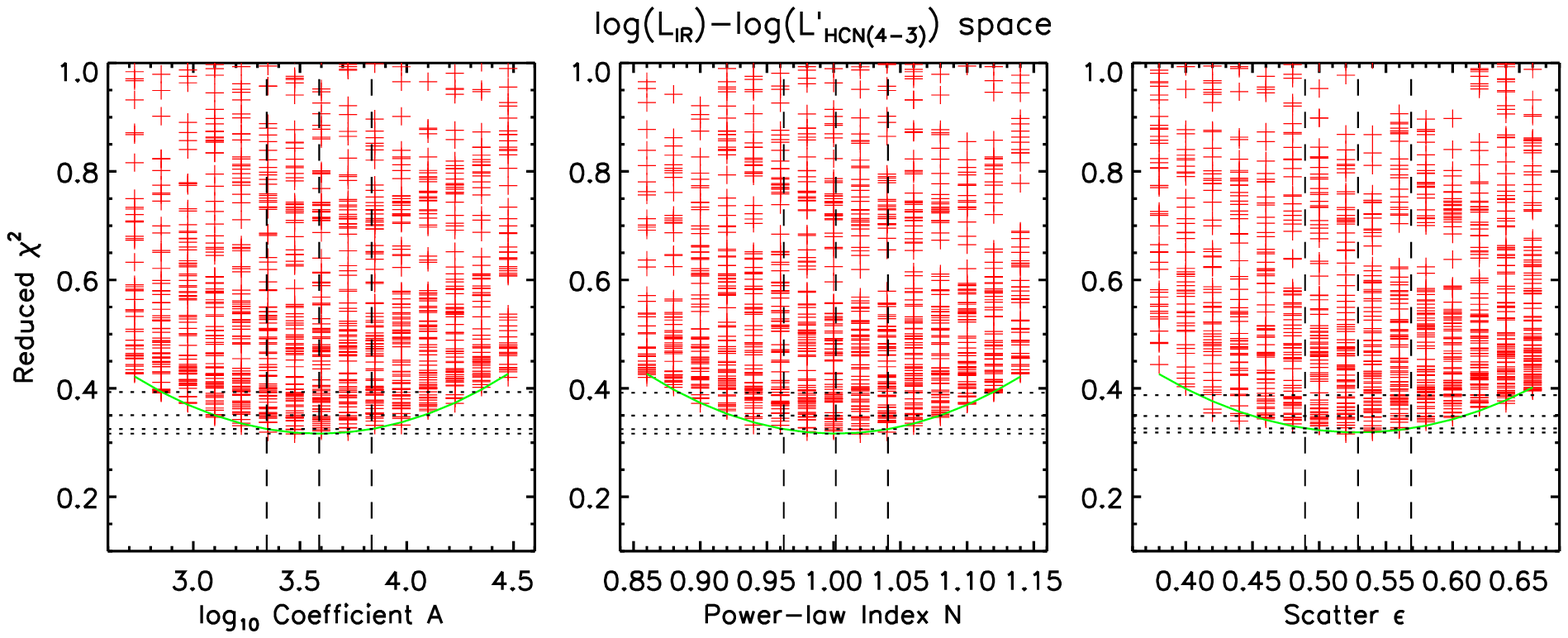}
\includegraphics[scale=0.75]{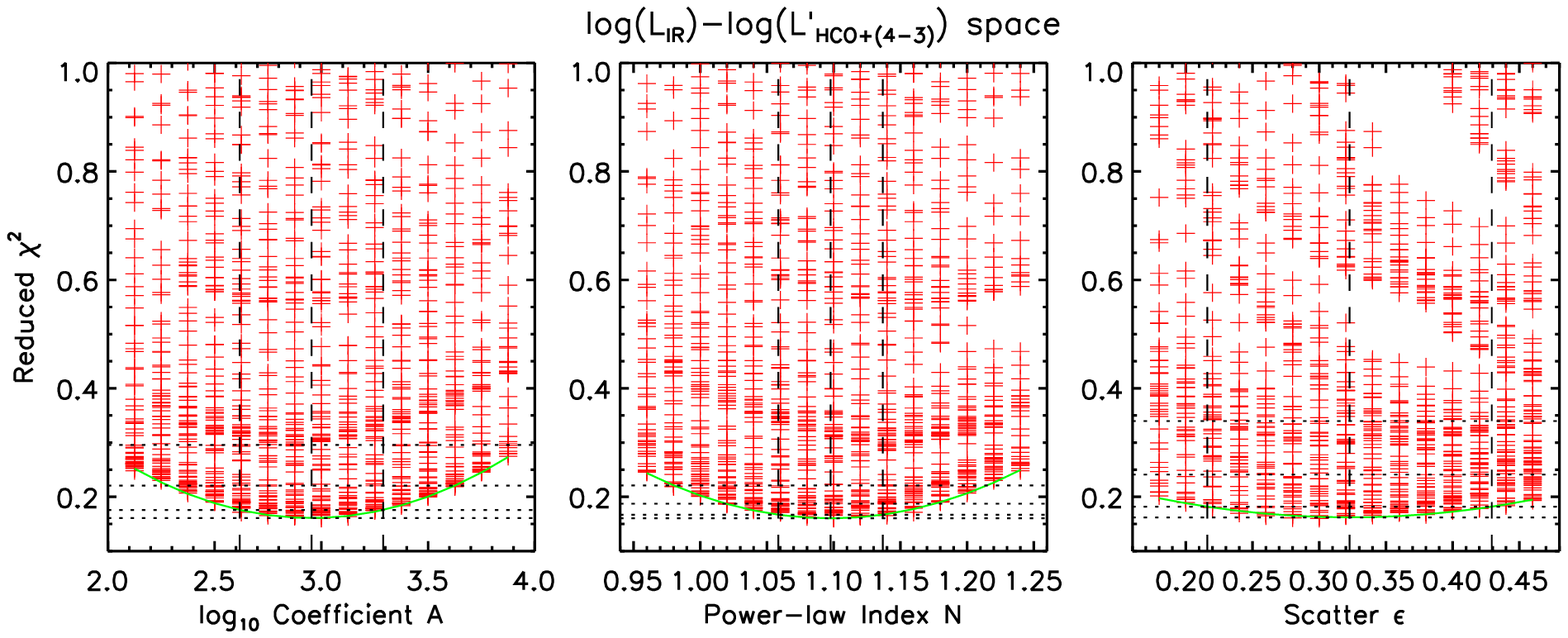}
\caption{Reduced $\chi^2$ for the three parameters $\{A, N, \epsilon\}$ in the MC fitting of $L_{\rm IR}-L^\prime_{\rm HCN(4-3)}$ relation (top row) and $L_{\rm IR}-L^\prime_{\rm HCO^+(4-3)}$ relation (bottom row) , marginalized over the other two. Red crosses show the $\chi^2$ obtained for each sampled combination of parameters. The best-fit $\chi^2$ is obtained by fitting a quadratic function to the minimum $\chi^2$ at each parameter value sampled. The best-fit quadratic function is shown in green line and the best-fit $\chi^2$ together with the \mbox{1 $\sigma$}, \mbox{2 $\sigma$}, and \mbox{3 $\sigma$} levels are shown as horizontal dotted lines. The vertical dashed lines represent the best-fit parameter and its \mbox{1 $\sigma$} uncertainty which is estimated by a bootstrapping method. \label{fig:mc}}
\end{figure*}
%%%%%%%%%%%%%%%%%%%%%%%

%%%%%%%%%- Fig-12 -%%%%%%%%%
\begin{figure*}[htbp]
\centering
\includegraphics[scale=0.75]{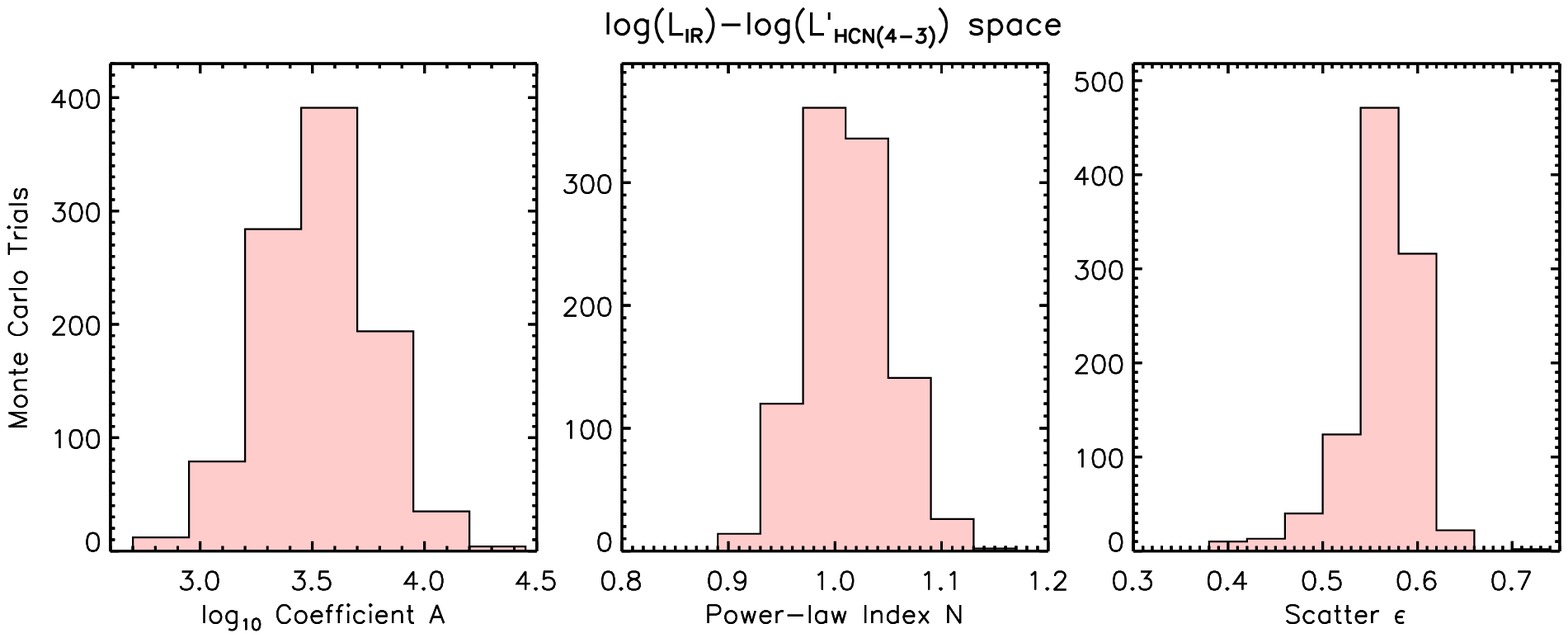}
\includegraphics[scale=0.75]{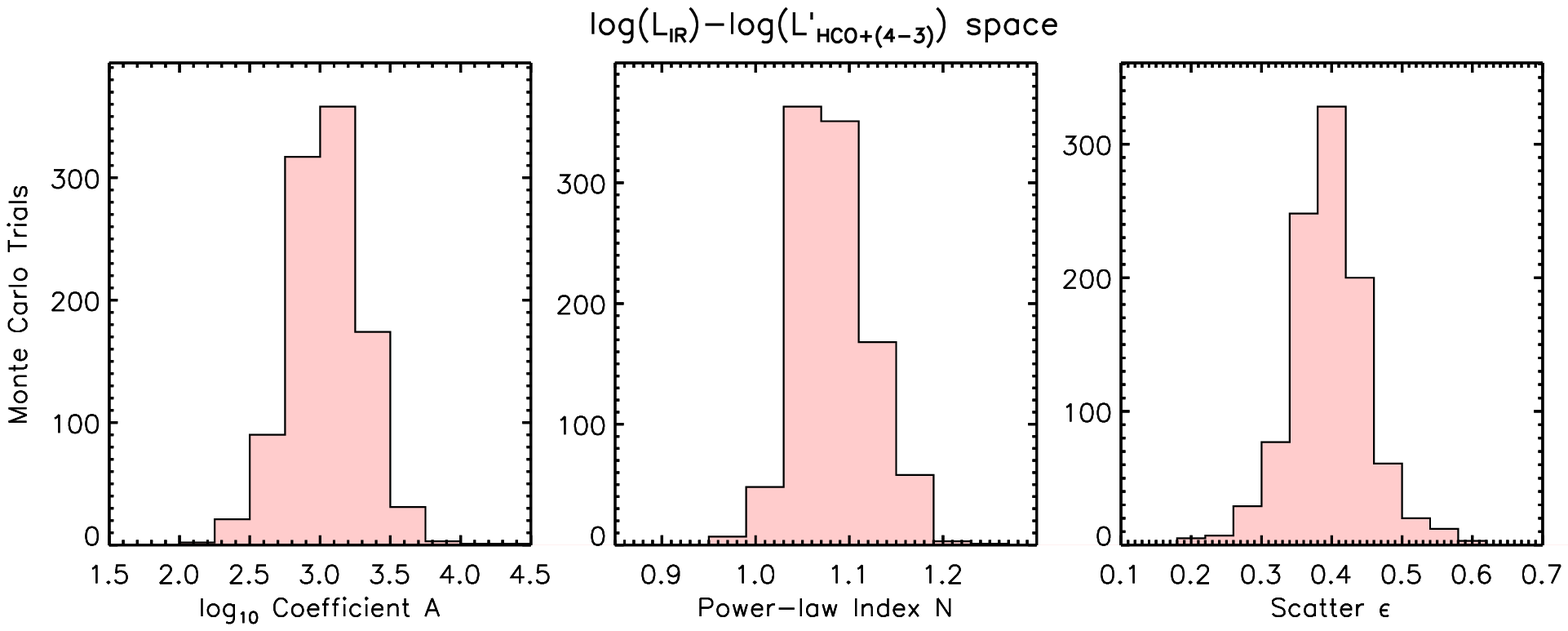}
\caption{Distribution of the best-fit parameters $\{A, N, \epsilon\}$ for the 1000 bootstrapping iterations to estimate the uncertainty for the parameters we determined based on the MC fitting to the $L_{\rm IR}-L^\prime_{\rm HCN(4-3)}$ relation (top row) and the $L_{\rm IR}-L^\prime_{\rm HCO^+(4-3)}$ relation (bottom row). \label{fig:bootstrap}}
\end{figure*}
%%%%%%%%%%%%%%%%%%%%%%%

\section{The fitting method using a Monte Carlo approach}\label{appendix}
Based on \citet{blanc09} and \citet{leroy13}, we fit the data using a Monte Carlo approach which allows us to include upper limits in the fit and incorporate the intrinsic scatter in the $L_{\rm IR}-L^\prime_{\rm dense}$ relation as a free parameter. In the following, we describe this approach.

1. We generate 1000 MC realizations of the data for each set of parameters $\{A, N, \epsilon\}$. For each realization, we take the observed $L^\prime_{\rm dense}$ as the true value and calculate the corresponding true $L_{\rm IR}$ using Equation (\ref{eq:mc}), drawing a new value from $\mathcal{N}(0,\epsilon)$ for each data point to introduce the intrinsic scatter. We apply the observational uncertainties in $L^\prime_{\rm gas}$ and $L_{\rm IR}$ by offseting the data points by random amounts. The uncertainty in $L^\prime_{\rm dense}$ is derived from statistical measurement errors and the systematic uncertainties in flux calibration (see Sect.~\ref{subsec:jcmt}), while the uncertainty in $L_{\rm IR}$ includes the statistical measurement errors and the errors introduced by the flux calibration and the TIR calibration from combined luminosities (see Section~\ref{subsec:lir}). For the non-detection of \hcnfour\ and \hcopfour\ emission, we use the measured values of these data points together with their error bars in the fitting procedure and exclude data with velocity-integrated intensity $I_{\rm dense}\leq 0$, given that our data are mainly limited by the sensitivity of the dense-gas observations.

2. We grid our observed data in ${\rm log}_{\rm 10} L_{\rm IR}-{\rm log}_{\rm 10} L^\prime_{\rm dense}$ space using cells 0.75 dex wide in both dimensions. We then compare the distribution of the gridded data with the model data from the MC realizations in the ${\rm log}_{\rm 10} L_{\rm IR}-{\rm log}_{\rm 10} L^\prime_{\rm dense}$ plane by counting the number of data points falling in each cell for each combination of $\{A, N, \epsilon\}$. After renormalizing the MC grid to have the same amount of data as the observed grid, we calculate a goodness-of-fit estimate which is referred to as $\chi^2$ following \citet{blanc09}:
\begin{equation}
\chi^2 = \displaystyle\sum_i\dfrac{(N_{\rm obs}^i-N_{\rm model}^i)^2}{N_{\rm model}^i}
\end{equation}
where the sum is over all grid cells, and $N_{\rm obs}^i$ and $N_{\rm model}^i$ are the number of observed and model data points respectively in grid cell $i$. 

3. We take a bootstrapping approach to estimate the errors in the parameters $\{A, N, \epsilon\}$ by randomly re-sampling the data points in each grid cell and performing the above MC analysis. The bootstrap procedure is repeated 1000 times for each solution and we measure the resulting standard deviation of the parameter values $\{A, N, \epsilon\}$.

Figure~\ref{fig:mc} shows the reduced $\chi^2$ for the three parameters $\{A, N, \epsilon\}$ in the fit, marginalized over the other two. Similar to Figure~14 of \citet{blanc09}, the best-fit value for each parameter is obtained by fitting a quadratic function to the minimum $\chi^2$ for each parameter value sampled. We adopt the \mbox{1 $\sigma$} dispersion of the $\chi^2$ distributions obtained through a bootstrapping approach for the estimate of the uncertainty in the parameters $\{A, N, \epsilon\}$ (see Fig.~\ref{fig:bootstrap}).

%% This command is needed to show the entire author+affilation list when
%% the collaboration and author truncation commands are used.  It has to
%% go at the end of the manuscript.
%\allauthors

%% Include this line if you are using the \added, \replaced, \deleted
%% commands to see a summary list of all changes at the end of the article.
%\listofchanges
\bibliographystyle{apj}
\bibliography{reference}{}

\begin{thebibliography}{}
\expandafter\ifx\csname natexlab\endcsname\relax\def\natexlab#1{#1}\fi

\bibitem[{{Aalto} {et~al.}(1995){Aalto}, {Booth}, {Black}, \&
  {Johansson}}]{aalto95}
{Aalto}, S., {Booth}, R.~S., {Black}, J.~H., \& {Johansson}, L.~E.~B. 1995,
  \aap, 300, 369

\bibitem[{{Aalto} {et~al.}(2012){Aalto}, {Garcia-Burillo}, {Muller}, {Winters},
  {van der Werf}, {Henkel}, {Costagliola}, \& {Neri}}]{aalto12}
{Aalto}, S., {Garcia-Burillo}, S., {Muller}, S., {et~al.} 2012, \aap, 537, A44

\bibitem[{{Aalto} {et~al.}(2015){Aalto}, {Mart{\'{\i}}n}, {Costagliola},
  {Gonz{\'a}lez-Alfonso}, {Muller}, {Sakamoto}, {Fuller},
  {Garc{\'{\i}}a-Burillo}, {van der Werf}, {Neri}, {Spaans}, {Combes}, {Viti},
  {M{\"u}hle}, {Armus}, {Evans}, {Sturm}, {Cernicharo}, {Henkel}, \&
  {Greve}}]{aalto15}
{Aalto}, S., {Mart{\'{\i}}n}, S., {Costagliola}, F., {et~al.} 2015, \aap, 584,
  A42

\bibitem[{{Aladro} {et~al.}(2011){Aladro}, {Mart{\'{\i}}n},
  {Mart{\'{\i}}n-Pintado}, {Mauersberger}, {Henkel}, {Oca{\~n}a Flaquer}, \&
  {Amo-Baladr{\'o}n}}]{aladro11}
{Aladro}, R., {Mart{\'{\i}}n}, S., {Mart{\'{\i}}n-Pintado}, J., {et~al.} 2011,
  \aap, 535, A84

\bibitem[{{Aniano} {et~al.}(2011){Aniano}, {Draine}, {Gordon}, \&
  {Sandstrom}}]{aniano11}
{Aniano}, G., {Draine}, B.~T., {Gordon}, K.~D., \& {Sandstrom}, K. 2011, \pasp,
  123, 1218

\bibitem[{{Baan} {et~al.}(2008){Baan}, {Henkel}, {Loenen}, {Baudry}, \&
  {Wiklind}}]{baan08}
{Baan}, W.~A., {Henkel}, C., {Loenen}, A.~F., {Baudry}, A., \& {Wiklind}, T.
  2008, \aap, 477, 747

\bibitem[{{Balog} {et~al.}(2014){Balog}, {M{\"u}ller}, {Nielbock}, {Altieri},
  {Klaas}, {Blommaert}, {Linz}, {Lutz}, {Mo{\'o}r}, {Billot}, {Sauvage}, \&
  {Okumura}}]{balog14}
{Balog}, Z., {M{\"u}ller}, T., {Nielbock}, M., {et~al.} 2014, Experimental
  Astronomy, 37, 129

\bibitem[{{Barvainis} {et~al.}(1997){Barvainis}, {Maloney}, {Antonucci}, \&
  {Alloin}}]{barvainis97}
{Barvainis}, R., {Maloney}, P., {Antonucci}, R., \& {Alloin}, D. 1997, \apj,
  484, 695

\bibitem[{{Bayet} {et~al.}(2012){Bayet}, {Davis}, {Bell}, \& {Viti}}]{bayet12}
{Bayet}, E., {Davis}, T.~A., {Bell}, T.~A., \& {Viti}, S. 2012, \mnras, 424,
  2646

\bibitem[{{Bigiel} {et~al.}(2008){Bigiel}, {Leroy}, {Walter}, {Brinks}, {de
  Blok}, {Madore}, \& {Thornley}}]{bigiel08}
{Bigiel}, F., {Leroy}, A., {Walter}, F., {et~al.} 2008, \aj, 136, 2846

\bibitem[{{Bigiel} {et~al.}(2015){Bigiel}, {Leroy}, {Blitz}, {Bolatto}, {da
  Cunha}, {Rosolowsky}, {Sandstrom}, \& {Usero}}]{bigiel15}
{Bigiel}, F., {Leroy}, A.~K., {Blitz}, L., {et~al.} 2015, \apj, 815, 103

\bibitem[{{Bigiel} {et~al.}(2016){Bigiel}, {Leroy}, {Jim{\'e}nez-Donaire},
  {Pety}, {Usero}, {Cormier}, {Bolatto}, {Garcia-Burillo}, {Colombo},
  {Gonz{\'a}lez-Garc{\'{\i}}a}, {Hughes}, {Kepley}, {Kramer}, {Sandstrom},
  {Schinnerer}, {Schruba}, {Schuster}, {Tomicic}, \& {Zschaechner}}]{bigiel16}
{Bigiel}, F., {Leroy}, A.~K., {Jim{\'e}nez-Donaire}, M.~J., {et~al.} 2016,
  \apjl, 822, L26

\bibitem[{{Blanc} {et~al.}(2009){Blanc}, {Heiderman}, {Gebhardt}, {Evans}, \&
  {Adams}}]{blanc09}
{Blanc}, G.~A., {Heiderman}, A., {Gebhardt}, K., {Evans}, II, N.~J., \&
  {Adams}, J. 2009, \apj, 704, 842

\bibitem[{{Bolatto} {et~al.}(2013){Bolatto}, {Wolfire}, \& {Leroy}}]{bolatto13}
{Bolatto}, A.~D., {Wolfire}, M., \& {Leroy}, A.~K. 2013, \araa, 51, 207

\bibitem[{{Braine} {et~al.}(2017){Braine}, {Shimajiri}, {Andr{\'e}},
  {Bontemps}, {Gao}, {Chen}, \& {Kramer}}]{braine17}
{Braine}, J., {Shimajiri}, Y., {Andr{\'e}}, P., {et~al.} 2017, \aap, 597, A44

\bibitem[{{Buckle} {et~al.}(2009){Buckle}, {Hills}, {Smith}, {Dent}, {Bell},
  {Curtis}, {Dace}, {Gibson}, {Graves}, {Leech}, {Richer}, {Williamson},
  {Withington}, {Yassin}, {Bennett}, {Hastings}, {Laidlaw}, {Lightfoot},
  {Burgess}, {Dewdney}, {Hovey}, {Willis}, {Redman}, {Wooff}, {Berry},
  {Cavanagh}, {Davis}, {Dempsey}, {Friberg}, {Jenness}, {Kackley}, {Rees},
  {Tilanus}, {Walther}, {Zwart}, {Klapwijk}, {Kroug}, \& {Zijlstra}}]{buckle09}
{Buckle}, J.~V., {Hills}, R.~E., {Smith}, H., {et~al.} 2009, \mnras, 399, 1026

\bibitem[{{Bussmann} {et~al.}(2008){Bussmann}, {Narayanan}, {Shirley},
  {Juneau}, {Wu}, {Solomon}, {Vanden Bout}, {Moustakas}, \&
  {Walker}}]{bussmann08}
{Bussmann}, R.~S., {Narayanan}, D., {Shirley}, Y.~L., {et~al.} 2008, \apjl,
  681, L73

\bibitem[{{Carilli} \& {Walter}(2013)}]{carilli13}
{Carilli}, C.~L., \& {Walter}, F. 2013, \araa, 51, 105

\bibitem[{{Chanial} {et~al.}(2007){Chanial}, {Flores}, {Guiderdoni}, {Elbaz},
  {Hammer}, \& {Vigroux}}]{chanial07}
{Chanial}, P., {Flores}, H., {Guiderdoni}, B., {et~al.} 2007, \aap, 462, 81

\bibitem[{{Chen} {et~al.}(2017){Chen}, {Braine}, {Gao}, {Koda}, \&
  {Gu}}]{chen17}
{Chen}, H., {Braine}, J., {Gao}, Y., {Koda}, J., \& {Gu}, Q. 2017, \apj, 836,
  101

\bibitem[{{Chen} {et~al.}(2015){Chen}, {Gao}, {Braine}, \& {Gu}}]{chen15}
{Chen}, H., {Gao}, Y., {Braine}, J., \& {Gu}, Q. 2015, \apj, 810, 140

\bibitem[{{Chu} {et~al.}(2017){Chu}, {Sanders}, {Larson}, {Mazzarella},
  {Howell}, {D{\'{\i}}az-Santos}, {Xu}, {Paladini}, {Schulz}, {Shupe},
  {Appleton}, {Armus}, {Billot}, {Chan}, {Evans}, {Fadda}, {Frayer}, {Haan},
  {Ishida}, {Iwasawa}, {Kim}, {Lord}, {Murphy}, {Petric}, {Privon}, {Surace},
  \& {Treister}}]{chu17}
{Chu}, J.~K., {Sanders}, D.~B., {Larson}, K.~L., {et~al.} 2017, \apjs, 229, 25

\bibitem[{{Combes} {et~al.}(2013){Combes}, {Garc{\'{\i}}a-Burillo}, {Braine},
  {Schinnerer}, {Walter}, \& {Colina}}]{combes13}
{Combes}, F., {Garc{\'{\i}}a-Burillo}, S., {Braine}, J., {et~al.} 2013, \aap,
  550, A41

\bibitem[{{Costagliola} {et~al.}(2011){Costagliola}, {Aalto}, {Rodriguez},
  {Muller}, {Spoon}, {Mart{\'{\i}}n}, {Per{\'e}z-Torres}, {Alberdi},
  {Lindberg}, {Batejat}, {J{\"u}tte}, {van der Werf}, \&
  {Lahuis}}]{costagliola11}
{Costagliola}, F., {Aalto}, S., {Rodriguez}, M.~I., {et~al.} 2011, \aap, 528,
  A30

\bibitem[{{Daddi} {et~al.}(2010){Daddi}, {Elbaz}, {Walter}, {Bournaud},
  {Salmi}, {Carilli}, {Dannerbauer}, {Dickinson}, {Monaco}, \&
  {Riechers}}]{daddi10}
{Daddi}, E., {Elbaz}, D., {Walter}, F., {et~al.} 2010, \apjl, 714, L118

\bibitem[{{Dalcanton} {et~al.}(2009){Dalcanton}, {Williams}, {Seth}, {Dolphin},
  {Holtzman}, {Rosema}, {Skillman}, {Cole}, {Girardi}, {Gogarten},
  {Karachentsev}, {Olsen}, {Weisz}, {Christensen}, {Freeman}, {Gilbert},
  {Gallart}, {Harris}, {Hodge}, {de Jong}, {Karachentseva}, {Mateo}, {Stetson},
  {Tavarez}, {Zaritsky}, {Governato}, \& {Quinn}}]{dalcanton09}
{Dalcanton}, J.~J., {Williams}, B.~F., {Seth}, A.~C., {et~al.} 2009, \apjs,
  183, 67

\bibitem[{{Davis} {et~al.}(2013){Davis}, {Bayet}, {Crocker}, {Topal}, \&
  {Bureau}}]{davis13}
{Davis}, T.~A., {Bayet}, E., {Crocker}, A., {Topal}, S., \& {Bureau}, M. 2013,
  \mnras, 433, 1659

\bibitem[{{Elmegreen}(2015)}]{elmegreen15}
{Elmegreen}, B.~G. 2015, \apjl, 814, L30

\bibitem[{{Elmegreen}(2018)}]{elmegreen18}
---. 2018, \apj, 854, 16

\bibitem[{{Engelbracht} {et~al.}(2008){Engelbracht}, {Rieke}, {Gordon},
  {Smith}, {Werner}, {Moustakas}, {Willmer}, \& {Vanzi}}]{engelbracht08}
{Engelbracht}, C.~W., {Rieke}, G.~H., {Gordon}, K.~D., {et~al.} 2008, \apj,
  678, 804

\bibitem[{{Evans}(2008)}]{evans08}
{Evans}, II, N.~J. 2008, in Astronomical Society of the Pacific Conference
  Series, Vol. 390, Pathways Through an Eclectic Universe, ed. J.~H. {Knapen},
  T.~J. {Mahoney}, \& A.~{Vazdekis}, 52

\bibitem[{{Evans} {et~al.}(2014){Evans}, {Heiderman}, \&
  {Vutisalchavakul}}]{evans14}
{Evans}, II, N.~J., {Heiderman}, A., \& {Vutisalchavakul}, N. 2014, \apj, 782,
  114

\bibitem[{{Galametz} {et~al.}(2013){Galametz}, {Kennicutt}, {Calzetti},
  {Aniano}, {Draine}, {Boquien}, {Brandl}, {Croxall}, {Dale}, {Engelbracht},
  {Gordon}, {Groves}, {Hao}, {Helou}, {Hinz}, {Hunt}, {Johnson}, {Li},
  {Murphy}, {Roussel}, {Sandstrom}, {Skibba}, \& {Tabatabaei}}]{galametz13}
{Galametz}, M., {Kennicutt}, R.~C., {Calzetti}, D., {et~al.} 2013, \mnras, 431,
  1956

\bibitem[{{Gao}(1996)}]{gao96}
{Gao}, Y. 1996, PhD thesis, STATE UNIVERSITY OF NEW YORK AT STONY BROOK

\bibitem[{{Gao} {et~al.}(2007){Gao}, {Carilli}, {Solomon}, \& {Vanden
  Bout}}]{gao07}
{Gao}, Y., {Carilli}, C.~L., {Solomon}, P.~M., \& {Vanden Bout}, P.~A. 2007,
  \apjl, 660, L93

\bibitem[{{Gao} \& {Solomon}(2004{\natexlab{a}})}]{gao04a}
{Gao}, Y., \& {Solomon}, P.~M. 2004{\natexlab{a}}, \apjs, 152, 63

\bibitem[{{Gao} \& {Solomon}(2004{\natexlab{b}})}]{gao04b}
---. 2004{\natexlab{b}}, \apj, 606, 271

\bibitem[{{Garc{\'{\i}}a-Burillo} {et~al.}(2012){Garc{\'{\i}}a-Burillo},
  {Usero}, {Alonso-Herrero}, {Graci{\'a}-Carpio}, {Pereira-Santaella},
  {Colina}, {Planesas}, \& {Arribas}}]{garcia12}
{Garc{\'{\i}}a-Burillo}, S., {Usero}, A., {Alonso-Herrero}, A., {et~al.} 2012,
  \aap, 539, A8

\bibitem[{{Garc{\'{\i}}a-Burillo} {et~al.}(2014){Garc{\'{\i}}a-Burillo},
  {Combes}, {Usero}, {Aalto}, {Krips}, {Viti}, {Alonso-Herrero}, {Hunt},
  {Schinnerer}, {Baker}, {Boone}, {Casasola}, {Colina}, {Costagliola},
  {Eckart}, {Fuente}, {Henkel}, {Labiano}, {Mart{\'{\i}}n}, {M{\'a}rquez},
  {Muller}, {Planesas}, {Ramos Almeida}, {Spaans}, {Tacconi}, \& {van der
  Werf}}]{garcia14}
{Garc{\'{\i}}a-Burillo}, S., {Combes}, F., {Usero}, A., {et~al.} 2014, \aap,
  567, A125

\bibitem[{{Gazak} {et~al.}(2014){Gazak}, {Davies}, {Bastian}, {Kudritzki},
  {Bergemann}, {Plez}, {Evans}, {Patrick}, {Bresolin}, \&
  {Schinnerer}}]{gazak14}
{Gazak}, J.~Z., {Davies}, B., {Bastian}, N., {et~al.} 2014, \apj, 787, 142

\bibitem[{{Goldsmith} \& {Kauffmann}(2017)}]{goldsmith17}
{Goldsmith}, P.~F., \& {Kauffmann}, J. 2017, \apj, 841, 25

\bibitem[{{Graci{\'a}-Carpio} {et~al.}(2006){Graci{\'a}-Carpio},
  {Garc{\'{\i}}a-Burillo}, {Planesas}, \& {Colina}}]{gracia06}
{Graci{\'a}-Carpio}, J., {Garc{\'{\i}}a-Burillo}, S., {Planesas}, P., \&
  {Colina}, L. 2006, \apjl, 640, L135

\bibitem[{{Graci{\'a}-Carpio} {et~al.}(2008){Graci{\'a}-Carpio},
  {Garc{\'{\i}}a-Burillo}, {Planesas}, {Fuente}, \& {Usero}}]{gracia08}
{Graci{\'a}-Carpio}, J., {Garc{\'{\i}}a-Burillo}, S., {Planesas}, P., {Fuente},
  A., \& {Usero}, A. 2008, \aap, 479, 703

\bibitem[{{Greve} {et~al.}(2009){Greve}, {Papadopoulos}, {Gao}, \&
  {Radford}}]{greve09}
{Greve}, T.~R., {Papadopoulos}, P.~P., {Gao}, Y., \& {Radford}, S.~J.~E. 2009,
  \apj, 692, 1432

\bibitem[{{Greve} {et~al.}(2014){Greve}, {Leonidaki}, {Xilouris}, {Wei{\ss}},
  {Zhang}, {van der Werf}, {Aalto}, {Armus}, {D{\'{\i}}az-Santos}, {Evans},
  {Fischer}, {Gao}, {Gonz{\'a}lez-Alfonso}, {Harris}, {Henkel}, {Meijerink},
  {Naylor}, {Smith}, {Spaans}, {Stacey}, {Veilleux}, \& {Walter}}]{greve14}
{Greve}, T.~R., {Leonidaki}, I., {Xilouris}, E.~M., {et~al.} 2014, \apj, 794,
  142

\bibitem[{{Harada} {et~al.}(2010){Harada}, {Herbst}, \& {Wakelam}}]{harada10}
{Harada}, N., {Herbst}, E., \& {Wakelam}, V. 2010, \apj, 721, 1570

\bibitem[{{Harada} {et~al.}(2013){Harada}, {Thompson}, \& {Herbst}}]{harada13}
{Harada}, N., {Thompson}, T.~A., \& {Herbst}, E. 2013, \apj, 765, 108

\bibitem[{{Heiderman} {et~al.}(2010){Heiderman}, {Evans}, {Allen}, {Huard}, \&
  {Heyer}}]{heiderman10}
{Heiderman}, A., {Evans}, II, N.~J., {Allen}, L.~E., {Huard}, T., \& {Heyer},
  M. 2010, \apj, 723, 1019

\bibitem[{{Helou} {et~al.}(1985){Helou}, {Soifer}, \&
  {Rowan-Robinson}}]{helou85}
{Helou}, G., {Soifer}, B.~T., \& {Rowan-Robinson}, M. 1985, \apjl, 298, L7

\bibitem[{{Imanishi} \& {Nakanishi}(2013{\natexlab{a}})}]{imanishi13b}
{Imanishi}, M., \& {Nakanishi}, K. 2013{\natexlab{a}}, \aj, 146, 91

\bibitem[{{Imanishi} \& {Nakanishi}(2013{\natexlab{b}})}]{imanishi13a}
---. 2013{\natexlab{b}}, \aj, 146, 47

\bibitem[{{Imanishi} \& {Nakanishi}(2014)}]{imanishi14}
---. 2014, \aj, 148, 9

\bibitem[{{Imanishi} {et~al.}(2016){Imanishi}, {Nakanishi}, \&
  {Izumi}}]{imanishi16}
{Imanishi}, M., {Nakanishi}, K., \& {Izumi}, T. 2016, \aj, 152, 218

\bibitem[{{Imanishi} {et~al.}(2004){Imanishi}, {Nakanishi}, {Kuno}, \&
  {Kohno}}]{imanishi04}
{Imanishi}, M., {Nakanishi}, K., {Kuno}, N., \& {Kohno}, K. 2004, \aj, 128,
  2037

\bibitem[{{Izumi} {et~al.}(2016){Izumi}, {Kohno}, {Aalto}, {Espada}, {Fathi},
  {Harada}, {Hatsukade}, {Hsieh}, {Imanishi}, {Krips}, {Mart{\'{\i}}n},
  {Matsushita}, {Meier}, {Nakai}, {Nakanishi}, {Schinnerer}, {Sheth},
  {Terashima}, \& {Turner}}]{izumi16}
{Izumi}, T., {Kohno}, K., {Aalto}, S., {et~al.} 2016, \apj, 818, 42

\bibitem[{{Jackson} {et~al.}(1995){Jackson}, {Paglione}, {Carlstrom}, \&
  {Rieu}}]{jackson95}
{Jackson}, J.~M., {Paglione}, T.~A.~D., {Carlstrom}, J.~E., \& {Rieu}, N.-Q.
  1995, \apj, 438, 695

\bibitem[{{Jenness} {et~al.}(2013){Jenness}, {Chapin}, {Berry}, {Gibb},
  {Tilanus}, {Balfour}, {Tilanus}, \& {Currie}}]{jenness13}
{Jenness}, T., {Chapin}, E.~L., {Berry}, D.~S., {et~al.} 2013, {SMURF:
  SubMillimeter User Reduction Facility}, Astrophysics Source Code Library,
  ascl:1310.007

\bibitem[{{Jenness} {et~al.}(2015){Jenness}, {Currie}, {Tilanus}, {Cavanagh},
  {Berry}, {Leech}, \& {Rizzi}}]{jenness15}
{Jenness}, T., {Currie}, M.~J., {Tilanus}, R.~P.~J., {et~al.} 2015, \mnras,
  453, 73

\bibitem[{{Juneau} {et~al.}(2009){Juneau}, {Narayanan}, {Moustakas}, {Shirley},
  {Bussmann}, {Kennicutt}, \& {Vanden Bout}}]{juneau09}
{Juneau}, S., {Narayanan}, D.~T., {Moustakas}, J., {et~al.} 2009, \apj, 707,
  1217

\bibitem[{{Kauffmann} {et~al.}(2017){Kauffmann}, {Goldsmith}, {Melnick},
  {Tolls}, {Guzman}, \& {Menten}}]{kauffmann17}
{Kauffmann}, J., {Goldsmith}, P.~F., {Melnick}, G., {et~al.} 2017, \aap, 605,
  L5

\bibitem[{{Kelly}(2007)}]{kelly07}
{Kelly}, B.~C. 2007, \apj, 665, 1489

\bibitem[{{Kennicutt} \& {Evans}(2012)}]{kennicutt12}
{Kennicutt}, R.~C., \& {Evans}, N.~J. 2012, \araa, 50, 531

\bibitem[{{Kennicutt} {et~al.}(2011){Kennicutt}, {Calzetti}, {Aniano},
  {Appleton}, {Armus}, {Beir{\~a}o}, {Bolatto}, {Brandl}, {Crocker}, {Croxall},
  {Dale}, {Donovan Meyer}, {Draine}, {Engelbracht}, {Galametz}, {Gordon},
  {Groves}, {Hao}, {Helou}, {Hinz}, {Hunt}, {Johnson}, {Koda}, {Krause},
  {Leroy}, {Li}, {Meidt}, {Montiel}, {Murphy}, {Rahman}, {Rix}, {Roussel},
  {Sandstrom}, {Sauvage}, {Schinnerer}, {Skibba}, {Smith}, {Srinivasan},
  {Vigroux}, {Walter}, {Wilson}, {Wolfire}, \& {Zibetti}}]{kennicutt11}
{Kennicutt}, R.~C., {Calzetti}, D., {Aniano}, G., {et~al.} 2011, \pasp, 123,
  1347

\bibitem[{{Kennicutt}(1998)}]{kennicutt98}
{Kennicutt}, Jr., R.~C. 1998, \apj, 498, 541

\bibitem[{{Knudsen} {et~al.}(2007){Knudsen}, {Walter}, {Weiss}, {Bolatto},
  {Riechers}, \& {Menten}}]{knudsen07}
{Knudsen}, K.~K., {Walter}, F., {Weiss}, A., {et~al.} 2007, \apj, 666, 156

\bibitem[{{Kohno} {et~al.}(1996){Kohno}, {Kawabe}, {Tosaki}, \&
  {Okumura}}]{kohno96}
{Kohno}, K., {Kawabe}, R., {Tosaki}, T., \& {Okumura}, S.~K. 1996, \apjl, 461,
  L29

\bibitem[{{Kohno} {et~al.}(2001){Kohno}, {Matsushita}, {Vila-Vilar{\'o}},
  {Okumura}, {Shibatsuka}, {Okiura}, {Ishizuki}, \& {Kawabe}}]{kohno01}
{Kohno}, K., {Matsushita}, S., {Vila-Vilar{\'o}}, B., {et~al.} 2001, in
  Astronomical Society of the Pacific Conference Series, Vol. 249, The Central
  Kiloparsec of Starbursts and AGN: The La Palma Connection, ed. J.~H.
  {Knapen}, J.~E. {Beckman}, I.~{Shlosman}, \& T.~J. {Mahoney}, 672

\bibitem[{{Krips} {et~al.}(2008){Krips}, {Neri}, {Garc{\'{\i}}a-Burillo},
  {Mart{\'{\i}}n}, {Combes}, {Graci{\'a}-Carpio}, \& {Eckart}}]{krips08}
{Krips}, M., {Neri}, R., {Garc{\'{\i}}a-Burillo}, S., {et~al.} 2008, \apj, 677,
  262

\bibitem[{{Krips} {et~al.}(2011){Krips}, {Mart{\'{\i}}n}, {Eckart}, {Neri},
  {Garc{\'{\i}}a-Burillo}, {Matsushita}, {Peck}, {Stoklasov{\'a}}, {Petitpas},
  {Usero}, {Combes}, {Schinnerer}, {Humphreys}, \& {Baker}}]{krips11}
{Krips}, M., {Mart{\'{\i}}n}, S., {Eckart}, A., {et~al.} 2011, \apj, 736, 37

\bibitem[{{Kroupa}(2001)}]{kroupa01}
{Kroupa}, P. 2001, \mnras, 322, 231

\bibitem[{{Krumholz} \& {McKee}(2005)}]{krumholz05}
{Krumholz}, M.~R., \& {McKee}, C.~F. 2005, \apj, 630, 250

\bibitem[{{Krumholz} \& {Thompson}(2007)}]{krumholz07}
{Krumholz}, M.~R., \& {Thompson}, T.~A. 2007, \apj, 669, 289

\bibitem[{{Kuno} {et~al.}(2007){Kuno}, {Sato}, {Nakanishi}, {Hirota}, {Tosaki},
  {Shioya}, {Sorai}, {Nakai}, {Nishiyama}, \& {Vila-Vilar{\'o}}}]{kuno07}
{Kuno}, N., {Sato}, N., {Nakanishi}, H., {et~al.} 2007, \pasj, 59, 117

\bibitem[{{Lada} {et~al.}(2012){Lada}, {Forbrich}, {Lombardi}, \&
  {Alves}}]{lada12}
{Lada}, C.~J., {Forbrich}, J., {Lombardi}, M., \& {Alves}, J.~F. 2012, \apj,
  745, 190

\bibitem[{{Lada} {et~al.}(2010){Lada}, {Lombardi}, \& {Alves}}]{lada10}
{Lada}, C.~J., {Lombardi}, M., \& {Alves}, J.~F. 2010, \apj, 724, 687

\bibitem[{{Lepp} \& {Dalgarno}(1996)}]{lepp96}
{Lepp}, S., \& {Dalgarno}, A. 1996, \aap, 306, L21

\bibitem[{{Leroy} {et~al.}(2008){Leroy}, {Walter}, {Brinks}, {Bigiel}, {de
  Blok}, {Madore}, \& {Thornley}}]{leroy08}
{Leroy}, A.~K., {Walter}, F., {Brinks}, E., {et~al.} 2008, \aj, 136, 2782

\bibitem[{{Leroy} {et~al.}(2013){Leroy}, {Walter}, {Sandstrom}, {Schruba},
  {Munoz-Mateos}, {Bigiel}, {Bolatto}, {Brinks}, {de Blok}, {Meidt}, {Rix},
  {Rosolowsky}, {Schinnerer}, {Schuster}, \& {Usero}}]{leroy13}
{Leroy}, A.~K., {Walter}, F., {Sandstrom}, K., {et~al.} 2013, \aj, 146, 19

\bibitem[{{Leroy} {et~al.}(2015{\natexlab{a}}){Leroy}, {Bolatto}, {Ostriker},
  {Rosolowsky}, {Walter}, {Warren}, {Donovan Meyer}, {Hodge}, {Meier}, {Ott},
  {Sandstrom}, {Schruba}, {Veilleux}, \& {Zwaan}}]{leroy15b}
{Leroy}, A.~K., {Bolatto}, A.~D., {Ostriker}, E.~C., {et~al.}
  2015{\natexlab{a}}, \apj, 801, 25

\bibitem[{{Leroy} {et~al.}(2015{\natexlab{b}}){Leroy}, {Walter}, {Martini},
  {Roussel}, {Sandstrom}, {Ott}, {Weiss}, {Bolatto}, {Schuster}, \&
  {Dessauges-Zavadsky}}]{leroy15}
{Leroy}, A.~K., {Walter}, F., {Martini}, P., {et~al.} 2015{\natexlab{b}}, \apj,
  814, 83

\bibitem[{{Liu} {et~al.}(2015{\natexlab{a}}){Liu}, {Gao}, {Isaak}, {Daddi},
  {Yang}, {Lu}, \& {van der Werf}}]{dliu15}
{Liu}, D., {Gao}, Y., {Isaak}, K., {et~al.} 2015{\natexlab{a}}, \apjl, 810, L14

\bibitem[{{Liu} \& {Gao}(2010)}]{liu10}
{Liu}, F., \& {Gao}, Y. 2010, \apj, 713, 524

\bibitem[{{Liu} {et~al.}(2015{\natexlab{b}}){Liu}, {Gao}, \& {Greve}}]{lliu15}
{Liu}, L., {Gao}, Y., \& {Greve}, T.~R. 2015{\natexlab{b}}, \apj, 805, 31

\bibitem[{{Liu} {et~al.}(2016){Liu}, {Kim}, {Yoo}, {Liu}, {Tatematsu}, {Qin},
  {Zhang}, {Wu}, {Wang}, {Goldsmith}, {Juvela}, {Lee}, {T{\'o}th}, {Mardones},
  {Garay}, {Bronfman}, {Cunningham}, {Li}, {Lo}, {Ristorcelli}, \&
  {Schnee}}]{liu16}
{Liu}, T., {Kim}, K.-T., {Yoo}, H., {et~al.} 2016, \apj, 829, 59

\bibitem[{{Lu} {et~al.}(2014){Lu}, {Zhao}, {Xu}, {Gao}, {Armus}, {Mazzarella},
  {Isaak}, {Petric}, {Charmandaris}, {D{\'{\i}}az-Santos}, {Evans}, {Howell},
  {Appleton}, {Inami}, {Iwasawa}, {Leech}, {Lord}, {Sanders}, {Schulz},
  {Surace}, \& {van der Werf}}]{lu14}
{Lu}, N., {Zhao}, Y., {Xu}, C.~K., {et~al.} 2014, \apjl, 787, L23

\bibitem[{{Mao} {et~al.}(2010){Mao}, {Schulz}, {Henkel}, {Mauersberger},
  {Muders}, \& {Dinh-V-Trung}}]{mao10}
{Mao}, R.-Q., {Schulz}, A., {Henkel}, C., {et~al.} 2010, \apj, 724, 1336

\bibitem[{{Markwardt}(2009)}]{markwardt09}
{Markwardt}, C.~B. 2009, in Astronomical Society of the Pacific Conference
  Series, Vol. 411, Astronomical Data Analysis Software and Systems XVIII, ed.
  D.~A. {Bohlender}, D.~{Durand}, \& P.~{Dowler}, 251

\bibitem[{{Mart{\'{\i}}n} {et~al.}(2006){Mart{\'{\i}}n}, {Mauersberger},
  {Mart{\'{\i}}n-Pintado}, {Henkel}, \& {Garc{\'{\i}}a-Burillo}}]{martin06}
{Mart{\'{\i}}n}, S., {Mauersberger}, R., {Mart{\'{\i}}n-Pintado}, J., {Henkel},
  C., \& {Garc{\'{\i}}a-Burillo}, S. 2006, \apjs, 164, 450

\bibitem[{{Matsushita} {et~al.}(2015){Matsushita}, {Trung}, {Boone}, {Krips},
  {Lim}, \& {Muller}}]{matsushita15}
{Matsushita}, S., {Trung}, D.-V., {Boone}, F., {et~al.} 2015, \apj, 799, 26

\bibitem[{{Meijerink} {et~al.}(2007){Meijerink}, {Spaans}, \&
  {Israel}}]{meijerink07}
{Meijerink}, R., {Spaans}, M., \& {Israel}, F.~P. 2007, \aap, 461, 793

\bibitem[{{Mills} \& {Battersby}(2017)}]{mills17}
{Mills}, E.~A.~C., \& {Battersby}, C. 2017, \apj, 835, 76

\bibitem[{{Mills} {et~al.}(2013){Mills}, {G{\"u}sten}, {Requena-Torres}, \&
  {Morris}}]{mills13}
{Mills}, E.~A.~C., {G{\"u}sten}, R., {Requena-Torres}, M.~A., \& {Morris},
  M.~R. 2013, \apj, 779, 47

\bibitem[{{Murphy} {et~al.}(2011){Murphy}, {Condon}, {Schinnerer}, {Kennicutt},
  {Calzetti}, {Armus}, {Helou}, {Turner}, {Aniano}, {Beir{\~a}o}, {Bolatto},
  {Brandl}, {Croxall}, {Dale}, {Donovan Meyer}, {Draine}, {Engelbracht},
  {Hunt}, {Hao}, {Koda}, {Roussel}, {Skibba}, \& {Smith}}]{murphy11}
{Murphy}, E.~J., {Condon}, J.~J., {Schinnerer}, E., {et~al.} 2011, \apj, 737,
  67

\bibitem[{{Nagao} {et~al.}(2011){Nagao}, {Maiolino}, {Marconi}, \&
  {Matsuhara}}]{nagao11}
{Nagao}, T., {Maiolino}, R., {Marconi}, A., \& {Matsuhara}, H. 2011, \aap, 526,
  A149

\bibitem[{{Nakai} {et~al.}(1987){Nakai}, {Hayashi}, {Handa}, {Sofue},
  {Hasegawa}, \& {Sasaki}}]{nakai87}
{Nakai}, N., {Hayashi}, M., {Handa}, T., {et~al.} 1987, \pasj, 39, 685

\bibitem[{{Narayanan} {et~al.}(2008){Narayanan}, {Cox}, {Shirley}, {Dav{\'e}},
  {Hernquist}, \& {Walker}}]{narayanan08}
{Narayanan}, D., {Cox}, T.~J., {Shirley}, Y., {et~al.} 2008, \apj, 684, 996

\bibitem[{{Naylor} {et~al.}(2010){Naylor}, {Bradford}, {Aguirre}, {Bock},
  {Earle}, {Glenn}, {Inami}, {Kamenetzky}, {Maloney}, {Matsuhara}, {Nguyen}, \&
  {Zmuidzinas}}]{naylor10}
{Naylor}, B.~J., {Bradford}, C.~M., {Aguirre}, J.~E., {et~al.} 2010, \apj, 722,
  668

\bibitem[{{Nguyen-Luong} {et~al.}(2016){Nguyen-Luong}, {Nguyen}, {Motte},
  {Schneider}, {Fujii}, {Louvet}, {Hill}, {Sanhueza}, {Chibueze}, \&
  {Didelon}}]{nguyen16}
{Nguyen-Luong}, Q., {Nguyen}, H.~V.~V., {Motte}, F., {et~al.} 2016, \apj, 833,
  23

\bibitem[{{Nguyen-Q-Rieu} {et~al.}(1989){Nguyen-Q-Rieu}, {Nakai}, \&
  {Jackson}}]{nguyen89}
{Nguyen-Q-Rieu}, {Nakai}, N., \& {Jackson}, J.~M. 1989, \aap, 220, 57

\bibitem[{{Olivares E.} {et~al.}(2010){Olivares E.}, {Hamuy}, {Pignata},
  {Maza}, {Bersten}, {Phillips}, {Suntzeff}, {Filippenko}, {Morrel},
  {Kirshner}, \& {Matheson}}]{olivares10}
{Olivares E.}, F., {Hamuy}, M., {Pignata}, G., {et~al.} 2010, \apj, 715, 833

\bibitem[{{Omont}(2007)}]{omont07}
{Omont}, A. 2007, Reports on Progress in Physics, 70, 1099

\bibitem[{{Onodera} {et~al.}(2010){Onodera}, {Kuno}, {Tosaki}, {Kohno},
  {Nakanishi}, {Sawada}, {Muraoka}, {Komugi}, {Miura}, {Kaneko}, {Hirota}, \&
  {Kawabe}}]{onodera10}
{Onodera}, S., {Kuno}, N., {Tosaki}, T., {et~al.} 2010, \apjl, 722, L127

\bibitem[{{Onus} {et~al.}(2018){Onus}, {Krumholz}, \& {Federrath}}]{onus18}
{Onus}, A., {Krumholz}, M.~R., \& {Federrath}, C. 2018, ArXiv e-prints,
  arXiv:1801.09952

\bibitem[{{Origlia} {et~al.}(2004){Origlia}, {Ranalli}, {Comastri}, \&
  {Maiolino}}]{origlia04}
{Origlia}, L., {Ranalli}, P., {Comastri}, A., \& {Maiolino}, R. 2004, \apj,
  606, 862

\bibitem[{{Papadopoulos}(2007)}]{papadopoulos07}
{Papadopoulos}, P.~P. 2007, \apj, 656, 792

\bibitem[{{Poznanski} {et~al.}(2009){Poznanski}, {Butler}, {Filippenko},
  {Ganeshalingam}, {Li}, {Bloom}, {Chornock}, {Foley}, {Nugent}, {Silverman},
  {Cenko}, {Gates}, {Leonard}, {Miller}, {Modjaz}, {Serduke}, {Smith}, {Swift},
  \& {Wong}}]{poznanski09}
{Poznanski}, D., {Butler}, N., {Filippenko}, A.~V., {et~al.} 2009, \apj, 694,
  1067

\bibitem[{{Privon} {et~al.}(2015){Privon}, {Herrero-Illana}, {Evans},
  {Iwasawa}, {Perez-Torres}, {Armus}, {D{\'{\i}}az-Santos}, {Murphy},
  {Stierwalt}, {Aalto}, {Mazzarella}, {Barcos-Mu{\~n}oz}, {Borish}, {Inami},
  {Kim}, {Treister}, {Surace}, {Lord}, {Conway}, {Frayer}, \&
  {Alberdi}}]{privon15}
{Privon}, G.~C., {Herrero-Illana}, R., {Evans}, A.~S., {et~al.} 2015, \apj,
  814, 39

\bibitem[{{Radburn-Smith} {et~al.}(2011){Radburn-Smith}, {de Jong}, {Seth},
  {Bailin}, {Bell}, {Brown}, {Bullock}, {Courteau}, {Dalcanton}, {Ferguson},
  {Goudfrooij}, {Holfeltz}, {Holwerda}, {Purcell}, {Sick}, {Streich}, {Vlajic},
  \& {Zucker}}]{radburn11}
{Radburn-Smith}, D.~J., {de Jong}, R.~S., {Seth}, A.~C., {et~al.} 2011, \apjs,
  195, 18

\bibitem[{{Reiter} {et~al.}(2011){Reiter}, {Shirley}, {Wu}, {Brogan},
  {Wootten}, \& {Tatematsu}}]{reiter11}
{Reiter}, M., {Shirley}, Y.~L., {Wu}, J., {et~al.} 2011, \apjs, 195, 1

\bibitem[{{Riechers} {et~al.}(2011){Riechers}, {Walter}, {Carilli}, {Cox},
  {Weiss}, {Bertoldi}, \& {Menten}}]{riechers11a}
{Riechers}, D.~A., {Walter}, F., {Carilli}, C.~L., {et~al.} 2011, \apj, 726, 50

\bibitem[{{Riechers} {et~al.}(2010){Riechers}, {Wei{\ss}}, {Walter}, \&
  {Wagg}}]{riechers10}
{Riechers}, D.~A., {Wei{\ss}}, A., {Walter}, F., \& {Wagg}, J. 2010, \apj, 725,
  1032

\bibitem[{{Roberts-Borsani} {et~al.}(2017){Roberts-Borsani},
  {Jim{\'e}nez-Donaire}, {Dapr{\`a}}, {Alatalo}, {Aretxaga},
  {{\'A}lvarez-M{\'a}rquez}, {Baker}, {Fujimoto}, {Gallardo}, {Gralla},
  {Hilton}, {Hughes}, {Jim{\'e}nez}, {Laporte}, {Marriage}, {Nati}, {Rivera},
  {Sievers}, {Wei{\ss}}, {Wilson}, {Wollack}, \& {Yun}}]{roberts17}
{Roberts-Borsani}, G.~W., {Jim{\'e}nez-Donaire}, M.~J., {Dapr{\`a}}, M.,
  {et~al.} 2017, \apj, 844, 110

\bibitem[{{Salak} {et~al.}(2013){Salak}, {Nakai}, {Miyamoto}, {Yamauchi}, \&
  {Tsuru}}]{salak13}
{Salak}, D., {Nakai}, N., {Miyamoto}, Y., {Yamauchi}, A., \& {Tsuru}, T.~G.
  2013, \pasj, 65, 66

\bibitem[{{Sanders} {et~al.}(2003){Sanders}, {Mazzarella}, {Kim}, {Surace}, \&
  {Soifer}}]{sanders03}
{Sanders}, D.~B., {Mazzarella}, J.~M., {Kim}, D.-C., {Surace}, J.~A., \&
  {Soifer}, B.~T. 2003, \aj, 126, 1607

\bibitem[{{Sandstrom} {et~al.}(2013){Sandstrom}, {Leroy}, {Walter}, {Bolatto},
  {Croxall}, {Draine}, {Wilson}, {Wolfire}, {Calzetti}, {Kennicutt}, {Aniano},
  {Donovan Meyer}, {Usero}, {Bigiel}, {Brinks}, {de Blok}, {Crocker}, {Dale},
  {Engelbracht}, {Galametz}, {Groves}, {Hunt}, {Koda}, {Kreckel}, {Linz},
  {Meidt}, {Pellegrini}, {Rix}, {Roussel}, {Schinnerer}, {Schruba}, {Schuster},
  {Skibba}, {van der Laan}, {Appleton}, {Armus}, {Brandl}, {Gordon}, {Hinz},
  {Krause}, {Montiel}, {Sauvage}, {Schmiedeke}, {Smith}, \&
  {Vigroux}}]{sandstrom13}
{Sandstrom}, K.~M., {Leroy}, A.~K., {Walter}, F., {et~al.} 2013, \apj, 777, 5

\bibitem[{{Sch{\"o}ier} {et~al.}(2005){Sch{\"o}ier}, {van der Tak}, {van
  Dishoeck}, \& {Black}}]{schoier05}
{Sch{\"o}ier}, F.~L., {van der Tak}, F.~F.~S., {van Dishoeck}, E.~F., \&
  {Black}, J.~H. 2005, \aap, 432, 369

\bibitem[{{Schruba}(2013)}]{schruba13}
{Schruba}, A. 2013, in IAU Symposium, Vol. 292, Molecular Gas, Dust, and Star
  Formation in Galaxies, ed. T.~{Wong} \& J.~{Ott}, 311--318

\bibitem[{{Seaquist} \& {Frayer}(2000)}]{seaquist00}
{Seaquist}, E.~R., \& {Frayer}, D.~T. 2000, \apj, 540, 765

\bibitem[{{Shi} {et~al.}(2011){Shi}, {Helou}, {Yan}, {Armus}, {Wu}, {Papovich},
  \& {Stierwalt}}]{shi11}
{Shi}, Y., {Helou}, G., {Yan}, L., {et~al.} 2011, \apj, 733, 87

\bibitem[{{Shi} {et~al.}(2018){Shi}, {Yan}, {Armus}, {Gu}, {Helou}, {Qiu},
  {Gwyn}, {Stierwalt}, {Fang}, {Chen}, {Zhou}, {Wu}, {Zheng}, {Zhang}, {Gao},
  \& {Wang}}]{shi18}
{Shi}, Y., {Yan}, L., {Armus}, L., {et~al.} 2018, \apj, 853, 149

\bibitem[{{Shimajiri} {et~al.}(2017){Shimajiri}, {Andr{\'e}}, {Braine},
  {K{\"o}nyves}, {Schneider}, {Bontemps}, {Ladjelate}, {Roy}, {Gao}, \&
  {Chen}}]{shimajiri17}
{Shimajiri}, Y., {Andr{\'e}}, P., {Braine}, J., {et~al.} 2017, \aap, 604, A74

\bibitem[{{Solomon} {et~al.}(1997){Solomon}, {Downes}, {Radford}, \&
  {Barrett}}]{solomon97}
{Solomon}, P.~M., {Downes}, D., {Radford}, S.~J.~E., \& {Barrett}, J.~W. 1997,
  \apj, 478, 144

\bibitem[{{Solomon} {et~al.}(1992){Solomon}, {Radford}, \&
  {Downes}}]{solomon92}
{Solomon}, P.~M., {Radford}, S.~J.~E., \& {Downes}, D. 1992, \nat, 356, 318

\bibitem[{{Solomon} \& {Vanden Bout}(2005)}]{solomon05}
{Solomon}, P.~M., \& {Vanden Bout}, P.~A. 2005, \araa, 43, 677

\bibitem[{{Sorai} {et~al.}(2000){Sorai}, {Nakai}, {Kuno}, {Nishiyama}, \&
  {Hasegawa}}]{sorai00}
{Sorai}, K., {Nakai}, N., {Kuno}, N., {Nishiyama}, K., \& {Hasegawa}, T. 2000,
  \pasj, 52, 785

\bibitem[{{Spergel} {et~al.}(2007){Spergel}, {Bean}, {Dor{\'e}}, {Nolta},
  {Bennett}, {Dunkley}, {Hinshaw}, {Jarosik}, {Komatsu}, {Page}, {Peiris},
  {Verde}, {Halpern}, {Hill}, {Kogut}, {Limon}, {Meyer}, {Odegard}, {Tucker},
  {Weiland}, {Wollack}, \& {Wright}}]{spergel07}
{Spergel}, D.~N., {Bean}, R., {Dor{\'e}}, O., {et~al.} 2007, \apjs, 170, 377

\bibitem[{{Spilker} {et~al.}(2014){Spilker}, {Marrone}, {Aguirre}, {Aravena},
  {Ashby}, {B{\'e}thermin}, {Bradford}, {Bothwell}, {Brodwin}, {Carlstrom},
  {Chapman}, {Crawford}, {de Breuck}, {Fassnacht}, {Gonzalez}, {Greve},
  {Gullberg}, {Hezaveh}, {Holzapfel}, {Husband}, {Ma}, {Malkan}, {Murphy},
  {Reichardt}, {Rotermund}, {Stalder}, {Stark}, {Strandet}, {Vieira},
  {Wei{\ss}}, \& {Welikala}}]{spilker14}
{Spilker}, J.~S., {Marrone}, D.~P., {Aguirre}, J.~E., {et~al.} 2014, \apj, 785,
  149

\bibitem[{{Usero} {et~al.}(2015){Usero}, {Leroy}, {Walter}, {Schruba},
  {Garc{\'{\i}}a-Burillo}, {Sandstrom}, {Bigiel}, {Brinks}, {Kramer},
  {Rosolowsky}, {Schuster}, \& {de Blok}}]{usero15}
{Usero}, A., {Leroy}, A.~K., {Walter}, F., {et~al.} 2015, \aj, 150, 115

\bibitem[{{Wang} {et~al.}(2011){Wang}, {Zhang}, \& {Shi}}]{wang11}
{Wang}, J., {Zhang}, Z., \& {Shi}, Y. 2011, \mnras, 416, L21

\bibitem[{{Warren} {et~al.}(2010){Warren}, {Wilson}, {Israel}, {Serjeant},
  {Bendo}, {Brinks}, {Clements}, {Irwin}, {Knapen}, {Leech}, {Matthews},
  {M{\"u}hle}, {Mortimer}, {Petitpas}, {Sinukoff}, {Spekkens}, {Tan},
  {Tilanus}, {Usero}, {van der Werf}, {Vlahakis}, {Wiegert}, \&
  {Zhu}}]{warren10}
{Warren}, B.~E., {Wilson}, C.~D., {Israel}, F.~P., {et~al.} 2010, \apj, 714,
  571

\bibitem[{{Wei{\ss}} {et~al.}(2007){Wei{\ss}}, {Downes}, {Neri}, {Walter},
  {Henkel}, {Wilner}, {Wagg}, \& {Wiklind}}]{weiss07}
{Wei{\ss}}, A., {Downes}, D., {Neri}, R., {et~al.} 2007, \aap, 467, 955

\bibitem[{{Wei{\ss}} {et~al.}(2003){Wei{\ss}}, {Henkel}, {Downes}, \&
  {Walter}}]{weiss03}
{Wei{\ss}}, A., {Henkel}, C., {Downes}, D., \& {Walter}, F. 2003, \aap, 409,
  L41

\bibitem[{{Wei{\ss}} {et~al.}(2005){Wei{\ss}}, {Walter}, \&
  {Scoville}}]{weiss05}
{Wei{\ss}}, A., {Walter}, F., \& {Scoville}, N.~Z. 2005, \aap, 438, 533

\bibitem[{{Wild} {et~al.}(1992){Wild}, {Harris}, {Eckart}, {Genzel}, {Graf},
  {Jackson}, {Russell}, \& {Stutzki}}]{wild92}
{Wild}, W., {Harris}, A.~I., {Eckart}, A., {et~al.} 1992, \aap, 265, 447

\bibitem[{{Wilson} {et~al.}(2012){Wilson}, {Warren}, {Israel}, {Serjeant},
  {Attewell}, {Bendo}, {Butner}, {Chanial}, {Clements}, {Golding}, {Heesen},
  {Irwin}, {Leech}, {Matthews}, {M{\"u}hle}, {Mortier}, {Petitpas},
  {S{\'a}nchez-Gallego}, {Sinukoff}, {Shorten}, {Tan}, {Tilanus}, {Usero},
  {Vaccari}, {Wiegert}, {Zhu}, {Alexander}, {Alexander}, {Azimlu}, {Barmby},
  {Brar}, {Bridge}, {Brinks}, {Brooks}, {Coppin}, {C{\^o}t{\'e}},
  {C{\^o}t{\'e}}, {Courteau}, {Davies}, {Eales}, {Fich}, {Hudson}, {Hughes},
  {Ivison}, {Knapen}, {Page}, {Parkin}, {Rigopoulou}, {Rosolowsky}, {Seaquist},
  {Spekkens}, {Tanvir}, {van der Hulst}, {van der Werf}, {Vlahakis}, {Webb},
  {Weferling}, \& {White}}]{wilson12}
{Wilson}, C.~D., {Warren}, B.~E., {Israel}, F.~P., {et~al.} 2012, \mnras, 424,
  3050

\bibitem[{{Wu} {et~al.}(2005){Wu}, {Evans}, {Gao}, {Solomon}, {Shirley}, \&
  {Vanden Bout}}]{wu05}
{Wu}, J., {Evans}, II, N.~J., {Gao}, Y., {et~al.} 2005, \apjl, 635, L173

\bibitem[{{Wu} {et~al.}(2010){Wu}, {Evans}, {Shirley}, \& {Knez}}]{wu10}
{Wu}, J., {Evans}, II, N.~J., {Shirley}, Y.~L., \& {Knez}, C. 2010, \apjs, 188,
  313

\bibitem[{{Wu} {et~al.}(2014){Wu}, {Tully}, {Rizzi}, {Dolphin}, {Jacobs}, \&
  {Karachentsev}}]{wu14}
{Wu}, P.-F., {Tully}, R.~B., {Rizzi}, L., {et~al.} 2014, \aj, 148, 7

\bibitem[{{Yamada} {et~al.}(2007){Yamada}, {Wada}, \& {Tomisaka}}]{yamada07}
{Yamada}, M., {Wada}, K., \& {Tomisaka}, K. 2007, \apj, 671, 73

\bibitem[{{Zhang} {et~al.}(2014){Zhang}, {Gao}, {Henkel}, {Zhao}, {Wang},
  {Menten}, \& {G{\"u}sten}}]{zhang14}
{Zhang}, Z.-Y., {Gao}, Y., {Henkel}, C., {et~al.} 2014, \apjl, 784, L31

\end{thebibliography}

\end{document}